\documentstyle[11pt,epsf,mssymb]{article}

\begin{document}

\begin {titlepage}
\title {Field Theory reformulated without self-energy parts.\\  Divergence-free classical
electrodynamics} 
\author {M. de Haan \\
Service de Physique Th\'eorique et Math\'ematique\\Universit\'e
libre de Bruxelles\thanks{Campus Plaine
CP 231, Boulevard du Triomphe, 1050
Bruxelles, Belgique. email: mdehaan@ulb.ac.be}, Brussels, Belgium. }

\maketitle

\begin{abstract}

A manifestly gauge-invariant hamiltonian formulation of classical electrodynamics has
been shown to be relativistic invariant by the construction of the adequate generators of the
Poincare Lie algebra \cite{BP74}. 
The original formulation in terms of reduced distribution functions for the particles and the fields is
applied here to the case of two charges interacting through a classical electrodynamical field.
On the other hand, we have been able in previous work to
introduce irreversibility at the fundamental level of description
\cite{dH04a} by reformulating field theory without self-energy parts by integrating  
all processes associated with self-energy in a kinetic operator, while keeping the equivalence with
the original description.  
In this paper, the two approaches are combined to provide a formalism that enables
the use of methods of statistical physics \cite{RB75} to tackle the problem of the divergence of
the self-mass. 
Our approach leads to expressions that are finite even for point-like charged particles: 
the limit of a infinite cutoff can be taken in an harmless way on self consistent equations.
In order to check our theory, we recover the power dissipated by radiation in
geometries where the usual mass divergence does not play a role.

\end{abstract}

\end{titlepage}

\def\theequation{\thesection.\arabic{equation}}

\section {Introduction}

The derivation of an equation of motion of an electron that includes its reaction to the self-field 
has been initiated by Abraham and Lorentz hundred years ago and is still 
 a controversal matter \cite{FR03}, \cite{RC03}.
The main problem is the presence of divergences associated with point-like charged
particles. 
A way of removing them has to be devised without entering in
trouble with the special theory of relativity (see Ref. \cite{EP04} for a recent review and a relevant
bibliography).
The derivation of the self-force based on energy conservation \cite{JJ98} avoids that problem:
the power emitted in the radiating field is  due to the work of the radiative
reaction force.

In the usual derivation, a given motion is prescribed to the  charge
and  the potentials of Li\'enard-Wiechert associated with it are computed.
The self-fields are then derived and their expression used to get the reaction on the motion of the
charged particle, leading to an (infinite) term interpretated as a self-mass. 
But an infinite self-mass prevents the acceleration, hence the paradox. 
An approach that uses a finite expansion in the charge  does not allow 
to correct the situation.
In a theory without divergence, a natural cut-off for the wave numbers, proportionnal to the inverse
of the classical radius of the electron, should appear but that property cannot be obtained in a simple
expansion in the charge since the cut-off value should be proportionnal to the square of the inverse
of the charge.
Moreover, the use of an effective frequency dependent mass $M(\omega)$
\cite{JJ98} leads to an equation that is not free of runaway solutions.

In different contexts, methods of statistical mechanics \cite{RB75} enable, through resummations of
formally divergent class of diagrams, to get relevant finite physical results.
Here we present a formalism adapting those methods for problems in classical electrodynamics.

A manifestly gauge-invariant hamiltonian formulation  has been developed for
systems composed by point particles and  fields, the state of which is now described 
by reduced distribution functions \cite{BP74}. 
The dynamical variables are the positions and mechanical momentums for the charged point
particles  and the transverse components of the electric and magnetic fields. 
Potentials  are absent in that formulation.  
The Coulomb interaction takes into account the longitudinal part of the electric field. 
A generalized Liouville equation for the reduced distribution functions is derived.
It provides a statistical description that takes into account the Lorentz force between the particles
and the Maxwell equations for the fields.  
The formalism formally ressembles a statistical description of charged particles in Coulomb gauge
but with a different interpretation and the guarantee of satisfying the principle of special relativity. 
The relativistic invariance is proved  by the explicit construction of
the generators of the Poincar\'e Lie algebra.

While Balescu and Poulain have developed their formalism for an arbitrary number of particles,
described by reduced distribution functions, we can apply it as such in the simplest case of two
charged point particles.
They thus interact through the Coulomb interaction and the classical transverse electrodynamical
field (electric and magnetic).
An alternative possibility is to consider a single charge in interaction with a Coulomb
potential (due for instance to an infinitely massive particle) at the origin of the coordinates 
but the translation invariance is then immediately broken. 
At the final stage, for the sake of interpretation, we will consider the limit of our expressions when
one of the two particles becomes very massive.
Working with two particles avoids the consideration of an external force to
accelerate the particles: the relativistic and gauge invariance is therefore  preserved.
A prescribed motion for the charged particle is therefore avoided. 
The consideration of an incident transverse field is also relevant to the problem but is not treated
here.

The Balescu-Poulain formulation seems therefore an adequate starting point to deal with classical
electrodynamics thanks to its intrinsic properties: namely relativistic invariance, explicit gauge
invariance. 
The formulation is statistical:  the particles and the modes of the field are described by
distribution functions.
The distribution functions associated  with the particles can be spatially extended.
A particle does not interact with the electric longitudinal field its generates: the Coulomb
interaction is considered only between different particles.
The present paper starts with the results of the last section of the paper of R. Balescu and M. Poulain
\cite{BP74}.

A theory of  subdynamics has been introduced thirty years ago by the Brussels group (see e.g.
\cite{PGHR73}, \cite{RB75}) for a dynamics provided by the Liouville-von Neuman equation.
A setback of that approach is a limitation on the class of possible initial conditions  since they have
to belong to the subdynamics.
To avoid the trap, we have introduced the so-called single subdynamics approach \cite{dHG00b} based
on the existence of self-energy contributions to the dynamics.
In that way, we obtain a reformulation of field theory that excludes self-energy contributions in the
dynamics. However, being able to accommodate also initial conditions 
outside the scope of the original dynamics: our dynamics is larger that the initial formulation.
Since the formal properties of the subdynamics do not depend on a particular realisation  of the
operators, we have picked up all the formal properties without a need to redemonstrate them.
The so-called single subdynamics approach has been illustrated first in a quantal non relativistic
framework \cite{dHG98}, \cite{dHG00a}, \cite{dHG02}, \cite{dHG03}, \cite{dH04a}.
All the claimed formal properties have been subject to an explicit check for exactly soluble
models.
These basis  ideas can also be applied in the present classical relativistic context.

The adequate way of dealing with the self-field is provided automatically by the single subdynamics
approach. 
The dynamics is first extended to be able to distinguish the self-field contributions from the other.
A subdynamics, inspired by the formalism developed at Brussels \cite{RB75} enables the obtention
of dynamical equations of motion in which the self-field does no longer appear.
It has been proven \cite{dHG03} that the description exactly contains the original
description and that the effects of the self-field are now taken into account in the new generator of
motion. 
The relevant subdynamics  incorporates all the features of usual CED.
That description therefore includes not only the original dynamics
but could also include a more general class of initial conditions, enabling the inclusion of irreversibility 
at a fundamental level.

Here, we deliberately restrict ourselves to the derivation of the  closed irreversible evolution
equations  for the interacting charged particles, in the absence of incident field from an outside
source, and  to the obtention of the emitted fields (velocity and acceleration fields) at the lowest
non-vanishing order.

Our paper is structured as follows.
First, we present the Balescu-Poulain's formalism and derive the evolution equations for all reduced
distribution functions defining the state of the system. 
The  basic idea for constructing the single subdynamics in CED is the use of a distinction
between real and virtual fields (the virtual field forms the self-field).
We propose an extension of the dynamics suitable for our purpose and the constitutive relations that
connects the original and extended dynamics are displayed.
The elements ot the extended dynamics bear a tilde accent.

The kinetic operator $\tilde\Theta$,  considered in section 3, describes the closed evolution
of the distribution functions that do not  involve the self-field. 
Their elements are evaluated from the corresponding
vacuum-vacuum elements of the subdynamics operator  $\tilde \Sigma(t)$.
The first non-vanishing contribution appears at the second order in the interaction with the
transverse fields, without considering, in the first step, the influence of the Coulomb interaction
between the charged particles.  
The various steps of the derivation are illustrated  and the final expression for $\tilde\Theta$ is given
in Appendix B.
All the elements are known to examine the putative second order mass correction  for the charged
particles, that is found to vanish.  

A non-vanishing contributions to the kinetic operator, reflecting the presence of the effect of the
transverse self-field, requires to consider either a non-vanishing incident transverse field, either  
a Coulomb interaction between the charged particles or either the mutual influence of the transverse
emitted  field: the particle has to be accelerated to receive a radiation reaction force.

To get a better insight of the previous result, we take another road in $\S$4.
The kinetic operator can indeed be also evaluated from the knowledge of the self-field determined
by the so-called creation operator. 
The value of the self-field at the location
of the particle induces its self-interaction. 
Since the equivalence conditions require the equality
of the emitted and self-field, the creation operator provides us moreover with the expression of the
emitted field. 
Correlation-vacuum elements of the resolvent are considered for evaluating the
elements of the subdynamics.
A simple computation enables to get explicitly the expression of the common value of the
Fourier transform of the emitted and self-field. 

To obtain a source of acceleration and to prepare an easy comparison with the usal
approaches,  the first order effect of the Coulomb acceleration will also be computed in $\S$5 and 
$\S$6 from two different ways: the direct consideration of the kinetic operator and the recourse to
the creation operator for the self-field.  

The direct computation of  the kinetic operator is performed in the next
section $\S$5  from the vacuum-vacuum elements of the resolvent acting in absence of field (field
vacuum).  
All relative orders of the vertices have to be considered: the
Coulomb interaction can {\it a priori} take  place before, after or between the two interactions
with the transverse field. 
Only the last two circonstances lead to a non-vanishing contribution.
Indeed, when the Coulomb interaction takes  place after the two interactions with the transverse
field, we receive as factor, as expected, the previous vanishing second order contribution to the
kinetic operator. 
The computation, although lenghty, is straightforward.
 
For a consistency check, in $\S$6 we  consider the creation operator at first
order in field-particle interaction and  first order  in the Coulomb interaction.
This enables to get the effect of the acceleration, due to the Coulomb interaction, to the self-field,
hence to the retroaction of the emitted field  on the accelerated particle.
From the equivalence conditions, we deduce for all points
the field emitted during the acceleration of the particle.
If we use that expression in the kinetic equation, we recover the previous result.
 From its expression at the localisation of the particle, the power emitted can be computed.

Our expressions are analysed in $\S$7. 
We consider a situation in which the distribution functions of the charged particles are
infinitely sharp in configuration and momentum space, with an absence of free field. 
One particle is then considered as infinitely heavy and we use  the referential in which the heavy
particle is at rest. 
In the geometry where the position and velocity vectors are orthogonal, the power dissipated by
the field due to the motion of the light particle can be computed exactly: all integrals can be
performed. In other geometries, we do not avoid the usual divergence.
This is natural since our approach contains the usual formalism and no resummation has been
performed yet.
The usual result is  explicitly recovered as a particular case in small velocities circonstances.
Indeed, under the equivalence conditions, both theories provide the same equations for the
motion of the charged particles.

The finiteness of the theory through resummations is considered in $\S8$.
We start from the divergent contribution for the self-field at
the lowest order in the charge. 
We regulate it by introducing a cut-off function, with a cut-off value
$K_c$ to smoothen the contributions from very high wave numbers.
That contribution is the first term of a series that can be formally resummed, in a
self consistent way.
We then show that the limit $K_c\to \infty$ can be performed in an harmless way.
The effective cut-off resulting from  the solutions of the  non-linear equations is naturally linked to
the classical radius of the electron. 
Our final expressions do not admit a simple expansion in the charge, since the dependence
of  the effective cut-off in the charge $e$ is in $\frac1{e^2}$. 
This result is obtained through an analysis that uses several steps.

Some conclusions and perspectives are considered in the last section $\S$9.

\section{ Electrodynamics in the manifestly gauge invariant Balescu-Poulain formalism}   

\setcounter{equation}{0}

In this section,  we define the model for the description of the two charges
in interaction with the electromagnetic field.
We use the approach by R. Balescu and M. Poulain \cite{BP74}. 
The only difference is that for the description of matter, we do not deal with a reduced
formalism but keep the two-particle distribution function.
Although the transposition is straightforward, we will present it in the following subsection, using
their notations (and expressions whenever possible). 

\subsection{The Balescu-Poulain formalism}
The state of the system is described by a {\it distribution vector} ${\cal F}$, i.e. by a collection of
 functions describing two different particles and the reduced distribution of  $m$ field oscillators,
describing the transverse field components which are the only ones that appear explicitly:
\begin{equation}
{\cal F}=\{f_{11[m]}(x^{(1)}, x^{(2)};\chi^{[1]},\dots,\chi^{[m]};{\bf k}^{[1]},\dots, {\bf k}^{[m]})\}
\qquad; m=0,1,2,\dots
\label{2.1}
\end{equation}
For $m=0$, the system does not contain fields variables.
An obvious convention in the notation is implicit for $m=0$.
Here $x^{(j)}$ denotes the coordinates $({\bf q}^{(j)},{\bf p}^{(j)})$ of particle $j$, and
$\chi^{[j]}$ denotes the variables describing a given field oscillator associated with the
wavevector ${\bf k}^{[j]}$: $(\eta_{\alpha}^{[j]}, \xi_{\alpha}^{[j]}, \alpha=1,2)$ \footnote
{In contrast with \cite{dH91}, the reduction is not performed up to the level of each polarized
mode,  in the same way that  reduced distribution functions for the particles are not considered
to only one component of the velocity. This procedure ensures more easily the rotationnal invariance of
the treatment.}
that are the action  ($\eta_{\alpha}^{[j]}$) and angle variables ($\xi_{\alpha}^{[j]}$) associated with the
oscillator characterized by the wave number ${\bf k}^{[j]}$.

If two mutually orthogonal unit vectors, or "polarization vectors", ${\bf e}^{\alpha}({\bf k})$ associated with a given wavevector ${\bf k}$ are introduced such that, together with the unit
vector $\frac{{\bf k}}{k}$, they form a right-handed cartesian frame, the electromagnetic fields are
expressed as follows in these variables.
\begin{equation}
{\bf E}^{{\bot }}({\bf x})=\frac1{(2\pi)^{\frac32}}\sum_{\alpha=1,2} \sum_{a=\pm 1}
\int d^3k\, k^{\frac12} {\bf e}^{\alpha}({\bf k}) \eta_{\alpha}^{\frac12}({\bf k})
\exp \{ia[{\bf k}.{\bf x}-2\pi\xi_{\alpha}({\bf k})]\},
\label{2.2}
\end{equation}
\begin{equation}
{\bf B}({\bf x})=\frac1{(2\pi)^{\frac32}}\sum_{\alpha=1,2} \sum_{a=\pm 1}
\int d^3k\, k^{\frac12}(-1)^{\alpha'} {\bf e}^{\alpha'}({\bf k}) \eta_{\alpha}^{\frac12}({\bf k})
\exp \{ia[{\bf k}.{\bf x}-2\pi\xi_{\alpha}({\bf k})]\},
\label{2.3}
\end{equation}
where $\alpha'=2$ for $\alpha=1$ and $\alpha'=1$ for $\alpha=2$.

The dynamical functions of the system are described by a set ${\cal B}$:
\begin{equation}
{\cal B}=\{b_{11[m]}(x^{(1)}, x^{(2)};\chi^{[1]},\dots,\chi^{[m]};{\bf k}^{[1]},\dots, {\bf k}^{[m]})\}
\qquad; m=0,1,2,\dots
\label{2.4}
\end{equation}

The average value of an element $b_{11[m]}$ of ${\cal B}$ is calculated by the following
formula:
\begin{eqnarray}
<b_{11[m]}>&=&\int d^3{\bf k}^{[1]}\dots d^3{\bf k}^{[m]} \int d^4\chi^{[1]} \dots
d^4\chi^{[m]} \int d^6x^{(1)}\, d^6x^{(2)}
\nonumber \\ &\times&b_{11[m]}(x^{(1)}, x^{(2)};\chi^{[1]},\dots,\chi^{[m]};{\bf k}^{[1]},\dots, {\bf k}^{[m]})
\nonumber \\ &\times&f_{11[m]}(x^{(1)}, x^{(2)};\chi^{[1]},\dots,\chi^{[m]};{\bf k}^{[1]},\dots, {\bf
k}^{[m]}).
\label{2.5}
\end{eqnarray}

It is well known that to each generator $G$ of the Poincar\'e Lie algeba corresponds an infinite
hierarchy of equations describing the transformation properties of the reduced distribution
functions \cite{CJS63}.  These equations can be written compactly as 
\begin{equation}
\partial_g  {\cal F} ={\cal L}_G {\cal F},
\label{2.6}
\end{equation}
where ${\cal L}_G$ is a matrix operator.
The components of this equation are written as 
\begin{equation}
\partial_g f_{11[m]}=\sum_{m'=0}^{\infty} <11[m]|{\cal L}_G|11[m']>f_{11[m']}.
\label{2.7}
\end{equation}
The matrix elements entering these equations are obtained as in \cite{BPB71} and listed below,
considering separately the three contributions corresponding to the splitting of the liouvillians in
three terms, describing respectively free particles $L_G^{0P}$, free field $L_G^{0F}$ and
interactions $L_G'$.

For the free particles, we have:
\begin{equation}
 <11[m]|{\cal L}_G^{OP}|11[m']>=\delta_{mm'}\left(L_G^{0(1)}+L_G^{0(2)}
\right).
\label{2.8}
\end{equation}
In this work, we consider only the generator corresponding to the time
translation  ($g=t$, $G=H$) and get, using Einstein convention for the summation:
\begin{equation}
L_H^{0(j)}=-v_r^{(j)}\frac{\partial}{\partial q_r^{(j)}},
\label{2.9}
\end{equation}
where the velocity $v_r^{(j)}$ is connected with the mechanical momentum $p_r^{(j)}$ in the usual
way (in the  units chosen,  $c=1$ and  the div (divergence) of the electric field vector is $4\pi$ the
charge density):
\begin{equation}
v_r^{(j)}=\frac{p_r^{(j)}}{(m_j^2 +p_s^{(j)}p_s^{(j)})^{\frac12}}.
\label{2.10}
\end{equation}
For the free field, we have:
\begin{equation}
 <11[m]|{\cal L}_G^{0F}|11[m']>=\delta_{mm'} \sum_{i=1}^m L_G^{0[i]},
\label{2.11}
\end{equation}
\begin{equation}
L_H^{0[i]}=-\frac{1}{2\pi} k^{[i]}\sum_{\alpha=1}^2\frac{\partial}{\partial \xi_{\alpha}^{[i]}}.
\label{2.12}
\end{equation}
For the interaction,
\begin{eqnarray}
 &&<11[m]|{\cal L}'_G|11[m']>=\delta_{mm'}\sum_{i=1}^m\left(L_G^{'1[i]}+L_G^{'2[i]}\right)
+\delta_{m',m} L_G^{'(12)}
\nonumber \\&+&
\delta_{m',m+1}\int'd^3k^{[m+1]} \int d\gamma^{[m+1]}
\left(L_G^{'1[m+1]}+L_G^{'2[m+1]}\right),
\label{2.13}
\end{eqnarray}
where $\int d\gamma^{[m+1]}$ stands for
\begin{equation}
\int d\gamma^{[m+1]}\dots
=\int_0^{\infty}d\eta_1^{[m+1]}\int_0^{\infty}d\eta_2^{[m+1]}
\int_0^1d\xi_1^{[m+1]}\int_0^1d\xi_2^{[m+1]}\dots
\label{2.14}
\end{equation}
The prime on the ${\bf k}$ integral means that the values ${\bf k}^{[m+1]}={\bf k}^{[1]},\dots,{\bf k}^{[m]}$ must be excluded through a principal-part procedure. We have for the
interaction of particle $j$, bearing the charge $e_j$, with the $i$ labeled mode:
\begin{eqnarray}
L_H^{'j[i]}&=&-e_j \frac1{(2\pi)^{\frac32}}
\sum_{\alpha=1,2} 
\sum_{a=\pm 1}
\left(
\frac{\eta_{\alpha}^{[i]}}
{k^{[i]}}
\right)^{\frac12} 
\exp\{ia[{\bf k}^{[i]}.{\bf q}^{(j)}-2\pi\xi_{\alpha}^{[i]}]\}
\nonumber \\&\times&
\left[ 
[k^{[i]} e_r^{(\alpha)[i]}-g^{st} v_s^{(j)}
(e_t^{(\alpha)[i]} k_r^{[i]}-e_r^{(\alpha)[i]} k_t^{[i]})]
\frac{\partial}
{\partial p_r^{(j)}}
\right.\nonumber \\ &&\left.
-({\bf v}^{(j)}.{\bf e}^{(\alpha)[i]})
\left( 
2\pi \frac{\partial}
{\partial \eta_{\alpha}^{[i]}}
-\frac{ia}{2\eta_{\alpha}^{[i]}}\frac{\partial}{\partial \xi_{\alpha}^{[i]}}
\right)
\right].
\label{2.15}
\end{eqnarray}
The elements of the metric tensor $g$ have been chosen as $g_{rs}=g^{rs}=-\delta_{rs}$, $i,r=1\to
3$.
The last matrix element of interest for us describes the Coulomb interaction between the two
charged particles:
\begin{equation}
L_H^{'(12)}=e_1e_2\left(\frac{\partial | {\bf q}^{(1)} -{\bf q}^{(2)} |^{-1}}
{\partial {\bf q}^{(2)}}
\right).\left(\frac{\partial}{\partial {\bf p}^{(1)}}-\frac{\partial}
{\partial {\bf p}^{(2)}}
\right).
\label{2.16}
\end{equation}

\subsection{Enlargement of dynamics}

We proceed now to an enlargement of dynamics as in previous publications \cite{dHG00b},
\cite{dHG03}: 
we multiply the number of
variables on physical ground in such a way that the original dynamics (\ref{2.7}) be included as a
particular case. 
The choice of a particular enlargement is determined by opportunity  linked
to  physical considerations and the properties to be examined \cite{dHG04}.
Since all enlargements provide an equivalent alternative description, that degree of freedom is
welcome.  
In the present paper, focalized on the self-force on each particle, our choice is to define the self-field
with respect to each particle.
If the interest bears on the field far from the two particles, defining the self-field with respect to
both charged particles would be an alternative useful option.

The elements of the enlarged dynamics will be noted by a supplementary
upper index tilde ``$\,\tilde{}\,$", as well for the variables as for the evolution operator.
 Our aim is indeed to eliminate explicit self-interaction processes from the evolution,
while taking their effect into account.
We distinguish formally between 5 varieties of oscillators, based on the recognition of self-energy
parts in the evolution.
To each oscillator $[i]$, we associate a discrete index that determines which
interactions are possible for the oscillator (the index $j$ takes the two values 1 and 2).

\noindent
$[i(s_j)]$ will be the label of an oscillator which has  previously interacted with the particle $j$
and will further interact with it in a future, without interaction with the other particle ($j'\not=j$), and
without playing a role in a measurement: by definition, such oscillator does not play a role in the
computation of the mean values.

\noindent
$[i(e_j)]$ will be the label of an oscillator which has previously interacted with the particle ($j$)
and will no longer interact with it directly: its next interaction should involve  the other particle
($j'$), or it should contribute in the computation of mean values.

\noindent
$[i(f)]$ will be the label of an oscillator mode which has not previously interacted with the particles
(1) or (2).  Its excitation has its origin outside the two charges and such an oscillator is free of
constraints on its interactions: 
either with one of the particle or with an external devise. 
It provide a contribution in the computation of mean values.

The free evolution of those oscillators is the same as in the original dynamics and does not involve a
change in their nature.

The vertices for the computation of $<11[m]|\tilde{{\cal L}}'_H|11[m]>$  involve
$L_H^{'j[i]}$  for all $i:1\to m$. the numerical value will be preserved for the non-vanishing elements.
We have to take into account the (possible) change of nature of the oscillator after the
interaction. 
We introduce indices corresponding to the transition of nature of the field 
($i(e_1f)$ means that a free oscillator $i(f)$ becomes of the emitted $e_1$  variety) and we
have the non-vanishing possibilities:
$\tilde L_H^{'1[i(s_1f)]}$, $\tilde L_H^{'2[i(s_2f)]}$,  
$\tilde L_H^{'1[i(e_1f)]}$, $\tilde L_H^{'2[i(e_2f)]}$, 
$\tilde L_H^{'1[i(e_1e_2)]}$, $\tilde L_H^{'2[i(e_2e_1)]}$,  
$\tilde L_H^{'1[i(s_1e_2)]}$, $\tilde L_H^{'2[i(s_2e_1)]}$,  
$\tilde L_H^{'1[i(s_1s_1)]}$, $\tilde L_H^{'2[i(s_2s_2)]}$, 
$\tilde L_H^{'1[i(e_1s_1]}$, $\tilde L_H^{'2[i(e_2s_2)]}$ 
while the elements  
$\tilde L_H^{'1[i(e_1e_1]}$, $\tilde L_H^{'2[i(e_2e_2)]}$, 
$\tilde L_H^{'1[i(s_1e_1]}$, $\tilde L_H^{'2[i(s_2e_2)]}$, 
 $\tilde L_H^{'1[i(s_1s_2)]}$, $\tilde L_H^{'2[i(s_2s_1)]}$, 
 $\tilde L_H^{'1[i(e_1s_2)]}$, $\tilde L_H^{'2[i(e_2s_1)]}$ 
vanish by construction. 

Other elements, such as 
$\tilde L_H^{'1[i(ff]}$, $\tilde L_H^{'2[i(ff]}$, 
$\tilde L_H^{'1[i(fs_1)]}$, $\tilde L_H^{'2[i(fs_2)]}$,
$\tilde L_H^{'1[i(fe_1]}$, $\tilde L_H^{'2[i(fe_2)]}$, 
$\tilde L_H^{'1[i(fs_2)]}$, $\tilde L_H^{'2[i(fs_1)}$,
$\tilde L_H^{'1[i(fe_2]}$, $\tilde L_H^{'2[i(fe_1)]}$, 
$\tilde L_H^{'1[i(e_2e_2]}$, $\tilde L_H^{'2[i(e_1e_1)]}$, 
$\tilde L_H^{'1[i(s_2e_2]}$, $\tilde L_H^{'2[i(s_1e_1)]}$, 
$\tilde L_H^{'1[i(e_2e_1]}$, $\tilde L_H^{'2[i(e_1e_2)]}$,
$\tilde L_H^{'1[i(s_2s_1)]}$, $\tilde L_H^{'2[i(s_1s_2)]}$,
$\tilde L_H^{'1[i(e_2s_1)]}$, $\tilde L_H^{'2[i(e_1s_2)]}$,
$\tilde L_H^{'1[i(s_2e_1]}$, $\tilde L_H^{'2[i(s_1e_2)]}$, 
$\tilde L_H^{'1[i(s_2f)]}$, $\tilde L_H^{'2[i(s_1f)]}$,  
$\tilde L_H^{'1[i(e_2f)]}$, $\tilde L_H^{'2[i(e_1f)]}$,
$\tilde L_H^{'1[i(s_2s_2)]}$, $\tilde L_H^{'2[i(s_1s_1)]}$, 
$\tilde L_H^{'1[i(e_2s_2)]}$, $\tilde L_H^{'2[i(e_1s_1)]}$ 
vanish obviously since the final label of the oscillator does not bear the name of the interacting
particle.

The vertices for the computation of $<11[m]|\tilde{{\cal L}}'_H|11[m']>$ ($m'\not=m$) involve a
$(m+1)^{th}$ oscillator mode and its disparition from the explicit description. The value of  the
vertices involved, corresponding to $\tilde L_H^{'j[m+1]}$ is the same as the value of $L_H^{'j[m+1]}$:
we have to consider the non-vanishing possibilities for the nature of the $(m+1)^{th}$ oscillator.
The oscillator on which the integration is performed is considered belonging to the self variety and 
$<11[m]|\tilde{{\cal L}}'_H|11[m']>$ will thus involve the following elements:
$\tilde L_H^{'1[m+1(s_1f)]}$, $\tilde L_H^{'2[m+1(s_2f)]}$, $\tilde L_H^{'1[m+1(s_1e_2)]}$, $\tilde
L_H^{'2[m+1(s_2e_1)]}$,
$\tilde L_H^{'1[m+1(s_1s_1)]}$,
$\tilde L_H^{'2[m+1(s_2s_2)]}$, while the elements of $<11[m]|\tilde{{\cal L}}'_H|11[m']>$ involving the
$\tilde L_H^{'1[m+1(s_1e_1)]}$ and $\tilde L_H^{'2[m+1(s_2e_2)]}$ vanish by construction. The other
oscillators $(1\to m)$ are unchanged by the transition vertex.

\subsection{Constitutive relations-Equivalence conditions}

Matrix elements of the evolution operator for an enlarged dynamics involve now the
five varieties of oscillators.
We have to connect the elements of the extended dynamics to the original one. 
The simplest case involves one oscillator only, from the first equation of the hierarchy:
\begin{equation}
\partial_t\tilde f_{11[m]}=\sum_{m'=0}^{\infty} <11[m]|\tilde{{\cal L}}_H|11[m']>\tilde f_{11[m']}.
\label{2.17}
\end{equation}
For $m=0$, we take obviously $ f_{11[0]}=\tilde f_{11[0]}$.
That first equation means:
\begin{eqnarray}
&&\partial_t\tilde f_{11[0]}= <11[0]|\tilde{{\cal L}}_H|11[0]>\tilde f_{11[0]}
+<11[0]|\tilde{{\cal L}}_H|11[1(f)]>\tilde f_{11[1(f)]}
\nonumber \\&&+
<11[0]|\tilde{{\cal L}}_H|11[1(s_1)]>\tilde f_{11[1(s_1)]}
+<11[0]|\tilde{{\cal L}}_H|11[1(s_2)]>\tilde f_{11[1(s_2)]}
\nonumber \\&&+
<11[0]|\tilde{{\cal L}}_H|11[1(e_1)]>\tilde f_{11[1(e_1)]}
+<11[0]|\tilde{{\cal L}}_H|11[1(e_2)]>\tilde f_{11[1(e_2)]}.
\nonumber \\&&
\label{2.18}
\end{eqnarray}
$\tilde{{\cal L}}_H$ is composed of the parts part $\tilde{{\cal L}}^1_H$ and $\tilde{{\cal L}}^2_H$ 
according to the particles involved in the interaction.
$\tilde{{\cal L}}^1_H$ acts on $\tilde f_{11[1(f)]}$, $\tilde f_{11[1(s_1)]}$, $\tilde f_{11[1(e_2)]}$ while
$\tilde{{\cal L}}^2_H$ acts on $\tilde f_{11[1(f)]}$, $\tilde f_{11[1(s_2)]}$, $\tilde f_{11[1(e_1)]}$.
Since we have to recover the equation
\begin{equation}
\partial_t f_{11[0]}= <11[0]|{{\cal L}}_H|11[m']> f_{11[0]}+<11[0]|{{\cal L}}_H|11[m']> f_{11[1]},
\label{2.19}
\end{equation}
we are led to the constitutive relation  \cite{dHG00b}, \cite{dHG03}
\begin{equation}
f_{11[1]}=\tilde f_{11[1(f)]}+\tilde f_{11[1(e_1)]}+\tilde f_{11[1(e_2)]}.
\label{2.20}
\end{equation}
Indeed, if the conditions $\tilde f_{11[1(s_1)]}=\tilde f_{11[1(e_1)]}$ and $\tilde f_{11[1(s_2)]}=\tilde
f_{11[1(e_2)]}$ are satisfied at the initial time (equivalence conditions), they will remain satisfied for
all times and we recover (\ref{2.19}) as a particular solution of our set of equations.

Let us consider now the next equations of the hierarchy.
\begin{eqnarray}
\partial_t\tilde f_{11[1(f)]}&=& <11[1(f)]|\tilde{{\cal L}}_H|11[1(f)]>\tilde f_{11[1(f)]}
\nonumber \\&+&<11[1(f)]|\tilde{{\cal L}}_H|11[2(ff)]>\tilde f_{11[2(ff)]}
\nonumber \\&+&<11[1(f)]|\tilde{{\cal L}}_H|11[2(fs_1)]>\tilde f_{11[2(fs_1)]}
\nonumber \\&+&<11[1(f)]|\tilde{{\cal L}}_H|11[2(fs_2)]>\tilde f_{11[2(fs_2)]}
\nonumber \\&+&<11[1(f)]|\tilde{{\cal L}}_H|11[2(fe_1)]>\tilde f_{11[2(fe_1)]}
\nonumber \\&+&<11[1(f)]|\tilde{{\cal L}}_H|11[2(fe_2)]>\tilde f_{11[2(fe_2)]},
\label{2.21}
\end{eqnarray}
\begin{eqnarray}
\partial_t\tilde f_{11[1(s_1)]}&=& <11[1(s_1)]|\tilde{{\cal L}}_H|11[1(s_1)]>\tilde f_{11[1s_1]}
\nonumber \\&+&<11[1(s_1)]|\tilde{{\cal L}}_H|11[1(f)]>\tilde f_{11[1(f)]}
\nonumber \\&+&<11[1(s_1)]|\tilde{{\cal L}}_H|11[1(e_2)]>\tilde f_{11[1(e_2)]}
\nonumber \\&+&<11[1(s_1)]|\tilde{{\cal L}}_H|11[2(s_1f)]>\tilde f_{11[2(s_1f)]}
\nonumber \\&+&<11[1(s_1)]|\tilde{{\cal L}}_H|11[2(s_1f)]>\tilde f_{11[2(s_1s_1)]}
\nonumber \\&+&<11[1(s_1)]|\tilde{{\cal L}}_H|11[2(s_1s_2)]>\tilde f_{11[2(s_1s_2)]}
\nonumber \\&+&<11[1(s_1)]|\tilde{{\cal L}}_H|11[2(s_1e_1)]>\tilde f_{11[2(s_1e_1)]}
\nonumber \\&+&<11[1(s_1)]|\tilde{{\cal L}}_H|11[2(s_1e_2)]>\tilde f_{11[2(s_1e_2)]},
\label{2.22}
\end{eqnarray}
and a similar expression for $\partial_t\tilde f_{11[1(s_2)]}$.
From the equality of the matrix elements, we have also
$\partial_t\tilde f_{11[1(e_1)]}=\partial_t\tilde f_{11[1(s_1)]}$ and 
$\partial_t\tilde f_{11[1(e_2)]}=\partial_t\tilde f_{11[1(s_2)]}$.
Those relations have to be compatible with:
\begin{equation}
\partial_t f_{11[1]}= <11[1]|{{\cal L}}_H|11[1]> f_{11[1]}
+<11[1]|{{\cal L}}_H|11[2]> f_{11[2]}.
\label{2.24}
\end{equation}
We have, for the terms diagonal in the numbers of oscillators:
\begin{eqnarray}
&&\partial_t(\tilde f_{11[1(f)]}+\tilde f_{11[1(e_1)]}+\tilde f_{11[1(e_2)]})_{diag}
\nonumber \\&&=
<11[1(f)]|\tilde{{\cal L}}_H|11[1(f)]>\tilde f_{11[1(f)]}
+<11[1(s_1)]|\tilde{{\cal L}}_H|11[1(s_1)]>\tilde f_{11[1s1]}
\nonumber \\&&
+<11[1(s_1)]|\tilde{{\cal L}}_H|11[1(f)]>\tilde f_{11[1(f)]}
+<11[1(s_1)]|\tilde{{\cal L}}_H|11[1(e_2)]>\tilde f_{11[1(e_2)]}
\nonumber \\&&
+<11[1(s_2)]|\tilde{{\cal L}}_H|11[1(s_2)]>\tilde f_{11[1s2]}
+<11[1(s_2)]|\tilde{{\cal L}}_H|11[1(f)]>\tilde f_{11[1(f)]}
\nonumber \\&&
+<11[1(s_2)]|\tilde{{\cal L}}_H|11[1(e_1)]>\tilde f_{11[1(e_1)]}
\label{2.25}
\end{eqnarray}
For these terms that do not involve a field oscillator,
that equation is manifestly compatible with the previous one.
Let us consider the other contributions involving $\tilde{{\cal L}}^1_H$.
We have its action on  $\tilde f_{11[1s1]}$, $\tilde f_{11[1(f)]}$, $\tilde
f_{11[1(e_2)]}$, and this is compatible with the original equation, thanks to the constitutive relations 
and to the numerical identification of $\tilde f_{11[1s1]}$ with $\tilde f_{11[1e1]}$ inside the
equivalence relations.
The other terms can be treated in a similar way.

That relation (\ref{2.20}) can be easily generalized for two or more
oscillators:
\begin{eqnarray}
f_{11[2]}&=&\tilde f_{11[2(ff)]}+\tilde f_{11[2(e_1f)]}+\tilde f_{11[2(e_2f)]}
+\tilde f_{11[2(fe_1)]}+\tilde f_{11[2(e_1e_1)]}
\nonumber \\&+&\tilde f_{11[2(e_2e_1)]}
+\tilde f_{11[2(fe_2)]}+\tilde f_{11[2(e_1e_2)]}+\tilde f_{11[2(e_2e_2)]}
\label{2.26}
\end{eqnarray}
and similar expressions for the set of all elements $\{\tilde f_{11[i]}\}$.

The consideration of initial conditions that do not satisfy the equivalence conditions requires that
the generators of the Poincar\'e Lie algebra be constructed for the extended dynamics and that point
is beyond our aim in this paper.
However, we believe that the distinction between the self-field and the external field resists a Lorentz
transformation and therefore no problem should arise from the extension of dynamics.
Moreover, the subdynamics operator $\Pi$ has been proved (in another
realisation but within a similar framework) by R. Balescu and L. Brenig \cite{BB71a}
to be relativistically invariant.
Nevertheless, as far as the adequate construction of the Lie brackets for the ten generators of the
Lorentz group has not been performed, we have to restrict ourself to the equivalence case.
When the compatible (or equivalence) conditions are fulfilled, the new dynamics is simply a
reformulation of the original one.

\subsection{Factorisation properties}

The free fields variables are by definition not connected with the charged
particles variables.
Therefore, the initial conditions concerning the free fields  and particles variables 
can be chosen as independent.
The vacuum components $\{\tilde f_{11[i]}\}$ can be factorised into a part, describing the 
particles and the oscillators of the emitted $e_1$ and $e_2$ varieties, and a part describing the
free variety of the field, for instance an incident field that may be or not vanishingly small.
We shall therefore write for instance:
\begin{eqnarray}
\tilde f_{11[1(f)]}&=&\tilde f_{11[0]} \tilde f_{[1(f)]}
\nonumber \\
\tilde f_{11[2(ff)]}&=&\tilde f_{11[0]} \tilde f_{[2(ff)]}
\nonumber \\
\tilde f_{11[2(e_1f)]}&=&\tilde f_{11[1(e_1)]} \tilde f_{[1(f)]}
\label{2.27}
\end{eqnarray}
The free field $\tilde f_{[1(f)]}$ distribution function may be in particular the 
distribution function $\tilde f_{[1(f)]}^V$ describing the absence of field (field vacuum) and 
considered later on for the computation of the effect of the self-field
on the motion of the charged particles. 
In the extended dynamics, the natural choice is to consider for $\tilde f_{[1(f)]}^V$ a distribution
function corresponding to  a field of null amplitude and no phase dependance. 
In those circonstances,  the function $\tilde f_{11[1(s_j)]}$  
receive e.g. contributions form
$\{\tilde f_{11[0]} \tilde f_{[n(ff\dots f)]}\}$ directly through the creation operator 
$<11[1(s_j)]|\tilde C|11[n(ff\dots f)]>$ (Other contributions are written in $\S$4). 

Even in absence of field, the factorisation (\ref{2.27}) is not equivalent to a factorisation
$f_{11[1]}=f_{11[0]}\tilde f_{[1(f)]}^V$in the original dynamics.
In the equivalence conditions, we have numerically that the functions $\tilde f_{11[1(s_j)]}$ and 
$\tilde f_{11[1(e_j)]}$ coincide.
The constitutive relation (\ref{2.20}) requires that 
$f_{11[1]}=\tilde f_{11[1(f)]}+\tilde f_{11[1(e_1)]}+\tilde f_{11[1(e_2)]}$.
Therefore, in the original representation, we are not allowed to consider a factorisation 
$f_{11[1]}=\tilde f_{11[0]}\tilde f_{[1(f)]}^V$ corresponding to the absence of field at the time
considered.
If we impose at some time $f_{11[1]}=\tilde f_{11[0]}\tilde f_{[1(f)]}^V$, at the same time, we have
to consider $\tilde f_{11[1(f)]}= f_{11[0]}\tilde f_{[1(f)]}^V-\tilde f_{11[1(e_1)]}-\tilde f_{11[1(e_2)]}$.
Therefore, we have to admit the presence in $\tilde f_{11[1(f)]}$ of contributions $-\tilde
f_{11[1(e_1)]}-\tilde f_{11[1(e_2)]}$.
When computing the equation of evolution of the charged particles, those terms play a role
directly through, for instance, the element $<11[0]|\tilde{{\cal L}}_H|11[1(f)]>$ of the first equation
of the hierarchy (\ref{2.18}).
That contribution has to be combined with the contribution arising from the kinetic operator
$\tilde\Theta$.
Upon that imposition $f_{11[1]}=\tilde f_{11[0]}\tilde f_{[1(f)]}^V$, it is mandatory to consider also that
contribution to have a valid comparison with the usual results.

In conclusion, we have been able to provide a set of evolution equations for the reduced distribution
functions that enables the explicit identification of the self-field.

\section{ The kinetic operator}   
\subsection{General concept}
\setcounter{equation}{0}

We now proceed to the construction of a subynamics associated with the enlarged dynamics.
First of all, we have to define the vacuum and correlation states.
A correlation state contains at least one self oscillator while the vacuum (of correlation) is defined as
the set  $\{\tilde f_{11[i]}\}$ where all oscillators are of the free $f$ and emitted $e_1$ and $e_2$
varieties.
The construction of the subdynamics rests on that distinction.
All the formal results of the Brussels group, concerning its construction rules and its formal
properties, are  applied directly, with our specific realisation of the operators involved. 

Inside the subdynamics, the vacuum components obey close evolution equations,  namely  kinetic
equations. These equations are not time reversal invariant, hence their name.
For their determination, we can limit ourselves to the consideration of the vacuum-vacuum elements
of the superoperator $\tilde \Sigma(t)$ \cite{dH04a}, its $t=0$ value defining the $\tilde{\Pi}$
operator. 
We take for granted the usual properties of idempotency, factorised structure and
commutation  of $\tilde{\Pi}$ with the evolution operator $\tilde {{\cal L}}$.
The explicit verification of those properties requires the explicit knowledge of all the elements of
$\tilde \Sigma(t)$ \cite{dHG03}, and not only of the vacuum-vacuum ones.

In the enlarged dynamics, the evolution equation takes the form:
\begin{equation}
\partial_t\tilde f_{11[m]}=\sum_{m'=0}^{\infty} <11[m]|\tilde{{\cal L}}_H|11[m']>\tilde f_{11[m']},
\label{3.1}
\end{equation}
which involves all the varieties of the oscillators. 
The kinetic operator $\tilde \Theta$ associated with the
subdynamics provides in an exact way close equations involving the vacuum oscillators only:
\begin{equation}
\partial_tV\tilde {f}_{11[m]}=\sum_{m'=0}^{\infty} <11[m]|\tilde{\Theta}|11[m']>V\tilde {f}_{11[m']}.
\label{3.2}
\end{equation}
The value of that operator can be reached by the direct computation of the vacuum-vacuum
elements of the superoperator $\tilde \Sigma(t)$.

The hierarchical form of the equations (\ref{3.1})  ($m'\ge m$) enables the determination of the elements
of
$\tilde
\Theta$ in a successive way.
The elements of $\tilde \Theta$ that do not involve an oscillator are the same as those of
$\tilde{{\cal L}}_H$, and therefore also the same as ${{\cal L}}_H$.

We proceed to the computation of the first non diagonal element 
\newline\noindent$<11[0]|\tilde \Theta|11[1(f)]>$.
It is based on the evaluation of the corresponding element $<11[0]|\tilde \Sigma(t)|11[1(f)]>$.
Its evaluation is performed in a perturbative way.
The elements will be affected by a couple of upper indices which describes the number of Coulomb interaction and the power of interaction with the oscillators.
The simplest element is of course $<11[0]|\tilde \Sigma(t)|11[1(f)]>^{(0,1)}$, in which the Coulomb
interaction is not considered and only one interaction with the (free) oscillator takes place.
Such element involves no self oscillator and we have trivially: 
\begin{equation}
<11[0]|\tilde \Sigma(t)|11[1(f)]>^{(0,1)}=<11[0]|\exp {\tilde{{\cal L}}}_Ht |11[1(f)]>^{(0,1)}.
\label{3.3}
\end{equation}
The vacuum-vacuum elements of $\tilde \Sigma(0)$ are noted $\tilde{{\cal A}}$ and from the general
relation valid for vacuum-vacuum elements
\begin{equation}
<11[0]|\tilde \Sigma(t)|11[1(f)]>=<11[0]|e^{\tilde \Theta t}\tilde{{\cal A}}|11[1(f)]>,
\label{3.37}
\end{equation} 
we have $<11[0]|\tilde \Theta|11[1(f)]>^{(0,1)}=<11[0]|\tilde{{\cal L}}_H |11[1(f)]>$.

\subsection{Construction rules of $V\tilde \Sigma(t)V$} 

The general construction rules are illustrated on the
first non trivial element  $<11[0]|\tilde \Sigma|11[1(f)]>^{(0,2)}$, in which the Coulomb
interaction is not considered and  two interactions with an oscillator take place.
Such element involves one self oscillator if the two interactions involve the same particle.
If they involve different particles, only physical states are present in the contribution and we have
anew the equivalence of the  corresponding elements of $\tilde \Theta$ and $\tilde{{\cal L}}_H$.
We dispense ourself of a supplementary index and concentrate on the contribution involving a
self oscillator.
We have:
\begin{eqnarray}
&&<11[0]|\tilde \Sigma|11[1(f)]>^{(0,2)}=\frac{-1}{2\pi i} \int'_c dz\,e^{-izt}
\sum_{j=1,2}
\left(\frac{1}{z-i\tilde{{\cal L}_H^0}}
\right)_{11[0],11[0]}
\nonumber \\&\times&
i<11[0]|\tilde{{\cal L}}_H |11[1(s_j)]>
\left(\frac{1}{z-i\tilde{{\cal L}_H^0}}\right)_{11[1(s_j)],11[1(s_j)]}
\nonumber \\&\times&
i<11[1(s_j)]|\tilde{{\cal L}}_H |11[1(f)]>
\left(\frac{1}{z-i\tilde{{\cal L}_H^0}}
\right)_{11[1(f)],11[1(f)]}.
\label{3.4}
\end{eqnarray}
The prime on the integral sign means that only poles corresponding to propagators arising to
vacuum states (without self oscillators) have to be included in the path $c$. 
In the present case, the pole due to the intermediate propagator is thus excluded from the path.  
That selection of poles corresponds to the recipe to construct the subdynamics.
The accidental coincidence of poles due to the correlation and vacuum propagators 
is avoided by adding a positive imaginary infinitesimal $i\epsilon$
to the correlation propagators  when computing the residues. 
Another formulation of the recipe is the following:
a positive imaginary  infinitesimal $i\epsilon$ is first added to all propagators corresponding
to the correlation states and the path $c$ encloses then the real axis, above $-i\epsilon$.
When no resonance can occur, the $i\epsilon$ can be dropped.

\subsection{Fourier representation}

The evaluation is more easy in variables such that the free motion operator is diagonal.
For the free motion of particles, those variables are well known and correspond to the Fourier
transform of the original spatial variables. 
Therefore, we will replace the unknown $\tilde {f}_{11[0]}$ where the variables
$x^{(1)}, x^{(2)}$ are $({\bf q}^{(1)},{\bf p}^{(1)}), ({\bf q}^{(2)},{\bf p}^{(2)})$ by new functions
depending on variables $({\bf k}^{(1)},{\bf p}^{(1)}), ({\bf k}^{(2)},{\bf p}^{(2)})$.
We will not introduce a new symbol: the nature of the argument  specifies
the function under consideration. The transition between the two functions is provided by (we use
Balescu's choice for the normalisation factor):
\begin{eqnarray}
&&\tilde {f}_{11[0]}({\bf q}^{(1)},{\bf p}^{(1)},{\bf q}^{(2)},{\bf p}^{(2)})
\nonumber\\&&=
\frac1{(2\pi)^6}\int d^3k_1\,d^3k_2\,e^{i({\bf k}^{(1)}.{\bf q}^{(1)}+{\bf k}^{(2)}.{\bf q}^{(2)})}
\tilde {f}_{11[0]}({\bf k}^{(1)},{\bf p}^{(1)},{\bf k}^{(2)},{\bf p}^{(2)}),
\nonumber \\&&
\tilde {f}_{11[0]}({\bf k}^{(1)},{\bf p}^{(1)},{\bf k}^{(2)},{\bf p}^{(2)})
\nonumber \\&&
=\int d^3q_1\,d^3q_2\,e^{-i({\bf k}^{(1)}.{\bf q}^{(1)}+{\bf k}^{(2)}.{\bf q}^{(2)})}
\tilde {f}_{11[0]}({\bf q}^{(1)},{\bf p}^{(1)},{\bf q}^{(2)},{\bf p}^{(2)}).
\label{3.5}
\end{eqnarray}
All functions $\tilde {f}_{11[m]}$ have to be similarly replaced.

We have to perform a similar change with respect to the variables associated with the oscillators.
As the functions are periodic in the variables $\xi^{[m]}$, 
Fourier series are relevant.
The function $\tilde {f}_{11[1]}$ becomes a new function depending for the oscillator on the
new variables $(\eta_{\alpha}^{[j]}, m_{\alpha}^{[j]}, \alpha=1,2)$ ($m_{\alpha}^{[j]}$  discrete) in place
of the continuous variables
$(\eta_{\alpha}^{[j]}, \xi_{\alpha}^{[j]}, \alpha=1,2)$.
\begin{eqnarray}
&&\tilde {f}_{11[1]}({\bf k}^{(1)},{\bf p}^{(1)},{\bf k}^{(2)},{\bf p}^{(2)};\eta_{1}^{[1]},
m_{1}^{[1]},\eta_{2}^{[1]}, m_{2}^{[1]};{\bf k}^{[1]})
\nonumber \\&&=\int_0^1d\xi_1^{[1]}\int_0^1d\xi_2^{[1]}
e^{2\pi i(m_{1}^{[1]}\xi_{1}^{[1]}+m_{2}^{[1]}\xi_{2}^{[1]})}
\nonumber \\&&\times 
\tilde {f}_{11[1]}({\bf k}^{(1)},{\bf p}^{(1)},{\bf k}^{(2)},{\bf p}^{(2)};\eta_{1}^{[1]},
\xi_{1}^{[1]},\eta_{2}^{[1]}, \xi_{2}^{[1]};{\bf k}^{[1]}),
\end{eqnarray}
\begin{eqnarray}
&&\tilde {f}_{11[1]}({\bf k}^{(1)},{\bf p}^{(1)},{\bf k}^{(2)},{\bf p}^{(2)};\eta_{1}^{[1]},
\xi_{1}^{[1]},\eta_{2}^{[1]}, \xi_{2}^{[1]};{\bf k}^{[1]})
\nonumber \\&&=
\sum_{m_{1}^{[1]},m_{2}^{[1]}}
e^{-2\pi i(m_{1}^{[1]}\xi_{1}^{[1]}+m_{2}^{[1]}\xi_{2}^{[1]})}
\nonumber \\&&\times 
\tilde {f}_{11[1]}({\bf k}^{(1)},{\bf p}^{(1)},{\bf k}^{(2)},{\bf p}^{(2)};\eta_{1}^{[1]},
m_{1}^{[1]},\eta_{2}^{[1]}, m_{2}^{[1]};{\bf k}^{[1]}),
\label{3.6}
\end{eqnarray}
where the summations run on all integers, positive and negative.

In Fourier variables, the one particle and one oscillator free motion operators take a simple
diagonal form:
\begin{eqnarray}
&&L_H^{0(j)}=-ik_r^{(j)}v_r^{(j)}=-i{\bf k}^{(j)}.{\bf v}^{(j)}
\nonumber \\
&&L_H^{0[i]}=i k^{[i]}\sum_{\alpha=1}^2m_{\alpha}^{[i]}
\label{3.8}
\end{eqnarray}
while we have for $L_H^{'(12)}$
\begin{equation}
L_H^{'(12)}=i\frac{e_1e_2}{8\pi^2}\int d^3l\,l^{-2}e^{\frac{{\bf l}}{2}.(\frac{\partial}{\partial {\bf k}^{(1)}}-\frac{\partial}{\partial {\bf k}^{(2)}})}
{\bf l}.\left(\frac{\partial}{\partial {\bf p}^{(1)}}-\frac{\partial}{\partial {\bf p}^{(2)}}
\right)
\label{3.9}
\end{equation}
or (alternative more usual form)
\begin{equation}
L_H^{'(12)}=i\frac{e_1e_2}{2\pi^2}\int d^3l\,l^{-2}e^{{{\bf l}}.(\frac{\partial}{\partial {\bf k}^{(1)}}-\frac{\partial}{\partial {\bf k}^{(2)}})}
{\bf l}.\left(\frac{\partial}{\partial {\bf p}^{(1)}}-\frac{\partial}{\partial {\bf p}^{(2)}}
\right)
\label{3.9a}
\end{equation}
We have for $L_H^{'j[i]}$:
\begin{eqnarray}
&&L_H^{'j[i]}=-e_j \frac1{(2\pi)^{\frac32}}
\sum_{\alpha=1,2} 
\sum_{a=\pm 1}
\left(
\frac{\eta_{\alpha}^{[i]}}
{k^{[i]}}
\right)^{\frac12} 
\nonumber \\&&\qquad\quad\times
\left[ 
[k^{[i]} e_r^{(\alpha)[i]}-g^{st} v_s^{(j)}
(e_t^{(\alpha)[i]} k_r^{[i]}-e_r^{(\alpha)[i]} k_t^{[i]})]
\frac{\partial}
{\partial p_r^{(j)}}
\right.\nonumber \\ &&\qquad\quad\left.
-\pi({\bf v}^{(j)}.{\bf e}^{(\alpha)[i]})
\left( 
2 \frac{\partial}
{\partial \eta_{\alpha}^{[i]}}
-\frac{a}{\eta_{\alpha}^{[i]}}(m_{\alpha}^{[i]}-a)
\right)
\right]
\nonumber \\&&\qquad\quad\times
\exp a\left\{-{\bf k}^{[i]}.\frac{\partial}{\partial {\bf k}^{(j)}}-\frac{\partial}{\partial
m_{\alpha}^{[i]}}\right\}.
\label{3.10}
\end{eqnarray}
The only difference is the replacement of the variable $q_r^{(j)}$ by the partial derivative
$-i\frac{\partial}{\partial k_r^{(j)}}$ and a similar transposition for the angle variable of the field.
The notation $\exp-a\frac{\partial}{\partial m_{\alpha}^{[i]}}$ enables to take into account the
non-diagonality of $L_H^{'j[i]}$ with respect to the index $m_{\alpha}^{[i]}$: the transition is $\pm1$
according to the value of $a$. Another possibility is the introduction of the factor
$\sum_{m_{\alpha}^{'[i]}}\delta_{m_{\alpha}^{[i]},m_{\alpha}^{'[i]}+a}$, writing with a prime the
corresponding argument of the function on which the matrix element acts.

\subsection{First order kinetic operator}

In those variables, the operator $<11[0]|\tilde \Theta|11[1(f)]>^{(0,1)}=<11[0]|\tilde{{\cal L}}_H
|11[1(f)]>$ takes a simple form, due to the presence of a front factor
$\delta_{m_{1}^{[1]},0}\delta_{m_{2}^{[1]},0}$, $a^2=1$ and the property 
$$
\int_0^{\infty}d\eta_\alpha^{[1]}\,(\eta_\alpha^{[1]})^{\frac12}\left( 
2 \frac{\partial}{\partial\eta_{\alpha}^{[i]}}+\frac{1}{\eta_{\alpha}^{[i]}}\right)\dots
=2\int_0^{\infty}d\eta_\alpha^{[1]}\,\frac{\partial} {\partial \eta_{\alpha}^{[i]}}
\left((\eta_\alpha^{[1]})^{\frac12} \dots \right) =0,
$$ 
when acting on a regular function. 
\begin{eqnarray}
&&<11[0]|\tilde \Theta|11[1(f)]>^{(0,1)}=
-\sum_{j=1,2} e_j \frac1{(2\pi)^{\frac32}}
\int d^3k^{[1]}
\int_0^{\infty}d\eta_1^{[1]}\int_0^{\infty}d\eta_2^{[1]}
\nonumber \\&&\times
\sum_{m_{1}^{[1]},m_{2}^{[1]}}
\delta_{m_{1}^{[1]},0}\delta_{m_{2}^{[1]},0}
\sum_{\alpha=1,2} 
\sum_{a=\pm 1}
\left(
\frac{\eta_{\alpha}^{[1]}}
{k^{[1]}}
\right)^{\frac12}
\nonumber \\&&\times 
\left[ 
[k^{[1]} e_r^{(\alpha)[1]}-g^{st} v_s^{(j)}
(e_t^{(\alpha)[1]} k_r^{[1]}-e_r^{(\alpha)[1]} k_t^{[1]})]
\frac{\partial}
{\partial p_r^{(j)}}
\right]
\nonumber \\&&\times
\exp a\left\{-{\bf k}^{[1]}.\frac{\partial}{\partial {\bf k}^{(j)}}
-\frac{\partial}{\partial m_{\alpha}^{[1]}}\right\}.
\label{3.11}
\end{eqnarray}
\subsection{Recovering the Lorentz force}

For pedagogical reasons, we show in Appendix A that under conditions of independence 
(factorisation) of the field and one particle distribution function, 
describing a particle sharply located at ${\bf r}(t)$, this expression leads to 
\begin{equation}
\left. \partial_t \tilde f({\bf k},{\bf p},t)\right\vert_1=
-e<{\bf E}^{{\bot }}({\bf r}(t))+{\bf p}\times{\bf B}^{{\bot }}({\bf r}(t))>.\nabla_{{\bf p}}
\tilde f({\bf k},{\bf p},t). 
\label{3.11d}
\end{equation}
$e<{\bf E}^{{\bot }}({\bf r}(t))+{\bf v}\times{\bf B}^{{\bot }}({\bf r}(t))>$ is the usual
electromagnetic force acting on the particle. The minus sign is easily accounted for.
If the distribution function  corresponds to a well defined value of the velocity and an uniform
acceleration ${\bf a}$, it can be written $f({\bf k},{\bf v},t) \propto \delta({\bf v}-{\bf v}_0-{\bf a}t)$ and 
for the time dependence  due to the acceleration, we have $\left.\partial_t \tilde f({\bf k},{\bf
v},t)\right\vert_{1}=-{\bf a}.\nabla_{{\bf v}}\tilde f({\bf k},{\bf v},t)$.  
That expression clearly shows that the present formalism yields the well known
expression for the Lorentz force.
That relation (\ref{3.11d})  will be used in $\S$ 6.

\subsection{Evaluation of the second order $<11[0]|\tilde \Sigma|11[1(f)]>$}

While the evolution of the field will be considered in the next section, we
focus here on the reaction of the particles  to the presence of the field, namely the radiative
corrections to the direct interaction between the
particles and the field.  
To provide their contribution to the evolution of the two-particle distribution
function, only the elements of the kinetic operator that acts on the (reduced) distribution
function of the two charged particles and one mode of the field are relevant.
When acting on the distribution function corresponding to the absence of field  (defined in the
extended dynamics), they will determine the first
radiative correction to the free motion of the particles.

The operator $<11[0]|\tilde \Sigma|11[1(f)]>^{(0,2)}$  (\ref{3.4}) in the new variables is now determined.
The two interactions have to involve the same particle.
We use the  same convention for the index of particles as for the polarisation of the
oscillators: 
$j'$ is 2 when
$j$ is 1 and vice versa. We get:
\begin{eqnarray}
&&<11[0]|\tilde \Sigma|11[1(f)]>^{(0,2)}
\nonumber \\&&=\frac{-1}{2\pi i} \int'_c dz\,e^{-izt}
\sum_{j=1,2}
\left(\frac{1}{z-{\bf k}^{(j)}.{\bf v}^{(j)}-{\bf k}^{(j')}.{\bf v}^{(j')}}
\right)
(-i) e_j \frac1{(2\pi)^{\frac32}}
\nonumber \\&&\times
\int d^3k^{[1]}
\int_0^{\infty}d\eta_1^{[1]}\int_0^{\infty}d\eta_2^{[1]}
\sum_{m_{1}^{[1]},m_{2}^{[1]}}
\delta_{m_{1}^{[1]},0}\delta_{m_{2}^{[1]},0}
\sum_{\alpha=1,2} 
\sum_{a=\pm 1}
\nonumber \\&&\times
\left(
\frac{\eta_{\alpha}^{[1]}}
{k^{[1]}}
\right)^{\frac12} 
\left[ 
[k^{[1]} e_r^{(\alpha)[1]}-g^{st} v_s^{(j)}
(e_t^{(\alpha)[1]} k_r^{[1]}-e_r^{(\alpha)[1]} k_t^{[1]})]
\frac{\partial}
{\partial p_r^{(j)}}
\right]
\nonumber \\&&\times
\exp a\left\{-{\bf k}^{[1]}.\frac{\partial}{\partial {\bf
k}^{(j)}}-\frac{\partial}{\partial m_{\alpha}^{[1]}}\right\}
\nonumber \\&&\times
\left(\frac{1}{z-{\bf k}^{(j)}.{\bf v}^{(j)}-{\bf k}^{(j')}.{\bf v}^{(j')}
+k^{[1]}(m_{\alpha}^{[1]}+m_{\alpha'}^{[1]})}
\right)
\nonumber \\&&\times
(-i)e_j \frac1{(2\pi)^{\frac32}} 
\sum_{\beta=1,2} 
\sum_{b=\pm 1}
\left(
\frac{\eta_{\beta}^{[1]}}
{k^{[1]}}
\right)^{\frac12} 
\nonumber \\&&\times
\left[ 
[k^{[1]} e_{r'}^{(\beta)[1]}-g^{s't'} v_{s'}^{(j)}
(e_{t'}^{(\beta)[1]} k_{r'}^{[1]}-e_{r'}^{(\beta)[1]} k_{t'}^{[1]})]
\frac{\partial}
{\partial p_{r'}^{(j)}}
\right.\nonumber \\ &&\left.
-\pi({\bf v}^{(j)}.{\bf e}^{(\beta)[1]})
\left( 
2 \frac{\partial}
{\partial \eta_{\beta}^{[1]}}
-\frac{b}{\eta_{\beta}^{[1]}}(m_{\beta}^{[1]}-b)
\right)
\right]
\exp b\left\{-{\bf k}^{[1]}.\frac{\partial}{\partial {\bf k}^{(j)}}-\frac{\partial}{\partial
m_{\beta}^{[1]}}\right\}
\nonumber \\&&\times
\left(\frac{1}{z-{\bf k}^{(j)}.{\bf v}^{(j)}-{\bf k}^{(j')}.{\bf v}^{(j')}
+k^{[1]}(m_{\beta}^{[1]}+m_{\beta'}^{[1]})}
\right).
\label{3.12}
\end{eqnarray}
The displacement operators can be transfered at the right of the expression to provide
\begin{eqnarray}
&&<11[0]|\tilde \Sigma|11[1(f)]>^{(0,2)}
\nonumber \\&&=\frac{-1}{2\pi i} \int'_c dz\,e^{-izt}
\sum_{j=1,2}
\left(z-{\bf k}^{(j)}.{\bf v}^{(j)}-{\bf k}^{(j')}.{\bf v}^{(j')}
\right)^{-1}
(-i) e_j \frac1{(2\pi)^{\frac32}}
\nonumber \\&&\times
\int d^3k^{[1]}
\int_0^{\infty}d\eta_1^{[1]}\int_0^{\infty}d\eta_2^{[1]}
\sum_{m_{1}^{[1]},m_{2}^{[1]}}
\delta_{m_{1}^{[1]},0}\delta_{m_{2}^{[1]},0}
\sum_{\alpha=1,2} 
\sum_{a=\pm 1}
\nonumber \\&&\times
\left(
\frac{\eta_{\alpha}^{[1]}}
{k^{[1]}}
\right)^{\frac12} 
\left[ 
[k^{[1]} e_r^{(\alpha)[1]}-g^{st} v_s^{(j)}
(e_t^{(\alpha)[1]} k_r^{[1]}-e_r^{(\alpha)[1]} k_t^{[1]})]
\frac{\partial}
{\partial p_r^{(j)}}
\right]
\nonumber \\&&\times
\left(z-{\bf k}^{(j)}.{\bf v}^{(j)}+a{\bf k}^{[1]}.{\bf v}^{(j)}-{\bf
k}^{(j')}.{\bf v}^{(j')} +k^{[1]}(m_{\alpha}^{[1]}+m_{\alpha'}^{[1]}-a)
\right)^{-1}
\nonumber \\&&\times
(-i)e_j \frac1{(2\pi)^{\frac32}} 
\sum_{\beta=1,2} 
\sum_{b=\pm 1}
\left(
\frac{\eta_{\beta}^{[1]}}
{k^{[1]}}
\right)^{\frac12} 
\nonumber \\&&\times
\left[ 
[k^{[1]} e_{r'}^{(\beta)[1]}-g^{s't'} v_{s'}^{(j)}
(e_{t'}^{(\beta)[1]} k_{r'}^{[1]}-e_{r'}^{(\beta)[1]} k_{t'}^{[1]})]
\frac{\partial}
{\partial p_{r'}^{(j)}}
\right.\nonumber \\ &&\left.
-\pi({\bf v}^{(j)}.{\bf e}^{(\beta)[1]})
\left( 
2 \frac{\partial}
{\partial \eta_{\beta}^{[1]}}
-\frac{b}{\eta_{\beta}^{[1]}}(m_{\beta}^{[1]}-b-a\delta_{\alpha, \beta})
\right)
\right]
\nonumber \\&&\times
\left(z-{\bf k}^{(j)}.{\bf v}^{(j)}+b{\bf k}^{([1])}.{\bf v}^{(j)} +a{\bf k}^{[1]}.{\bf v}^{(j)} 
\right.\nonumber \\&&\left.
-{\bf k}^{(j')}.{\bf v}^{(j')} 
+k^{[1]}(m_{\beta}^{[1]}+m_{\beta'}^{[1]}-b-a)
\right)^{-1}
\nonumber \\&&\times
\exp \left\{-(a+b){\bf k}^{[1]}.\frac{\partial}{\partial {\bf k}^{(j)}}-a\frac{\partial}{\partial
m_{\alpha}^{[1]}}-b\frac{\partial}{\partial m_{\beta}^{[1]}}\right\}.
\label{3.14}
\end{eqnarray}
The summation over $m_{1}^{[1]}$, $m_{2}^{[1]}=0$ and the Kronecker delta functions of the
variables $m_{\alpha}^{[1]}=0$ and $m_{\alpha'}^{[1]}=0$ are also written at the right
of the expression.
In the explicit computation, a separation has to be performed between the contributions with
$\beta=\alpha$ and $\beta\not=\alpha$ on one hand, $a=b$ and $a\not=b$ on the other hand.
In our future expressions, the first sign $=$ or $\not=$ will refer to the polarisation index while the
second one to the relative value of $a$ and $b$.

From that expression, the computation of $<11[0]|\tilde \Sigma|11[1(f)]>^{(0,2)}$ is rather
straightforward: we have to proceed formally to the derivatives with respect to the mechanical
momentums of the particle. A further index can be introduced to reflect the factor on which they act.
The last operation is an integral by residue that
takes into account the poles due to the first and last propagators that correspond to the
vacuum of correlation.
For the sake of illustration, let us consider  one of the contributions that will matter when acting in
absence of free field:
\begin{eqnarray}
&&<11[0]|\tilde \Sigma|11[1(f)]>_{=,3,\not=}^{(0,2)}
\nonumber \\&&=\frac{-1}{2\pi i} \int'_c dz\,e^{-izt}
\sum_{j=1,2}
(-1) e_j^2 \frac1{(2\pi)^{3}}
\int d^3k^{[1]}
\int_0^{\infty}d\eta_1^{[1]}\int_0^{\infty}d\eta_2^{[1]}
\nonumber \\&&\times
\sum_{\alpha=1,2} 
\sum_{a=\pm 1}
\left(
\frac{\eta_{\alpha}^{[1]}}
{k^{[1]}}
\right)
\left(\frac{1}{z-{\bf k}^{(j)}.{\bf v}^{(j)}+a{\bf k}^{[1]}.{\bf v}^{(j)}-{\bf k}^{(j')}.{\bf v}^{(j')}
-a k^{[1]}}
\right)
\nonumber \\&&\times
\left\{
\left(\frac{1}{z-{\bf k}^{(j)}.{\bf v}^{(j)} -{\bf k}^{(j')}.{\bf v}^{(j')} }
\right)^2
\right.\nonumber \\&&\times\left.
\left[ 
[k^{[1]} e_r^{(\alpha)[1]}-g^{st} v_s^{(j)}
(e_t^{(\alpha)[1]} k_r^{[1]}-e_r^{(\alpha)[1]} k_t^{[1]})]
\frac{\partial}{\partial p_r^{(j)}}
\right]
\right.\nonumber \\&&\times
\left.
\left[ 
[k^{[1]} e_{r'}^{(\alpha)[1]}-g^{s't'} v_{s'}^{(j)}
(e_{t'}^{(\alpha)[1]} k_{r'}^{[1]}-e_{r'}^{(\alpha)[1]} k_{t'}^{[1]})]
\frac{\partial}
{\partial p_{r'}^{(j)}}
\right.\right.\nonumber \\&&
\left.\left.\qquad
-\pi({\bf v}^{(j)}.{\bf e}^{(\alpha)[1]})
\left( 
2 \frac{\partial}
{\partial \eta_{\alpha}^{[1]}}
\right)
\right]
\right.
\nonumber \\&&+\left.
\left(\frac{1}{z-{\bf k}^{(j)}.{\bf v}^{(j)} -{\bf k}^{(j')}.{\bf v}^{(j')}}
\right)^3
\right.
\nonumber \\&&\times
\left.
\left[ 
[k^{[1]} e_r^{(\alpha)[1]}-g^{st} v_s^{(j)}
(e_t^{(\alpha)[1]} k_r^{[1]}-e_r^{(\alpha)[1]} k_t^{[1]})]
\frac{\partial}{\partial p_r^{(j)}}
\right]
\right.\nonumber \\&&\times
\left.
\left[ 
k^{[1]} e_{r'}^{(\alpha)[1]}-g^{s't'} v_{s'}^{(j)}
(e_{t'}^{(\alpha)[1]} k_{r'}^{[1]}-e_{r'}^{(\alpha)[1]} k_{t'}^{[1]})
\right]
 k_{s'}^{(j)}
\left(\frac{\partial v_{s'}^{(j)}}{\partial p_{r'}^{(j)}}
\right)
\right\}
\nonumber \\&&\times
\sum_{m_{1}^{[1]},m_{2}^{[1]}}
\delta_{m_{1}^{[1]},0}\delta_{m_{2}^{[1]},0}.
\label{3.26}
\end{eqnarray}
In that expression, the partial derivative $\frac{\partial}{\partial p_r^{(j)}}$ acts on everything at its
right: the factors $v_{s'}^{(j)}$ and ${\bf v}^{(j)}$ (in the first term), the factors $v_{s'}^{(j)}$ and
$\left(\frac{\partial v_{s'}^{(j)}}{\partial p_{r'}^{(j)}}
\right)$ in the second term 
and, for both terms, the momentum dependence of
the distribution function on which $<11[0]|\tilde \Sigma|11[1(f)]>_{=,3,\not=}^{(0,2)}$ is applied.
The only singularity to be included in the close path $c$ is the pole of 
$\left(\frac{1}{z-{\bf k}^{(j)}.{\bf v}^{(j)} -{\bf k}^{(j')}.{\bf v}^{(j')}}\right)$.
No coincidence is possible with the pole of 
$\frac{1}{z-{\bf k}^{(j)}.{\bf v}^{(j)}+a{\bf k}^{[1]}.{\bf v}^{(j)}-{\bf k}^{(j')}.{\bf v}^{(j')}
-a k^{[1]}}$ ($v^{(j)}<1$) and no $i\epsilon$ has to be introduced here.
The obtention of $<11[0]|\tilde \Sigma|11[1(f)]>_{=,3,\not=}^{(0,2)}$ is therefore straightforward.
\begin{eqnarray}
&&<11[0]|\tilde \Sigma|11[1(f)]>_{=,3,\not=}^{(0,2)}
\nonumber \\&&= 
e^{-i[{\bf k}^{(1)}.{\bf v}^{(1)}+{\bf k}^{(2)}.{\bf v}^{(2)}]t}
\sum_{j=1,2}
(-1) e_j^2 \frac1{(2\pi)^{3}}
\int d^3k^{[1]}
\int_0^{\infty}d\eta_1^{[1]}\int_0^{\infty}d\eta_2^{[1]}
\nonumber \\& &\times
\sum_{\alpha=1,2} 
\sum_{a=\pm 1}
\left(
\frac{\eta_{\alpha}^{[1]}}
{k^{[1]}}
\right)
\left\{
\left[
(-it)\left(\frac{1}{a{\bf k}^{[1]}.{\bf v}^{(j)}-a k^{[1]}}\right)
\right.\right.\nonumber \\& &\left.\left.
-\left(\frac{1}{-a{\bf k}^{[1]}.{\bf v}^{(j)}+ ak^{[1]}}\right)^2
\right]
\right.\nonumber \\&&\times\left.
\left[ 
[k^{[1]} e_r^{(\alpha)[1]}-g^{st} v_s^{(j)}
(e_t^{(\alpha)[1]} k_r^{[1]}-e_r^{(\alpha)[1]} k_t^{[1]})]
\frac{\partial}{\partial p_r^{(j)}}
\right]
\right.\nonumber \\&&\times\left.
\left[ 
[k^{[1]} e_{r'}^{(\alpha)[1]}-g^{s't'} v_{s'}^{(j)}
(e_{t'}^{(\alpha)[1]} k_{r'}^{[1]}-e_{r'}^{(\alpha)[1]} k_{t'}^{[1]})]
\frac{\partial}
{\partial p_{r'}^{(j)}}
\right.\right.\nonumber \\&&-\left.\left.
\pi({\bf v}^{(j)}.{\bf e}^{(\alpha)[1]})
\left( 
2 \frac{\partial}
{\partial \eta_{\alpha}^{[1]}}
\right)
\right]
+
\left[
(-it)^2\left(\frac{1}{a{\bf k}^{[1]}.{\bf v}^{(j)}-a k^{[1]}}\right)
\right.\right.
\nonumber \\&&-\left.\left.
2(-it)\left(\frac{1}{a{\bf k}^{[1]}.{\bf v}^{(j)}-a k^{[1]}}\right)^2
+2\left(\frac{1}{a{\bf k}^{[1]}.{\bf v}^{(j)}-a k^{[1]}}\right)^3
\right]
\right.\nonumber \\&& \times\left.
\frac12\left[ 
[k^{[1]} e_r^{(\alpha)[1]}-g^{st} v_s^{(j)}
(e_t^{(\alpha)[1]} k_r^{[1]}-e_r^{(\alpha)[1]} k_t^{[1]})]
\frac{\partial}{\partial p_r^{(j)}}
\right]
\right.\nonumber \\&& \times\left.
\left[ 
k^{[1]} e_{r'}^{(\alpha)[1]}-g^{s't'} v_{s'}^{(j)}
(e_{t'}^{(\alpha)[1]} k_{r'}^{[1]}-e_{r'}^{(\alpha)[1]} k_{t'}^{[1]})
\right]
 k_{s'}^{(j)}
\left(\frac{\partial v_{s'}^{(j)}}{\partial p_{r'}^{(j)}}
\right)
\right\}
\nonumber \\&& \times
\sum_{m_{1}^{[1]},m_{2}^{[1]}}
\delta_{m_{1}^{[1]},0}\delta_{m_{2}^{[1]},0}.
\label{3.51}
\end{eqnarray}

\subsection{The second order evolution operator $\tilde \Theta$}

From the general properties of the subdynamics, we have the  links (\ref{3.37}) between the subdynamics
operator $\tilde \Sigma(t)$ and the evolution operator $\tilde \Theta$ for the vacuum-vacuum elements.
Since we have a limited choice of possibilities for the vacuum intermediate states, we have
\begin{eqnarray}
&&\left.\frac{\partial}{\partial t}<11[0]|\tilde \Sigma|11[1(f)]>_{=,3,\not=}^{(0,2)}\right|_{t=0}
\nonumber \\& &
=<11[0]|\tilde \Theta|11[1(f)]>_{=,3,\not=}^{(0,2)}<11[1(f)]|\tilde{{\cal A}}|11[1(f)]>^{(0,0)}
\nonumber \\&&+<11[0]|\tilde {{\cal L}}^0|11[0]><11[0]|\tilde{{\cal A}}|11[1(f)]>_{=,3,\not=}^{(0,2)}.
\label{3.38}
\end{eqnarray}
Using $<11[1(f)]|\tilde{{\cal A}}|11[1(f)]>^{(0,0)}=1$  
and $<11[0]|\tilde {{\cal L}}^0|11[0]>=-i[{\bf k}^{(j)}.{\bf v}^{(j)}+{\bf k}^{(j')}.{\bf v}^{(j')}]$, we can obtain
directly $<11[0]|\tilde\Theta|11[1(f)]>_{=,3,\not=}^{(0,2)}$.
\begin{eqnarray}
&&<11[0]|\tilde \Theta|11[1(f)]>_{=,3,\not=}^{(0,2)}
\nonumber \\&&= 
\sum_{j=1,2}
(-1) e_j^2 \frac1{(2\pi)^{3}}
\int d^3k^{[1]}
\int_0^{\infty}d\eta_1^{[1]}\int_0^{\infty}d\eta_2^{[1]}
\sum_{\alpha=1,2} 
\sum_{a=\pm 1}
\left(
\frac{\eta_{\alpha}^{[1]}}
{k^{[1]}}
\right)
\nonumber \\&&\times
\left\{
(-i)\left(\frac{1}{a{\bf k}^{[1]}.{\bf v}^{(j)}-a k^{[1]}}\right)
\left[ 
[k^{[1]} e_r^{(\alpha)[1]}-g^{st} v_s^{(j)}
(e_t^{(\alpha)[1]} k_r^{[1]}
\right.\right.\nonumber \\&&\left.\left.
-e_r^{(\alpha)[1]} k_t^{[1]})]
\frac{\partial}{\partial p_r^{(j)}}
\right]
\left[ 
[k^{[1]} e_{r'}^{(\alpha)[1]}-g^{s't'} v_{s'}^{(j)}
(e_{t'}^{(\alpha)[1]} k_{r'}^{[1]}-e_{r'}^{(\alpha)[1]} k_{t'}^{[1]})]
\frac{\partial}
{\partial p_{r'}^{(j)}}
\right.\right.\nonumber \\&&\left.\left.
-\pi({\bf v}^{(j)}.{\bf e}^{(\alpha)[1]})
\left( 
2 \frac{\partial}
{\partial \eta_{\alpha}^{[1]}}
\right)
\right]
+i
\left(\frac{1}{a{\bf k}^{[1]}.{\bf v}^{(j)}-a k^{[1]}}\right)^2
\right.\nonumber \\&& \times\left.
\left[ 
[k^{[1]} e_r^{(\alpha)[1]}-g^{st} v_s^{(j)}
(e_t^{(\alpha)[1]} k_r^{[1]}-e_r^{(\alpha)[1]} k_t^{[1]})]
\frac{\partial}{\partial p_r^{(j)}}
\right]
\right.\nonumber \\&& \times\left.
\left[ 
k^{[1]} e_{r'}^{(\alpha)[1]}-g^{s't'} v_{s'}^{(j)}
(e_{t'}^{(\alpha)[1]} k_{r'}^{[1]}-e_{r'}^{(\alpha)[1]} k_{t'}^{[1]})
\right]
 k_{s'}^{(j)}
\left(\frac{\partial v_{s'}^{(j)}}{\partial p_{r'}^{(j)}}
\right)
\right\}
\nonumber \\&&\times
\sum_{m_{1}^{[1]},m_{2}^{[1]}}
\delta_{m_{1}^{[1]},0}\delta_{m_{2}^{[1]},0}.
\label{3.52}
\end{eqnarray}
All  contributions are treated in a similar way and recombined such that 
the conservation of the norm is manifest (due to the front factor $\frac{\partial}{\partial p_r^{(j)}}$):
\begin{eqnarray}
&&<11[0]|\tilde \Theta|11[1(f)]>_{=\not=}^{(0,2)}
\nonumber \\&&= 
\sum_{j=1,2}
i e_j^2 \frac1{(2\pi)^{3}}
\frac{\partial}{\partial p_r^{(j)}}
\int d^3k^{[1]}
\int_0^{\infty}d\eta_1^{[1]}\int_0^{\infty}d\eta_2^{[1]}
\sum_{\alpha=1,2} 
\sum_{a=\pm 1}
\left(
\frac{\eta_{\alpha}^{[1]}}
{k^{[1]}}
\right)
\nonumber \\&&\times
\left[
\left(\frac{1}{a{\bf k}^{[1]}.{\bf v}^{(j)}-a k^{[1]}}\right)
\right]
\left[ 
k^{[1]} e_r^{(\alpha)[1]}-g^{st} v_s^{(j)}
(e_t^{(\alpha)[1]} k_r^{[1]}-e_r^{(\alpha)[1]} k_t^{[1]})
\right]
\nonumber \\&&\times
\left[ 
[k^{[1]} e_{r'}^{(\alpha)[1]}-g^{s't'} v_{s'}^{(j)}
(e_{t'}^{(\alpha)[1]} k_{r'}^{[1]}-e_{r'}^{(\alpha)[1]} k_{t'}^{[1]})]
\frac{\partial}
{\partial p_{r'}^{(j)}}
\right.\nonumber \\&&\left.
-\pi({\bf v}^{(j)}.{\bf e}^{(\alpha)[1]})
\left( 
2 \frac{\partial}
{\partial \eta_{\alpha}^{[1]}}
\right)
\right]
\sum_{m_{1}^{[1]},m_{2}^{[1]}}
\delta_{m_{1}^{[1]},0}\delta_{m_{2}^{[1]},0}
\nonumber \\&&-i
\sum_{j=1,2}
 e_j^2 \frac1{(2\pi)^{3}}
\frac{\partial}{\partial p_r^{(j)}}
\int d^3k^{[1]}
\int_0^{\infty}d\eta_1^{[1]}\int_0^{\infty}d\eta_2^{[1]}
\sum_{\alpha=1,2} 
\sum_{a=\pm 1}
\left(
\frac{\eta_{\alpha}^{[1]}}
{k^{[1]}}
\right)
\nonumber \\&&\times
\left[
\left(\frac{1}{a{\bf k}^{[1]}.{\bf v}^{(j)}-a k^{[1]}}\right)^2
\right]
\left[ 
k^{[1]} e_r^{(\alpha)[1]}-g^{st} v_s^{(j)}
(e_t^{(\alpha)[1]} k_r^{[1]}-e_r^{(\alpha)[1]} k_t^{[1]})
\right]
\nonumber \\&&\times\left[ 
k^{[1]} e_{r'}^{(\alpha)[1]}-g^{s't'} v_{s'}^{(j)}
(e_{t'}^{(\alpha)[1]} k_{r'}^{[1]}-e_{r'}^{(\alpha)[1]} k_{t'}^{[1]})
\right]
 k_{s'}^{(j)}
\left(\frac{\partial v_{s'}^{(j)}}{\partial p_{r'}^{(j)}}
\right)
\nonumber \\&&\times
\sum_{m_{1}^{[1]},m_{2}^{[1]}}
\delta_{m_{1}^{[1]},0}\delta_{m_{2}^{[1]},0}.
\label{3.52g}
\end{eqnarray}
Here, the commutation property of $[k^{[1]} e_r^{(\alpha)[1]}-g^{st} v_s^{(j)}
(e_t^{(\alpha)[1]} k_r^{[1]}-e_r^{(\alpha)[1]} k_t^{[1]})]$ and $\frac{\partial}{\partial p_r^{(j)}}$ has been
used.

For the sake of completeness, similar expressions are provided in appendix B for 
$<11[0]|\tilde \Theta|11[1(f)]>_{==}^{(0,2)}$, 
$<11[0]|\tilde \Theta|11[1(f)]>_{\not==}^{(0,2)}$ and
\newline\noindent
$<11[0]|\tilde \Theta|11[1(f)]>_{\not=\not=}^{(0,2)}$.

\subsection{Evolution in the absence of incident field}

The knowledge of the operator $<11[0]|\tilde \Theta|11[1(f)]>^{(0,2)}$ enables us to look
for the behaviour of the particles when evolving into the absence of external field.
Other matrix elements could also be considered to provide, for instance, the vertex (charge)
renormalisation due to the self-field but are outside our scope in this paper.

We can use the factorisation (\ref{2.18}):
\begin{eqnarray}
\tilde f_{11[1(f)]}&=&\tilde f_{11[0]} \tilde f_{[1(f)]},
\label{4.1}
\end{eqnarray}
and use for $\tilde f_{[1(f)]}$ the distribution function $\tilde f^V_{[1(f)]}$ corresponding to the
absence of field: namely  the limit for
$\eta_1\to 0$ and $\eta_2\to 0$ of:
\begin{equation}
\tilde {f}_{[1]}(\eta_{1}^{[1]},
m_{1}^{[1]},\eta_{2}^{[1]}, m_{2}^{[1]};{\bf k}^{[1]})
=\delta(\eta_{1}^{[1]}-\eta_1)\delta(\eta_{2}^{[1]}-\eta_2)
\delta^{Kr}_{m_{1}^{[1]},0}\delta^{Kr}_{m_{2}^{[1]},0}
\label{4.2}
\end{equation}
Since the variables $\eta_{1}^{[1]}$ and $\eta_{2}^{[1]}$ are integrated from $0$ to $\infty$,
the limit $\eta_1\to 0$ and $\eta_2\to 0$ has to be taken after that we have performed that
integration.
We identify in $<11[0]|\tilde \Theta|11[1(f)]>^{(0,2)}$ all  the terms
which could provide a non vanishing  contribution.
The summations over $m_{1}^{[1]}$ and $m_{2}^{[1]}$ provide a vanishing result if a
displacement operator on $m_{1}^{[1]}$ and $m_{2}^{[1]}$ is involved: we would then meet a product
of Kronecker's delta functions with incompatible arguments.
Therefore, the only possible non-vanishing ones would arise from the contribution 
$<11[0]|\tilde \Theta|11[1(f)]>_{=\not=}^{(0,2)}$.
Since that contribution involves a front factor
$\eta_{\alpha}^{[1]}$, the presence of $\delta(\eta_{1}^{[1]}-\eta_1)\delta(\eta_{2}^{[1]}-\eta_2)$
for  $\eta_1\to 0$ and $\eta_2\to 0$ provides a vanishing result except for the contributions in
which the derivative of the Dirac's delta function appears.
The second term in (\ref{3.52g})) provides therefore a vanishing result and we are left with
\begin{eqnarray}
&&<11[0]|\tilde \Theta|11[1(f)]>_{=\not=}^{(0,2)}\tilde f^V_{[1(f)]}
\nonumber \\&=& 
\sum_{j=1,2}
i e_j^2 \frac1{(2\pi)^{3}}
\frac{\partial}{\partial p_r^{(j)}}
\int d^3k^{[1]}
\int_0^{\infty}d\eta_1^{[1]}\int_0^{\infty}d\eta_2^{[1]}
\sum_{\alpha=1,2} 
\sum_{a=\pm 1}
\left(
\frac{\eta_{\alpha}^{[1]}}
{k^{[1]}}
\right)
\nonumber \\&\times&
\left(\frac{1}{a{\bf k}^{[1]}.{\bf v}^{(j)}-a k^{[1]}}\right)
\left[ 
k^{[1]} e_r^{(\alpha)[1]}-g^{st} v_s^{(j)}
(e_t^{(\alpha)[1]} k_r^{[1]}-e_r^{(\alpha)[1]} k_t^{[1]})
\right]
\nonumber \\&\times&
\left[ 
-\pi({\bf v}^{(j)}.{\bf e}^{(\alpha)[1]})
\left( 
2 \frac{\partial}
{\partial \eta_{\alpha}^{[1]}}
\right)
\right]
\nonumber \\&\times&
\sum_{m_{1}^{[1]},m_{2}^{[1]}}
\delta_{m_{1}^{[1]},0}\delta_{m_{2}^{[1]},0}
\delta(\eta_{1}^{[1]}-\eta_1)\delta(\eta_{2}^{[1]}-\eta_2)
\delta^{Kr}_{m_{1}^{[1]},0}\delta^{Kr}_{m_{2}^{[1]},0}.
\label{3.52ga}
\end{eqnarray}
The summations and integrations over the fields variables can be performed in a
staightforward way using the Kronecker's and Dirac's delta functions (after an integration by parts and
performing the limits $\eta_1\to 0$, $\eta_2\to 0$) and we have:
\begin{eqnarray}
&&<11[0]|\tilde \Theta|11[1(f)]>_{=\not=}^{(0,2)}\tilde f^V_{[1(f)]}
\nonumber \\&&= 
\sum_{j=1,2}
(-i )e_j^2 \frac1{(2\pi)^{3}}
\frac{\partial}{\partial p_r^{(j)}}
\int d^3k^{[1]}
\sum_{\alpha=1,2} 
\sum_{a=\pm 1}
\left(
\frac{\eta_{\alpha}^{[1]}}
{k^{[1]}}
\right)
\left(\frac{1}{a{\bf k}^{[1]}.{\bf v}^{(j)}-a k^{[1]}}\right)
\nonumber \\&&\times
\left[ 
k^{[1]} e_r^{(\alpha)[1]}-g^{st} v_s^{(j)}
(e_t^{(\alpha)[1]} k_r^{[1]}-e_r^{(\alpha)[1]} k_t^{[1]})
\right]
\left[ 
-2\pi({\bf v}^{(j)}.{\bf e}^{(\alpha)[1]})
\right].
\label{3.52gb}
\end{eqnarray}
Nevertheless, this last expression vanishes also by parity for the summation over $a$.

\subsection{A useful expression :$<11[0]|\tilde{{\cal A}}|11[1(f)]>^{(0,2)}\tilde f^V_{[1(f)]}$}

For future use, the non vanishing element of
$<11[0]|\tilde{{\cal A}}|11[1(f)]>^{(0,2)}\tilde f^V_{[1(f)]}$ is required:
\begin{eqnarray}
&&<11[0]|\tilde{{\cal A}}|11[1(f)]>^{(0,2)}\tilde f^V_{[1(f)]}
\nonumber \\&&= 
\sum_{j=1,2}
(-1) e_j^2 \frac1{(2\pi)^{3}}
\int d^3k^{[1]}
\sum_{\alpha=1,2} 
\sum_{a=\pm 1}
\left(
\frac{1}
{k^{[1]}}
\right)
\nonumber \\&&\times
\left[ 
[k^{[1]} e_r^{(\alpha)[1]}-g^{st} v_s^{(j)}
(e_t^{(\alpha)[1]} k_r^{[1]}-e_r^{(\alpha)[1]} k_t^{[1]})]
\frac{\partial}{\partial p_r^{(j)}}
\right]
\nonumber \\&&\times
\left(\frac{1}{-{\bf k}^{[1]}.{\bf v}^{(j)}+ k^{[1]}}\right)^2
2\pi({\bf v}^{(j)}.{\bf e}^{(\alpha)[1]})
\label{4.9}
\end{eqnarray}
The derivative with respect to $p_r^{(j)}$ acts of course on all possible dependences at its right.

\subsection{Physical interpretation}

The concept of renormalised mass has now to be extracted from the kinetic
equation, by combining radiative corrections with the free motion operator.
Our expression of the  second order element  $<11[0]|\tilde \Theta|11[1(f)]>_{=\not=}^{(0,2)}$
of the kinetic operator shows that it vanishes
for a free particle that is not accelerated by external fields nor a Coulomb interaction 
No mass correction is provided.  
Indeed, if a relativistic expression is used for the energy of the particle, the momentum and energy
conservations cannot be simultaneously satisfied by an emission act of a non-accelerated charged
particle. 
No resonant  process is possible and their absence implies that no $i\epsilon$ is required and the
propagor is odd in $a$. 

This result could be expected from general considerations from our knowledge of general properties
of the subdynamics \cite{PGHR73} when the propagator involved  cannot be resonant.
In Brussels terminology \cite{PGHR73}, the second order kinetic operator 
vanishes for parity reasons: 
it is well known that, at that second order,  the contribution to the kinetic
operator (called $\psi_2$) arises from a Dirac delta ``function"  and not from the principal part of the
(usually regularised by a $i\epsilon$) propagator.
No regularisation has been required here when acting in absence of field and  the kinetic
operator provides a vanishing contribution.

The effect of the coupling with the field vacuum is therefore to be searched in other terms.
Indeed, radiation emission is present when the particles are accelerated.
We consider as a first step in $\S$5 $ <11[0]|\tilde\Theta|11[1(f)]>^{(1,2)}$.
The acceleration provided by the Coulomb interaction will induce a back reaction on the motion of
the particles.

The next section $\S$4  is devoted to the computation of the field generated by the particle in their
free motion.

\section{Emitted field in a free motion}   

\setcounter{equation}{0}

In order to get a better understanding of our previous result (\ref{3.52gb}), we intend to analyse the
field emitted by the particles and their back reaction at the same order (0,2) in the interactions.
The contribution is the same as for two
particles moving independently. 
We will use the equivalence conditions that enable to get the emitted
field from the self-field.  
The subdynamics theory determines the last one through the so-called
creation operator
$\tilde C$.   
Since we know that $\tilde\Sigma(t)=\tilde C e^{\tilde \Theta t}\tilde{{\cal A}}$ for the
correlation-vacuum elements, we focus on the elements of $\tilde\Sigma(t)^{(0,1)}$  
that provides a contribution to 
\begin{eqnarray}
&&\tilde f_{11[1(is_j)]}=<11[1(s_j)]|\tilde C|11[1(f)]>\tilde f_{11[1(f)]}
\nonumber \\&&+
<11[1(s_j)]|\tilde C|11[1(ej')]>\tilde f_{11[1(ej')]} 
\nonumber \\&&+ <11[1(s_j)]|\tilde C|11[2(ff)]>\tilde f_{11[2(ff)]}+\dots
\label{4.10}
\end{eqnarray} 
Since numerically, in the equivalence conditions, we have the equality
$\tilde f_{11[1(ie_j)]}=\tilde f_{11[1(is_j)]}$, the lowest order contribution to $\tilde
f_{11[1(is_j)]}$ requires   $<11[1(s_j)]|\tilde \Sigma(t)|11[1(f)]>^{(0,1)}$
that determines  the lowest order contribution to the creation operator.
Only the terms that provide a contribution when acting in absence
of field are considered.

\subsection{Correlation-vacuum element of $\tilde \Sigma(t)$}

In place of (\ref{3.12}), we start from
\begin{eqnarray}
&&<11[1(s_j)]|\tilde \Sigma(t)|11[1(f)]>^{(0,1)}\tilde f^V_{[1(f)]}
\nonumber \\&&=\frac{-1}{2\pi i} \int'_c dz\,e^{-izt}
\left(\frac{1}{z-{\bf k}^{(j)}.{\bf v}^{(j)}-{\bf k}^{(j')}.{\bf v}^{(j')}+k^{[1]}(m_{\alpha}^{[1]}+m_{\alpha'}^{[1]})}
\right)
\nonumber \\&&\times
(-i)e_j \frac1{(2\pi)^{\frac32}} 
\sum_{\beta=1,2} 
\sum_{a=\pm 1}
\left(
\frac{\eta_{\beta}^{[1]}}
{k^{[1]}}
\right)^{\frac12} 
\nonumber \\&&\times
\left[ 
[k^{[1]} e_{r'}^{(\beta)[1]}-g^{s't'} v_{s'}^{(j)}
(e_{t'}^{(\beta)[1]} k_{r'}^{[1]}-e_{r'}^{(\beta)[1]} k_{t'}^{[1]})]
\frac{\partial}
{\partial p_{r'}^{(j)}}
\right.\nonumber \\ &&\left.
-\pi({\bf v}^{(j)}.{\bf e}^{(\beta)[1]})
\left( 
2 \frac{\partial}
{\partial \eta_{\beta}^{[1]}}
-\frac{a}{\eta_{\beta}^{[1]}}(m_{\beta}^{[1]}-a)
\right)
\right]
\exp a\left\{-{\bf k}^{[1]}.\frac{\partial}{\partial {\bf k}^{(j)}}-\frac{\partial}{\partial
m_{\beta}^{[1]}}\right\}
\nonumber \\&&\times
\left(\frac{1}{z-{\bf k}^{(j)}.{\bf v}^{(j)}-{\bf k}^{(j')}.{\bf v}^{(j')}
+k^{[1]}(m_{\beta}^{[1]}+m_{\beta'}^{[1]})}
\right)
\nonumber \\&&\times
\delta(\eta_{\beta}^{[1]}-\eta_\beta)\delta(\eta_{\beta'}^{[1]}-\eta_{\beta'})
\delta^{Kr}_{m_{\beta}^{[1]},0}\delta^{Kr}_{m_{\beta'}^{[1]},0},
\label{4.12}
\end{eqnarray}
where we have taken into account that the final state  ( $<11[1(s_j)]|$) is a field correlated
state, hence the presence of $k^{[1]}(m_{\alpha}^{[1]}+m_{\alpha'}^{[1]})$ in the first propagator.
Only the pole of the second propagator (due to a vacuum state) is enclosed by the path $c$.
Moving the displacement operators to the right and using afterwards
 $m_{1}^{[1]}+m_{2}^{[1]}=m_{\beta}^{[1]}+m_{\beta'}^{[1]}=m_{\alpha}^{[1]}+m_{\alpha'}^{[1]}=a$,
we get
\begin{eqnarray}
&&<11[1(s_j)]|\tilde \Sigma(t)|11[1(f)]>^{(0,1)}\tilde f^V_{[1(f)]}
\nonumber \\&&=\frac{-1}{2\pi i} \int'_c dz\,e^{-izt}
\sum_{\beta=1,2} 
\sum_{a=\pm 1}
\left(\frac{1}{z-{\bf k}^{(j)}.{\bf v}^{(j)}-{\bf k}^{(j')}.{\bf v}^{(j')}+ak^{[1]}}
\right)
(-i)e_j 
\nonumber \\&&\times
\frac1{(2\pi)^{\frac32}} 
\left(
\frac{\eta_{\beta}^{[1]}}
{k^{[1]}}
\right)^{\frac12} 
\left[ 
[k^{[1]} e_{r'}^{(\beta)[1]}-g^{s't'} v_{s'}^{(j)}
(e_{t'}^{(\beta)[1]} k_{r'}^{[1]}-e_{r'}^{(\beta)[1]} k_{t'}^{[1]})]
\frac{\partial}
{\partial p_{r'}^{(j)}}
\right.\nonumber \\&&\left.
-\pi({\bf v}^{(j)}.{\bf e}^{(\beta)[1]})
\left( 
2 \frac{\partial}
{\partial \eta_{\beta}^{[1]}}
\right)
\right]
\left(\frac{1}{z-({\bf k}^{(j)}-a{\bf k}^{[1]}).{\bf v}^{(j)}-{\bf k}^{(j')}.{\bf v}^{(j')}}
\right)
\nonumber \\&&\times
\exp -\left(
a{\bf k}^{[1]}.\frac{\partial}{\partial {\bf k}^{(j)}}
\right)
\delta(\eta_{\beta}^{[1]}-\eta_\beta)\delta(\eta_{\beta'}^{[1]}-\eta_{\beta'})
\delta^{Kr}_{m_{\beta}^{[1]},a}\delta^{Kr}_{m_{\beta'}^{[1]},0}.
\label{4.15}
\end{eqnarray}
This expression can be computed easily and 
is identified with the same order of 
$<11[1(s_j)]|\tilde Ce^{\tilde \Theta t}\tilde{{\cal A}}|11[1(f)]>$ $ \tilde f^V_{[1(f)]}$.

\subsection{First order creation operator}

Therefore,
\begin{eqnarray}
&&<11[1(s_j)]|\tilde C|11[1(f)]>^{(0,1)}\tilde f^V_{[1(f)]}
\nonumber \\&&=
\sum_{\beta=1,2} 
\sum_{a=\pm 1}
\left(\frac{1}{-a{\bf k}^{[1]}.{\bf v}^{(j)}+ak^{[1]}}
\right)
(-i)e_j \frac1{(2\pi)^{\frac32}} 
\left(
\frac{\eta_{\beta}^{[1]}}
{k^{[1]}}
\right)^{\frac12} 
\nonumber \\&&\times
\left[ 
[k^{[1]} e_{r'}^{(\beta)[1]}-g^{s't'} v_{s'}^{(j)}
(e_{t'}^{(\beta)[1]} k_{r'}^{[1]}-e_{r'}^{(\beta)[1]} k_{t'}^{[1]})]
\frac{\partial}
{\partial p_{r'}^{(j)}}
\right.\nonumber \\&&\left.
-\pi({\bf v}^{(j)}.{\bf e}^{(\beta)[1]})
\left( 
2 \frac{\partial}
{\partial \eta_{\beta}^{[1]}}
\right)
\right]
\exp -\left(
a{\bf k}^{[1]}.\frac{\partial}{\partial {\bf k}^{(j)}}
\right)
\nonumber \\&&\times
\delta(\eta_{\beta}^{[1]}-\eta_\beta)\delta(\eta_{\beta'}^{[1]}-\eta_{\beta'})
\delta^{Kr}_{m_{\beta}^{[1]},a}\delta^{Kr}_{m_{\beta'}^{[1]},0}
\nonumber \\&&+
\sum_{\beta=1,2} 
\sum_{a=\pm 1}
\frac{-1}{(-a{\bf k}^{[1]}.{\bf v}^{(j)}+ak^{[1]})^2}
(-i)e_j \frac1{(2\pi)^{\frac32}} 
\left(
\frac{\eta_{\beta}^{[1]}}
{k^{[1]}}
\right)^{\frac12} 
\nonumber \\&&\times
\left[ 
[k^{[1]} e_{r'}^{(\beta)[1]}-g^{s't'} v_{s'}^{(j)}
(e_{t'}^{(\beta)[1]} k_{r'}^{[1]}-e_{r'}^{(\beta)[1]} k_{t'}^{[1]})]
\right]
(k_{s}^{(j)}-ak_s^{[1]})\frac{\partial v_s^{(j)}}
{\partial p_{r'}^{(j)}}
\nonumber \\&&\times
\exp -\left(
a{\bf k}^{[1]}.\frac{\partial}{\partial {\bf k}^{(j)}}
\right)
\delta(\eta_{\beta}^{[1]}-\eta_\beta)\delta(\eta_{\beta'}^{[1]}-\eta_{\beta'})
\delta^{Kr}_{m_{\beta}^{[1]},a}\delta^{Kr}_{m_{\beta'}^{[1]},0}.
\label{4.18}
\end{eqnarray}
The limits $\eta_\beta\to 0$, $\eta_{\beta'}\to 0$ have to be performed after the integration over
$\eta_{\beta}^{[1]}$ and $\eta_{\beta'}^{[1]}$.

Using (3.6), the original variables $\xi^{[1]}$ are reintroduced in place of $m^{[1]}$.

\subsection{Field associated with free particles}

Since $\tilde f_{11[1(e_j)]}=\tilde f_{11[1(s_j)]}$, we
have at first order in the field interaction and zero${}^{th}$ order in the Coulomb interaction:
\begin{eqnarray}
&&\tilde f_{11[1(e_j)]}^{(0,1)}
=
\sum_{\beta=1,2} 
\sum_{a=\pm 1}
e^{-2\pi i(a\xi_{\beta}^{[1]})}
\left(\frac{1}{-a{\bf k}^{[1]}.{\bf v}^{(j)}+ak^{[1]}}
\right)
(-i)e_j \frac1{(2\pi)^{\frac32}} 
\nonumber \\&&\times
\left(
\frac{\eta_{\beta}^{[1]}}
{k^{[1]}}
\right)^{\frac12} 
\left[ 
[k^{[1]} e_{r'}^{(\beta)[1]}-g^{s't'} v_{s'}^{(j)}
(e_{t'}^{(\beta)[1]} k_{r'}^{[1]}-e_{r'}^{(\beta)[1]} k_{t'}^{[1]})]
\frac{\partial}
{\partial p_{r'}^{(j)}}
\right.\nonumber \\&&\left.
-\pi({\bf v}^{(j)}.{\bf e}^{(\beta)[1]})
\left( 
2 \frac{\partial}
{\partial \eta_{\beta}^{[1]}}
\right)
\right]
\exp -\left(
a{\bf k}^{[1]}.\frac{\partial}{\partial {\bf k}^{(j)}}
\right)
\nonumber \\&&\times
\delta(\eta_{\beta}^{[1]}-\eta_\beta)\delta(\eta_{\beta'}^{[1]}-\eta_{\beta'})
\tilde f_{11[0]}
\nonumber \\&&+
\sum_{\beta=1,2} 
\sum_{a=\pm 1}
e^{-2\pi i(a\xi_{\beta}^{[1]})}
\frac{-1}{(-a{\bf k}^{[1]}.{\bf v}^{(j)}+ak^{[1]})^2}
(-i)e_j \frac1{(2\pi)^{\frac32}} 
\left(
\frac{\eta_{\beta}^{[1]}}
{k^{[1]}}
\right)^{\frac12} 
\nonumber \\&&\times
\left[ 
[k^{[1]} e_{r'}^{(\beta)[1]}-g^{s't'} v_{s'}^{(j)}
(e_{t'}^{(\beta)[1]} k_{r'}^{[1]}-e_{r'}^{(\beta)[1]} k_{t'}^{[1]})]
\right]
(k_{s}^{(j)}-ak_s^{[1]})\frac{\partial v_s^{(j)}}
{\partial p_{r'}^{(j)}}
\nonumber \\&&\times
\exp -\left(
a{\bf k}^{[1]}.\frac{\partial}{\partial {\bf k}^{(j)}}
\right)
\delta(\eta_{\beta}^{[1]}-\eta_\beta)\delta(\eta_{\beta'}^{[1]}-\eta_{\beta'})
\tilde f_{11[0]}.
\label{4.19a}
\end{eqnarray}
Combining the form (\ref{2.2}) for the observables associated with the transverse electric field and the form
(\ref{2.5}), we get the average transverse component of the electric field
and we can write
\begin{eqnarray}
&&<{\bf E}_r^{{\bot }}({\bf x})>^{e_j(0,1)}
\nonumber \\&&=
\int d^3{\bf k}^{[1]}\int_0^{\infty}d\eta_1^{[1]}\int_0^{\infty}d\eta_2^{[1]}
\int_0^1d\xi_1^{[1]}\int_0^1d\xi_2^{[1]}\int d^6x^{(1)}\, d^6x^{(2)}
\nonumber \\ &&\times
\frac1{(2\pi)^{\frac32}}\sum_{\alpha=1,2} \sum_{a'=\pm 1}
{k^{[1]}}^{\frac12} {\bf e}_r^{\alpha}({\bf k}^{[1]}) \eta_{\alpha}^{\frac12}({\bf k}^{[1]})
\nonumber \\ &&\times
\exp \{ia'[{\bf k}^{[1]}.{\bf x}-2\pi\xi_{\alpha}({\bf k}^{[1]})]\}
f_{11[1ej]}^{(0,1)}(x^{(1)}, x^{(2)};\chi^{[1]};{\bf k}^{[1]}).
\label{4.20}
\end{eqnarray}
We now take for $\tilde f_{11[0]}$ in (\ref{4.19a}) a distribution function corresponding to sharp values
of the positions and momenta and (\ref{3.5}) enables to get the expression into the variables ${\bf
k}^{(j)}$,
${\bf p}^{(j)}$:
$$
\tilde {f}_{11[0]}({\bf q}^{(1)},{\bf p}^{(1)},{\bf q}^{(2)},{\bf p}^{(2)})
=\delta({\bf q}^{(1)}-{\bf q}_1)\delta({\bf q}^{(2)}-{\bf q}_2)
\delta({\bf p}^{(1)}-{\bf p}_1)\delta({\bf p}^{(2)}-{\bf p}_2),
$$
\begin{eqnarray}
&&\tilde {f}_{11[0]}({\bf k}^{(1)},{\bf p}^{(1)},{\bf k}^{(2)},{\bf p}^{(2)})
=
\int d^3q^{(1)}\,d^3q^{(2)}\,e^{-i({\bf k}^{(1)}.{\bf q}^{(1)}+{\bf k}^{(2)}.{\bf q}^{(2)})}
\nonumber \\&&\qquad\times
\delta({\bf q}^{(1)}-{\bf q}_1)\delta({\bf q}^{(2)}-{\bf q}_2)
\delta({\bf p}^{(1)}-{\bf p}_1)\delta({\bf p}^{(2)}-{\bf p}_2)
\nonumber \\&&\qquad=
e^{-i({\bf k}^{(1)}.{\bf q}_1+{\bf k}^{(2)}.{\bf q}_2)}
\delta({\bf p}^{(1)}-{\bf p}_1)\delta({\bf p}^{(2)}-{\bf p}_2).
\label{4.23}
\end{eqnarray}
Combining the previous terms
and performing all trivial integrations, we get:
\begin{eqnarray}
&&<{\bf E}_r^{{\bot }}({\bf x})>^{e_j(0,1)}
=(-i)(2\pi)e_j\frac1{(2\pi)^3}\int d^3{\bf k}^{[1]}
\sum_{\alpha=1,2} \sum_{a=\pm 1}
 {\bf e}_r^{\alpha}({\bf k}^{[1]}) 
\nonumber \\ &&\times
\exp \{ia[{\bf k}^{[1]}.({\bf x}-{\bf q}_j)]\}
\left(\frac{1}{a{\bf k}^{[1]}.{\bf v}_j-ak^{[1]}}
\right)
({\bf v}_j.{\bf e}^{(\alpha)[1]}).
\label{4.31}
\end{eqnarray}
The value $<{\bf E}_r^{{\bot }}({\bf x})>^{e_j(0,1)}$ is now determined by an expression
that involves the values of the position  ${\bf q}_j$ and momentum ${\bf p}_j$ of the charged
particle $j$ at the same time.

The summations over $a$ and over the polarisation vectors lead to
\begin{eqnarray}
&&<{\bf E}^{{\bot }}({\bf x})>^{e_j(0,1)}
=(-i)(2\pi)e_j\frac1{(2\pi)^3}\int d^3{\bf k}^{[1]}
\left[\exp \{i[{\bf k}^{[1]}.({\bf x}-{\bf q}_j)]\}
\right.\nonumber \\ &&\left.
-\exp \{-i[{\bf k}^{[1]}.({\bf x}-{\bf q}_j)]\}\right]
\left(\frac{1}{{\bf k}^{[1]}.{\bf v}_j-k^{[1]}}
\right)
\left({{\bf v}}_{j}-\frac{({\bf v}_j.{\bf k}^{[1]}){\bf k}^{[1]}}{(k^{[1]})^2}\right).
\label{4.33}
\end{eqnarray}
This expression behaves obviously as $\frac1{\vert{\bf x}-{\bf q}_j\vert^2}$ and does not
describe a propagating field. It presents a discontinuity and vanishes exactly at the location of the
particle ${\bf x}={\bf q}_j$ since the integrand is identically null for that value.
In view of the value provided by (\ref{3.11d}), this explains why the corresponding terms in the kinetic
operator $\Theta$ vanishes. 
In our approach, the contribution of each mode to the self-interaction of the
electron is computed first.
When no acceleration mechanism is provided, each contribution vanishes exactly.
In the usual approach, the emitted fields are computed from the Li\'enard-Wiechert potential and 
evaluated, via the Abraham-Lorentz model, at the localisation of the electron.

The expression for the electric field can be further analysed.
We can write (\ref{4.33}) as the sum of two contributions by replacing 
$\left(({\bf v}_{j}-\frac{({\bf v}_j.{\bf k}^{[1]}){\bf k}^{[1]}}{(k^{[1]})^2}\right)$ 
by the sum $\left[({\bf v}_{j}
-\frac{{\bf k}^{[1]}}{k^{[1]}})+(\frac{{\bf k}^{[1]}}{k^{[1]}}-\frac{({\bf v}_j.{\bf k}^{[1]}){\bf k}^{[1]}}{(k^{[1]})^2})\right]$.
The contribution of the second term is:
\begin{eqnarray}
&&<{\bf E}^{{\bot }}({\bf x})>_{b}^{e_j(0,1)}
=-(4\pi)e_j\frac1{(2\pi)^3}\int d^3{\bf k}^{[1]}
\sin [{\bf k}^{[1]}.({\bf x}-{\bf q}_j)]
\frac{{\bf k}^{[1]}}{(k^{[1]})^2}\nonumber \\
&&=(4\pi)e_j\frac1{(2\pi)^3}{\bf\nabla_x}\int d^3{\bf k}^{[1]}
\cos[{\bf k}^{[1]}.({\bf x}-{\bf q}_j)]
\frac{1}{(k^{[1]})^2}.
\label{4.34}
\end{eqnarray}

From the known relations $\int d^3{\bf k}^{[1]}\exp \{-i[{\bf k}^{[1]}.{\bf x}]\}\frac{1}{k^{[1]}}$
$=4\pi\frac1{x^2}$ and 
\newline\noindent
$\int d^3{\bf k}^{[1]}\exp \{i[{\bf k}^{[1]}.{\bf x}]\}\frac{1}{(k^{[1]})2}$
$=2\pi^2\frac1{x}$, the following identification is possible:
\begin{equation}
<{\bf E}^{{\bot }}({\bf x})>_{b}^{e_j(0,1)}
=e_j{\bf\nabla_x}\frac1{\vert{\bf x}-{\bf q}_j\vert}
=-<{\bf E}^{\parallel }({\bf x})>.
\label{4.35b}
\end{equation}
Therefore, the first contribution $<{\bf E}^{{\bot }}({\bf x})>_{a}^{e_j(0,1)}$ should be identified
with the complete electric field $<{\bf E}({\bf x})>^{e_j(0,1)}$:
\begin{eqnarray}
&&<{\bf E}^{{\bot }}({\bf x})>_a^{e_j(0,1)}
=(-i)(2\pi)e_j\frac1{(2\pi)^3}\int d^3{\bf k}^{[1]}
\left[\exp \{i[{\bf k}^{[1]}.({\bf x}-{\bf q}_j)]\}
\right.\nonumber \\ &&\left.
-\exp \{-i[{\bf k}^{[1]}.({\bf x}-{\bf q}_j)]\}\right]
\left(k^{[1]}{\bf v}_{j}-{\bf k}^{[1]}\right)\frac1{k^{[1]}}
\nonumber \\&&=(-i)(2\pi)e_j\frac1{(2\pi)^3}\int d^3{\bf k}^{[1]}
\left[\exp \{i[{\bf k}^{[1]}.({\bf x}-{\bf q}_j)]\}
\right.\nonumber \\ &&\left.
-\exp \{-i[{\bf k}^{[1]}.({\bf x}-{\bf q}_j)]\}\right]
\left(\frac{1}{k^{[1]}-{\bf k}^{[1]}.{\bf v}_j}
\right)
\left({\bf k}^{[1]}-k^{[1]}{\bf v}_{j}\right)\frac1{k^{[1]}}.
\label{4.36}
\end{eqnarray}
If the $x$ axis is placed along $({\bf x}-{\bf q}_j)$ and the $y$ axis along ${\bf v}_{\bot j}$,
defined by
${\bf v}_{\bot j}={\bf v}_{j}-\frac{[{\bf v}_{ j}.({\bf x}-{\bf q}_j)]({\bf x}-{\bf q}_j)}
{\vert{\bf x}-{\bf q}_j\vert^2}$, we show in Appendix C that:
\begin{equation}
<{\bf E}^{{\bot }}({\bf x})>_{a}^{e_j(0,1)}
=[1-v_{j}^2]
\frac1{(1-v_{jy}^2)^{\frac32}}\frac1{\vert{\bf x}-{\bf q}_j\vert^2}{\bf e}_x.
\label{4.36ifa}
\end{equation}
On the other hand (11.154) of \cite{JJ98} gives us the field in terms of the position of the
charges at the same time: ($r=\vert{\bf x}-{\bf q}_j\vert$, $\beta$ can be identified with $v_j$,
$q=e_j$, 
$cos\psi={{\bf n}}.\frac{{\bf v}_j}{v_j}$, ${\bf r}=r{\bf n}$, $\gamma^2=\frac1{1-\beta^2}$)
\begin{equation}
{\bf E}=\frac{q{\bf r}}{r^3\gamma^2(1-\beta^2\sin^2\psi)^{\frac32}}.
\label{4.33o}
\end{equation}
Therefore, our expressions (\ref{4.33}-\ref{4.36}) reproduce correctly usual results for the
complete and transverse electric field outside the location of the charged particle.

\subsection{Physical interpretation}

The field that has just been computed is the field generated by a particle in uniform motion since the
Coulomb interaction between the charged particles or an outside field is not taken into account in its 
contribution. 
That field is therefore equivalent to the field that can be deduced fom the static
Coulomb field through a Lorentz transformation and this corresponds indeed to our result.
This point has to be viewed as a confirmation of the correctness of our alternative approach.
The usual (relativistic) expression is recovered outside the location of the particle ${\bf x}={\bf q}_j$.

On physical grounds, a charged particle in free motion does not emit a field and should 
experiment no self-force.
Our expression  (\ref{4.33}) is in accordance with that property since the self-field vanishes exactly
at the location of the point charged particle.
That property holds as well for the transverse field $<{\bf E}^{{\bot }}({\bf x})>$ as for the complete 
one. 
This explains why the corresponding terms in the kinetic operator $\Theta$  (\ref{3.52gb})vanish.

\section{ The radiative reaction force due to the Coulomb interaction}   

\setcounter{equation}{0}

In order to get a contribution to the reactive force due to the self-interaction of the particles, a
mechanism of acceleration of the particles has to be provided.
We have chosen to consider the Coulomb interaction between the charged particles as responsible for
the acceleration.
Other mechanisms are possible, such as the presence of a non-vanishing free field, or the consideration of
the field emitted by the other particles but they are not treated here.
We have to evaluate the elements of $\tilde\Sigma(t)^{(1,2)}$ (corresponding to one Coulomb
interaction and two interactions with the transverse fields)  that provide a contribution to $<11[0]|\tilde
\Theta|11[1(f)]>^{(1,2)}$, when that operator acts on the field vacuum.
The Coulomb interaction between the two particles can occur as the first, the second or
the last interaction.
Since we know that $<11[0]|\tilde \Theta|11[1(f)]>^{(0,2)}$ provides a vanishing result when
acting on the field vacuum, we expect that the only non vanishing contribution arises when the
Coulomb interaction takes place between or after the interaction of the particles  with the
transverse field.

\subsection{The subdynamics operator}

Therefore, we first focus on (the lower index $FPF$ describes the order of the interactions):
\begin{eqnarray}
&&<11[0]|\tilde \Sigma|11[1(f)]>^{(1,2)}_{FPF}
\nonumber \\&&=
\frac{-1}{2\pi i} \int'_c dz\,e^{-izt}
\sum_{j=1,2}
\left(\frac{1}{z-{\bf k}^{(j)}.{\bf v}^{(j)}-{\bf k}^{(j')}.{\bf v}^{(j')}}
\right)
\nonumber \\&&\times
(-i) e_j \frac1{(2\pi)^{\frac32}}
\int d^3k^{[1]}
\int_0^{\infty}d\eta_1^{[1]}\int_0^{\infty}d\eta_2^{[1]}
\sum_{m_{1}^{[1]},m_{2}^{[1]}}
\delta_{m_{1}^{[1]},0}\delta_{m_{2}^{[1]},0}
\sum_{\alpha=1,2} 
\sum_{a=\pm 1}
\nonumber \\&&\times
\left(
\frac{\eta_{\alpha}^{[1]}}
{k^{[1]}}
\right)^{\frac12} 
\left[ 
[k^{[1]} e_r^{(\alpha)[1]}-g^{st} v_s^{(j)}
(e_t^{(\alpha)[1]} k_r^{[1]}-e_r^{(\alpha)[1]} k_t^{[1]})]
\frac{\partial}
{\partial p_r^{(j)}}
\right]
\nonumber \\&&\times
\exp a\left\{-{\bf k}^{[1]}.\frac{\partial}{\partial {\bf
k}^{(j)}}-\frac{\partial}{\partial m_{\alpha}^{[1]}}\right\}
\nonumber \\ &&\times
\left(\frac{1}{z-{\bf k}^{(j)}.{\bf v}^{(j)}-{\bf k}^{(j')}.{\bf v}^{(j')}
+k^{[1]}(m_{\alpha}^{[1]}+m_{\alpha'}^{[1]})}
\right)
\nonumber \\&&\times
e_je_{j'}\frac{-1}{8\pi^2}\int d^3 l\frac{1}{l^2}
{\bf l}.\left(\frac{\partial}{\partial {\bf p}^{(j)}}-\frac{\partial}
{\partial {\bf p}^{(j')}}
\right)
e^{\frac{{\bf l}}{2}.\left(\frac{\partial}{\partial {\bf k}^{(j)}}-\frac{\partial}
{\partial {\bf k}^{(j')}}
\right)}
\nonumber \\&&\times\left(\frac{1}{z-{\bf k}^{(j)}.{\bf v}^{(j)}-{\bf k}^{(j')}.{\bf v}^{(j')}
+k^{[1]}(m_{\alpha}^{[1]}+m_{\alpha'}^{[1]})}
\right)
\nonumber \\&&\times
(-i)e_j \frac1{(2\pi)^{\frac32}} 
\sum_{\beta=1,2} 
\sum_{b=\pm 1}
\left(
\frac{\eta_{\beta}^{[1]}}
{k^{[1]}}
\right)^{\frac12} 
\nonumber \\&&\times
\left[ 
[k^{[1]} e_{r'}^{(\beta)[1]}-g^{s't'} v_{s'}^{(j)}
(e_{t'}^{(\beta)[1]} k_{r'}^{[1]}-e_{r'}^{(\beta)[1]} k_{t'}^{[1]})]
\frac{\partial}
{\partial p_{r'}^{(j)}}
\right.\nonumber \\ &&\left.
-\pi({\bf v}^{(j)}.{\bf e}^{(\beta)[1]})
\left( 
2 \frac{\partial}
{\partial \eta_{\beta}^{[1]}}
-\frac{b}{\eta_{\beta}^{[1]}}(m_{\beta}^{[1]}-b)
\right)
\right]
\exp b\left\{-{\bf k}^{[1]}.\frac{\partial}{\partial {\bf k}^{(j)}}-\frac{\partial}{\partial
m_{\beta}^{[1]}}\right\}
\nonumber \\&&\times
\left(\frac{1}{z-{\bf k}^{(j)}.{\bf v}^{(j)}-{\bf k}^{(j')}.{\bf v}^{(j')}
+k^{[1]}(m_{\beta}^{[1]}+m_{\beta'}^{[1]})}
\right).
\label{5.1}
\end{eqnarray}
This expression is very similar to the expression of  $<11[0]|\tilde \Sigma|11[1(f)]>^{(0,2)}$ (\ref{3.14}), but
with the supplementary factors due to the Coulomb interaction: the matrix element (\ref{3.9}) and a
propagator.
The order of all the elements has to be strictly respected, on view of the presence of displacement and
derivation operators.
The contributions due to the different orders of interaction ($FFP$ and $PFF$) are evaluated from 
similar expressions.
We can proceed exactly as for the second order contribution $<11[0]|\tilde \Sigma|11[1(f)]>^{(0,2)}$.
A difference is the presence of a denominator that can be resonant.
The subdynamics theory has prescribed, from the beginning of its elaboration, that a propagator
corresponding to a correlation state has to be treated with an $i\epsilon$.
A second difference is the consideration of  $<11[0]|\tilde{{\cal A}}|11[1(f)]>^{(0,2)}\tilde f^V_{[1(f)]}$
(\ref{4.9}) for the extraction of $<11[0]|\tilde \Theta|11[1(f)]>\tilde f^V_{[1(f)]}$ from $<11[0]|\tilde
\Sigma(t)|11[1(f)]>\tilde f^V_{[1(f)]}$ 
through (\ref{3.37}).
Moreover, we have to take into account, in the final simplifications, that the matrix elements
associated with the Coulomb interaction and the field interaction do not commute.
The terms corresponding to that case are affected by a lower index $II$, the other ones by an index $I$.

\subsection{Kinetic operator}

Straightforward but very lenghty computations lead to the following expression (as expected, the
$PFF$ order of interaction has provided a vanishing result)
\begin{eqnarray}
&&<11[0]|\tilde \Theta|11[1(f)]>_I^{(1,2)}\tilde f^V_{[1(f)]}
\nonumber \\&&=
\sum_{j=1,2}
(-i) e_j^3 e_{j'}\frac1{(2\pi)^{3}}\frac{1}{4\pi}
\int d^3k^{[1]}\int d^3 l\,\frac{1}{l^2}
\sum_{\alpha=1,2} 
\sum_{a=\pm 1}
\nonumber \\&&\times
\frac{1}{k^{[1]}}
\left[ 
[k^{[1]} e_r^{(\alpha)[1]}-g^{st} v_s^{(j)}
(e_t^{(\alpha)[1]} k_r^{[1]}-e_r^{(\alpha)[1]} k_t^{[1]})]
\frac{\partial}
{\partial p_r^{(j)}}
\right]
\nonumber \\&&\times
\left(\frac{1}{i\epsilon+a{\bf k}^{[1]}.{\bf v}^{(j)}
-ak^{[1]}}
\right)^2
\left(\frac{1}{i\epsilon+(\frac12{\bf l}+a{\bf k}^{[1]}).{\bf v}^{(j)}
-\frac12{\bf l}.{\bf v}^{(j')} -ak^{[1]}}
\right)
\nonumber \\&&\times
l_v(-a k_u^{[1]})
\left(\frac{\partial v_u^{(j)}}{\partial p_v^{(j)}}\right)
2\pi({\bf v}^{(j)}.{\bf e}^{(\alpha)[1]})
e^{\frac{{\bf l}}{2}.\left(\frac{\partial}{\partial {\bf k}^{(j)}}-\frac{\partial}
{\partial {\bf k}^{(j')}}
\right)},
\label{8.1}
\end{eqnarray}
\begin{eqnarray}
&&<11[0]|\tilde \Theta|11[1(f)]>^{(1,2)}_{II}\tilde f^V_{[1(f)]}
\nonumber \\&&=
\sum_{j=1,2}
(-i) e_j^3 e_{j'}\frac1{(2\pi)^{3}}\frac{1}{4\pi}\int d^3k^{[1]}\int d^3 l\frac{1}{l^2}
\sum_{\alpha=1,2} 
\sum_{a=\pm 1}
\nonumber \\&&\times
\frac{1}{k^{[1]}}
\left[ 
[k^{[1]} e_r^{(\alpha)[1]}-g^{st} v_s^{(j)}
(e_t^{(\alpha)[1]} k_r^{[1]}-e_r^{(\alpha)[1]} k_t^{[1]})]
\frac{\partial}
{\partial p_r^{(j)}}
\right]
\nonumber \\&&\times
\left(\frac{1}{i\epsilon+(\frac12{\bf l}+a{\bf k}^{[1]}).{\bf v}^{(j)}
-\frac12{\bf l}.{\bf v}^{(j')} -ak^{[1]}}
\right)
\left(\frac{1}{i\epsilon+a{\bf k}^{[1]}.{\bf v}^{(j)}
-ak^{[1]}}
\right)
\nonumber \\&&\times
l_v
\frac{\partial v_u^{(j)}}{\partial p_v^{(j)}}
e_u^{(\alpha)[1]}
e^{\frac{{\bf l}}{2}.\left(\frac{\partial}{\partial {\bf k}^{(j)}}-\frac{\partial}
{\partial {\bf k}^{(j')}}
\right)}.
\label{8.1a}
\end{eqnarray}
The order of integration  respects the ordering of the apparition of the
vertices: in all remaining contributions, the integration over the field modes ${\bf k}^{[1]}$ has to
be performed after the integration over the wave number ${\bf l}$ exchanged by the Coulomb
interaction. An opposite order of integration would have been required in the contribution
involving the order $PFF$ of the vertices.

 For the sake of completion, let us compute
$\left(\frac{\partial v_u^{(j)}}{\partial p_u^{(j)}}\right)$. We have ${\bf p}=\frac{m{\bf v}}{(1-v^2)^{\frac12}}$, from which we deduce
$\left(\frac{\partial v_u^{(j)}}{\partial
p_v^{(j)}}\right)=\frac{\delta^{Kr}_{u,v}}{(m_j^2+(p^{(j)})^2)^{\frac12}}
-\frac{(p_u^{(j)}p_v^{(j)})}{(m_j^2+(p^{(j)})^2)^{\frac32}}$.

The partial derivative $\frac{\partial}{\partial p_r^{(j)}}$ can be placed in front of the matrix element
since the contribution when it acts on the factor $v_s^{(j)}$ provide a vanishing result.
The property can be checked explicitly.
Taking into account the value of the matrix tensor $g^{st}$ = $-\delta^{Kr}_{s,t}$, the first term of
the derivative, with a $\delta^{Kr}_{r,s}$, can be seen to involve the scalar product of the vectors
${\bf e}^{(\alpha)[1]}$ and ${\bf k}^{[1]}$ that vanishes by definition of the polarisation vector.
The second contribution, with the product $p_u^{(j)}p_v^{(j)}$, vanishes by symmetry.
This result is not unexpected and reflects that the magnetic force is orthogonal to the velocity
vector.

This form  (\ref{8.1}) shows clearly that the norm is not affected by that contribution to the equations of
motion. Indeed, the partial derivative $\frac{\partial}{\partial p_r^{(j)}}$ ensures that the whole
contribution vanishes when integrated over $p_r^{(j)}$. We have the same structure that for the
contribution (7.15) of \cite{BP74} or for the operator $<11[0]|\tilde \Theta|11[1(f)]>^{(0,2)}$.

Since the first propagator in  (\ref{8.1}) and the second one in  (\ref{8.1a}) cannot be resonant, the
$i\epsilon$ can be dropped from them. 

We do not yet analyse the possible divergence of these contributions to 
$<11[0]|\tilde\Theta|11[1(f)]>^{(1,2)}$.

\subsection{Evolution of the distribution function}

We consider anew the case in which the two particles $j$ and $j'$ are perfectly
localised with a well defined momentum (\ref{4.23}).
If we perform the trivial integrations, due to simplified form of the distribution function, we get:
\begin{eqnarray}
&&
\left. \partial_t \tilde {f}_{11[0]}({\bf q}^{(1)},{\bf p}^{(1)},{\bf q}^{(2)},{\bf p}^{(2)})
\right\vert_{\theta I}
\nonumber \\&&=
i\frac1{(2\pi)^3}\frac{1}{4\pi}\sum_{j=1,2} e_j^3 e_{j'}
\int d^3k^{[1]}\int d^3 l\,\frac{1}{l^2}
\sum_{\alpha=1,2} 
\sum_{a=\pm 1}
\frac{1}{k^{[1]}}
\nonumber \\&&\times
\left[ 
[k^{[1]} e_r^{(\alpha)[1]}-g^{st} v_s^{(j)}
(e_t^{(\alpha)[1]} k_r^{[1]}-e_r^{(\alpha)[1]} k_t^{[1]})]
\frac{\partial}
{\partial p_r^{(j)}}
\right]
\nonumber \\&&\times
\left(\frac{1}{{\bf k}^{[1]}.{\bf v}^{(j)}
-k^{[1]}}
\right)^2
\left(\frac{1}{i\epsilon+(\frac12{\bf l}+a{\bf k}^{[1]}).{\bf v}^{(j)}
-\frac12{\bf l}.{\bf v}^{(j')} -ak^{[1]}}
\right)
\nonumber \\&&\times
al_vk_u^{[1]}
\left(\frac{\partial v_u^{(j)}}{\partial p_v^{(j)}}\right)
({\bf v}^{(j)}.{\bf e}^{(\alpha)[1]})
e^{-i\frac{{\bf l}}{2}.({\bf q}_j-{\bf q}_{j'})}
\nonumber \\&&\times
\delta({\bf q}^{(j)}-{\bf q}_j)\delta({\bf q}^{(j')}-{\bf q}_{j'})
\delta({\bf p}^{(j)}-{\bf p}_j)\delta({\bf p}^{(j')}-{\bf p}_{j'}),
\label{8.9}
\end{eqnarray}
\begin{eqnarray}
&&
\left. \partial_t \tilde {f}_{11[0]}({\bf q}^{(1)},{\bf p}^{(1)},{\bf q}^{(2)},{\bf p}^{(2)})
\right\vert_{\theta II}
\nonumber \\&&=
i\frac1{(2\pi)^3} \frac{-1}{4\pi}\sum_{j=1,2}e_j^3 e_{j'}
\int d^3k^{[1]}\int d^3 l\frac{1}{l^2}
\sum_{\alpha=1,2} 
\sum_{a=\pm 1}
\nonumber \\&&\times\frac{1}
{k^{[1]}}
\left[ 
[k^{[1]} e_r^{(\alpha)[1]}-g^{st} v_s^{(j)}
(e_t^{(\alpha)[1]} k_r^{[1]}-e_r^{(\alpha)[1]} k_t^{[1]})]
\frac{\partial}
{\partial p_r^{(j)}}
\right]
\nonumber \\&&\times
\left(\frac{1}{i\epsilon+(\frac12{\bf l}+a{\bf k}^{[1]}).{\bf v}^{(j)}
-\frac12{\bf l}.{\bf v}^{(j')} -ak^{[1]}}
\right)
\left(\frac{1}{a{\bf k}^{[1]}.{\bf v}^{(j)}
-ak^{[1]}}
\right)
\nonumber \\&&\times
 l_v
\frac{\partial v_u^{(j)}}{\partial p_v^{(j)}}
e_u^{(\alpha)[1]}
e^{-i\frac{{\bf l}}{2}.({\bf q}_j-{\bf q}_{j'})}
\delta({\bf q}^{(j)}-{\bf q}_j)\delta({\bf q}^{(j')}-{\bf q}_{j'})
\nonumber \\&&\times
\delta({\bf p}^{(j)}-{\bf p}_j)\delta({\bf p}^{(j')}-{\bf p}_{j'}).
\label{8.9a}
\end{eqnarray}

\subsection{ Radiative reaction force}

The effect of the coupling of the Coulomb interaction with the field is to provide a supplementary
force, the  radiative reaction force, that changes the mean value of the momentum of one particle.
The expression of the $r$ component $F_r^{(j)}$ of the radiative reaction force can be obtained by
considering the relation $F_r^{(j)}=\frac{d}{dt}<p_r^{(j)}>$.
By a partial derivative, we get that 
the $r$ component $F_r^{(j)}$ is provided by minus the
coefficient of the expression (\ref{8.9}) when the partial derivative $\frac{\partial}{\partial p_r^{(j)}}$ is
removed and where the variables ${\bf q}^{(j)}$, ${\bf q}^{(j')}$, ${\bf p}^{(j)}$ and ${\bf p}^{(j')}$ are
replaced by their values obtained from the Dirac delta functions. 
\begin{eqnarray}
&&<F_r^{(j)}>_I
=
-i\frac1{(2\pi)^3}\frac{1}{4\pi} e_j^3 e_{j'}
\int d^3k^{[1]}\int d^3 l\,\frac{1}{l^2}
\sum_{\alpha=1,2} 
\sum_{a=\pm 1}
\frac{1}{k^{[1]}}
e^{-i\frac{{\bf l}}{2}.({\bf q}_j-{\bf q}_{j'})}
\nonumber \\&&\times
[k^{[1]} e_r^{(\alpha)[1]}-g^{st} v_{js}
(e_t^{(\alpha)[1]} k_r^{[1]}-e_r^{(\alpha)[1]} k_t^{[1]})]
\left(\frac{1}{{\bf k}^{[1]}.{\bf v}_{j}
-k^{[1]}}
\right)^2
\nonumber \\&&\times
\left(\frac{1}{i\epsilon+(\frac12{\bf l}+a{\bf k}^{[1]}).{\bf v}_{j}
-\frac12{\bf l}.{\bf v}_{j'} -ak^{[1]}}
\right)
al_vk_u^{[1]}
\nonumber \\&&\times
\left[
\frac{\delta^{Kr}_{u,v}}{(m_j^2+p_j^2)^{\frac12}}
-\frac{(p_{ju}p_{jv})}{(m_j^2+p_j^2)^{\frac32}}
\right]
({\bf v}_{j}.{\bf e}^{(\alpha)[1]}),
\label{8.10}
\end{eqnarray}
\begin{eqnarray}
&&<F_r^{(j)}>_{II}=
-i \frac1{(2\pi)^{3}}\frac{-1}{4\pi}e_j^3 e_{j'}
\int d^3k^{[1]}\int d^3 l\frac{1}{l^2}
\sum_{\alpha=1,2} 
\sum_{a=\pm 1}
e^{-i\frac{{\bf l}}{2}.({\bf q}_j-{\bf q}_{j'})}
\nonumber \\&&\times
\frac{1}{k^{[1]}}
\left[ 
[k^{[1]} e_r^{(\alpha)[1]}-g^{st} v_s^{(j)}
(e_t^{(\alpha)[1]} k_r^{[1]}-e_r^{(\alpha)[1]} k_t^{[1]})]
\right]
\nonumber \\&&\times
\left[
\left(\frac{1}{i\epsilon+(\frac12{\bf l}+a{\bf k}^{[1]}).{\bf v}^{(j)}
-\frac12{\bf l}.{\bf v}^{(j')} -ak^{[1]}}
\right)
\left(\frac{1}{a{\bf k}^{[1]}.{\bf v}^{(j)}
-ak^{[1]}}
\right)
\right]
\nonumber \\&&\times
l_v
\frac{\partial v_u^{(j)}}{\partial p_v^{(j)}}
e_u^{(\alpha)[1]}.
\label{8.10a}
\end{eqnarray}
We focus first on the first contribution.
The value of that radiative reaction force depends on the relative orientation of the vectors
position  and momentum.
We explicit the summations over $u$ and $v$.
We get then, using the value of the metric tensor $g^{st}$ to replace $g^{st} v_{js}e_t^{(\alpha)[1]}
k_r^{[1]}$ by $-{\bf v}.{\bf e}^{(\alpha)[1]}k_r^{[1]}$ and $g^{st} v_{js}e_r^{(\alpha)[1]} k_t^{[1]}$ by 
$-{\bf v}.{\bf k}^{[1]}e_r^{(\alpha)[1]}$:
\begin{eqnarray}
&&<F_r^{(j)}>_I
=
-i\frac1{(2\pi)^3}\frac{e_j^3 e_{j'}}{4\pi}
\int d^3k^{[1]}\int d^3 l\,\frac{1}{l^2}
\sum_{\alpha=1,2} 
\sum_{a=\pm 1}a
\frac{1}{k^{[1]}}
e^{-i\frac{{\bf l}}{2}.({\bf q}_j-{\bf q}_{j'})}
\nonumber \\&&\times
[(k^{[1]}-{\bf v}_j.{\bf k}^{[1]}) e_r^{(\alpha)[1]}
+{\bf v}_j.{\bf e}^{(\alpha)[1]}k_r^{[1]}
]
\nonumber \\&&\times
\sum_{u,v}l_vk_u^{[1]}
\left[
\frac{\delta^{Kr}_{u,v}}{(m_j^2+p_j^2)^{\frac12}}
-\frac{(p_{ju}p_{jv})}{(m_j^2+p_j^2)^{\frac32}}\right]
({\bf v}_{j}.{\bf e}^{(\alpha)[1]})
\nonumber \\&&\times
\left(\frac{1}{{\bf k}^{[1]}.{\bf v}_{j}
-k^{[1]}}
\right)^2
\left(\frac{1}{i\epsilon+(\frac12{\bf l}+a{\bf k}^{[1]}).{\bf v}_{j}
-\frac12{\bf l}.{\bf v}_{j'} -ak^{[1]}}
\right).
\label{8.11}
\end{eqnarray}
The reality of this expression can be checked by considering the symmetry $a\to -a$, ${\bf l}\to
-{\bf l}$.

\subsection{Emitted power}

We distinguish the components of the radiative reaction force in the direction parallel and
perpendicular to the velocity  ${\bf v}_j$ of the $j$ particle.
The power emitted is given by $<{{\bf F}}^{(j)}.{\bf v}_j>$.
As can be seen, the magnetic force, arising from $-g^{st} v_{js}
(e_t^{(\alpha)[1]} k_r^{[1]}-e_r^{(\alpha)[1]} k_t^{[1]})$ does not contribute. The force parallel to
${\bf q}_j-{\bf q}_{j'}$ provides a radiative correction to the Coulomb force that is not considered
here. We use
$
\sum_{\alpha=1,2} ({\bf v}_{j}.{\bf e}^{(\alpha)[1]})({\bf v}_{j}.{\bf e}^{(\alpha)[1]})
=v_{j}^2- \frac{({\bf v}_{j}.{\bf k}^{([1]})^2}{(k^{([1]})^2}$ to obtain
\begin{eqnarray}
&&<{{\bf F}}^{(j)}.{\bf v}_j>_I
=
-i\frac1{(2\pi)^3}\frac{e_j^3 e_{j'}}{4\pi}
\int d^3k^{[1]}\int d^3 l\,\frac{1}{l^2}
\sum_{a=\pm 1}a
\frac{1}{k^{[1]}}
e^{-i\frac{{\bf l}}{2}.({\bf q}_j-{\bf q}_{j'})}
\nonumber \\&&\times
k^{[1]}\left[v_{j}^2- \frac{({\bf v}_{j}.{\bf k}^{([1]})^2}{(k^{([1]})^2}\right]
\left[
\frac{{\bf l}.{\bf k}^{[1]}}{(m_j^2+p_j^2)^{\frac12}}
-\frac{({\bf l}.{\bf p}_{j})({\bf p}_{j}.{\bf k}^{[1]})}{(m_j^2+p_j^2)^{\frac32}}\right]
\nonumber \\&&\times
\left(\frac{1}{{\bf k}^{[1]}.{\bf v}_{j}
-k^{[1]}}
\right)^2
\left(\frac{1}{i\epsilon+(\frac12{\bf l}+a{\bf k}^{[1]}).{\bf v}_{j}
-\frac12{\bf l}.{\bf v}_{j'} -ak^{[1]}}
\right)
\label{8.12}
\end{eqnarray}
That expression is further analyzed in Appendix D, particularly in the  situation where the particle $j'$ is
much more heavy that the $j$ particle. In the referentiel in which the heavy particle is at rest at the
origin of coordinates, we treat the case where the vectors ${\bf q}_j$ and ${\bf v}_{j}$ are
orthogonal (the orbital situation).
In such a case, all integrals can be performed explicitly and the final result is
\begin{equation}
<{{\bf F}}^{(j)}.{\bf v}_j>_{Iorb}=
\frac43
e_j^3 e_{j'}\frac{m_j^2}{(m_j^2+p_j^2)^{\frac32}}
\frac{v_j^2}{q_j^3}
\frac{1}{(1-v_j^2)^{3}}
\label{8.31}
\end{equation}
\begin{equation}
<{{\bf F}}^{(j)}.{\bf v}_j>_{IIorb}
=
-\frac12e^3_je_{j'}
\frac{m_j^2}{(m_j^2+p_j^2)^{\frac32}}
\frac1{q_j^3}\frac1{v_{j}}
\left[\ln\frac{1-v_{j}}{1+v_{j}}+\frac{2v_{j}}{(1-v_{j}^2)}
\right]
\label{8.31a}
\end{equation}
These expressions enable to determine the component of the self electric field at the localisation of
the particle:
\begin{eqnarray}
&&<{\bf E}^{{\bot }}({\bf q}_j).{\bf v}_j>_{orb}
=
e^2_je_{j'}
\frac{m_j^2}{(m_j^2+p_j^2)^{\frac32}}
\frac1{q_j^3}
\nonumber \\&&\times
\left[\frac{-3+10v_j^2-3v_j^4}{3(1-v_j^2)^{3}}-\frac1{2v_{j}}\ln\frac{1-v_{j}}{1+v_{j}}
\right]
\label{8.31b}
\end{eqnarray}
For the geometry chosen, the radiative reaction force is  known exactly by an explicit expression.

\subsection{Non-relativistic limit of the emitted power}

The previous expression  can be developped in powers of $v_j^2$ to make the connection with the well
known result. We have to consider the expression up to order $v_j^2$. 
The result is:
\begin{eqnarray}
&&<{\bf E}^{{\bot }}({\bf q}_j).{\bf v}_j>_{NR}^{e_j(1,1)}
=\frac23
\frac{e^2_je_{j'}}{m_j}
\frac{v_j^2}{q_j^3}
\label{8.31ba}
\end{eqnarray}
The coulombian acceleration of the charge $j$ is provided by the dynamical function
$\frac{{\bf F}_c^{j}}{m_j}=\frac{e_j e_{j'}{\bf q}^{(j)}}{m_jq^{(j)3}}$.
The mean value of its time derivative, due to the free motion (\ref{2.9}), is
\begin{eqnarray}
&&\partial_t<\frac{{\bf F}_c^{j}}{m_j}>
=
\int d^3q^{(j)}\,\int d^3p^{(j)}\,\frac{e_j e_{j'}{\bf q}^{(j)}}{m_jq^{(j)3}}
(-v_r^{(j)}\frac{\partial}{\partial q_r^{(j)}})
\nonumber \\&&\qquad\qquad\quad\times
\delta({\bf q}^{(j)}-{\bf q}_j)
\delta({\bf p}^{(j)}-{\bf p}_j)
\nonumber \\&&\qquad\qquad\quad=
\left(
\frac{e_j e_{j'}}{m_jq_j^{3}}
\right)
{\bf v}_j
-\frac{3e_j e_{j'}{\bf q}_j}
{m_jq_j^{5}}
({{\bf q}_j}.{\bf v}_j)
\label{8.28a}
\end{eqnarray}
In the geometry where the vectors ${\bf q}_j$ and  ${\bf v}_j$ are perpendicular, we have then
$\frac{d}{dt}<{\bf a}_{cj}>=\frac{e_j e_{j'}{\bf v}_j}{m_jq_j^3}$. Therefore, we get the form
\begin{equation}
<{{\bf F}}^{(j)}.{\bf v}_j>_{NRorb}=\frac{2}{3}e_j^2<\frac{d}{dt}{\bf a}_{cj}>.{\bf v}_j
\label{8.29}
\end{equation}
If we restore the dimensions, we get
\begin{equation}
<{{\bf F}}^{(j)}.{\bf v}_j>_{NRIorb}=\frac{2}{3c^3}e_j^2\frac{d}{dt}<{\bf a}_{cj}>.{\bf v}_j
\label{8.30}
\end{equation}
The usual result, with a front factor $\frac23$, is recovered directly, without having met any
divergence for that contribution. This result is not astonishing. 
In the usual approach, the divergence appears as an infinite self-mass correction, in front of the
acceleration vector. 
Since we have considered the geometry where the Coulomb force (hence the acceleration) is
perpendicular to the velocity, that divergence has no influence on the emitted power.
In the general case, the contribution $<{{\bf F}}^{(j)}>_{II}$ would provide the usual divergence.

\section{ Emitted field due to the Coulomb interaction}   

\setcounter{equation}{0}

\subsection{Subdynamics operator}

The determination of the emitted field due to the Coulomb interaction requires the determination
of the creation operator in the first order in both the Coulomb and the transverse field
interactions. The starting expression is:
\begin{eqnarray}
&&<11[1(s_j)]|\tilde \Sigma(t)|11[1(f)]>^{(1,1)}_{PF}
\nonumber \\&&=\frac{-1}{2\pi i} \int'_c dz\,e^{-izt}
\left(\frac{1}{z-{\bf k}^{(j)}.{\bf v}^{(j)}-{\bf k}^{(j')}.{\bf v}^{(j')}
+k^{[1]}(m_{\alpha}^{[1]}+m_{\alpha'}^{[1]})}
\right)
\nonumber \\&&\times
(-i)e_je_{j'}\frac{-1}{2\pi^2}\int d^3 l\frac{1}{l^2}
{\bf l}.\left(\frac{\partial}{\partial {\bf p}^{(j)}}-\frac{\partial}
{\partial {\bf p}^{(j')}}
\right)
e^{{\bf l}.\left(\frac{\partial}{\partial {\bf k}^{(j)}}-\frac{\partial}
{\partial {\bf k}^{(j')}}
\right)}
\nonumber \\&&\times
\left(\frac{1}{z-{\bf k}^{(j)}.{\bf v}^{(j)}-{\bf k}^{(j')}.{\bf v}^{(j')}+k^{[1]}(m_{\alpha}^{[1]}+m_{\alpha'}^{[1]})}
\right)
\nonumber \\&&\times
(-i)e_j \frac1{(2\pi)^{\frac32}} 
\sum_{\beta=1,2} 
\sum_{b=\pm 1}
\left(
\frac{\eta_{\beta}^{[1]}}
{k^{[1]}}
\right)^{\frac12} 
\nonumber \\&&\times
\left[ 
[k^{[1]} e_{r'}^{(\beta)[1]}-g^{s't'} v_{s'}^{(j)}
(e_{t'}^{(\beta)[1]} k_{r'}^{[1]}-e_{r'}^{(\beta)[1]} k_{t'}^{[1]})]
\frac{\partial}
{\partial p_{r'}^{(j)}}
\right.\nonumber \\ &&\left.
-\pi({\bf v}^{(j)}.{\bf e}^{(\beta)[1]})
\left( 
2 \frac{\partial}
{\partial \eta_{\beta}^{[1]}}
-\frac{b}{\eta_{\beta}^{[1]}}(m_{\beta}^{[1]}-b)
\right)
\right]
\exp b\left\{-{\bf k}^{[1]}.\frac{\partial}{\partial {\bf k}^{(j)}}-\frac{\partial}{\partial
m_{\beta}^{[1]}}\right\}
\nonumber \\&&\times
\left(\frac{1}{z-{\bf k}^{(j)}.{\bf v}^{(j)}-{\bf k}^{(j')}.{\bf v}^{(j')}
+k^{[1]}(m_{\beta}^{[1]}+m_{\beta'}^{[1]})}
\right),
\label{4.37}
\end{eqnarray}
for the order $PF$
and a similar expression for the order $FP$.
For the order $PF$, only the first propagator, at the extreme right, corresponds to a vacuum state
while in the other order, $FP$, the first two propagators satisfy that condition and have to be
considered inside the path $c$.
We can proceed in a straightforward way, as in $\S$5.

Since the operator $\tilde{{\cal A}}$ (\ref{3.37}) can only deviate from unity when two field interactions take
place, the expression of 
$<11[1(s_j)]|\tilde \Sigma(0)|11[1(f)]>^{(1,1)}\tilde f^V_{[1(f)]}$ can be identified with the
corresponding term $<11[1(s_j)]|\tilde C|11[1(f)]>^{(1,1)}\tilde f^V_{[1(f)]}$.
As the equivalence conditions imply $\tilde f_{11[1(e_j)]}=\tilde f_{11[1(s_j)]}$, 
the distribution function for the emitted field at first order in the field interaction and first order in
the Coulomb interaction is determined in this way.

\subsection{Transverse emitted field}

We have therefore all the elements to deduce the emitted field (for sharp locations and momenta for
the particles)
\begin{eqnarray}
&&<{\bf E}_r^{{\bot }}({\bf x})>^{e_j(1,1)}
=\int d^3{\bf k}^{[1]}
\sum_{\alpha=1,2} \sum_{a=\pm 1}
{k^{[1]}}^{\frac12} {\bf e}_r^{\alpha}({\bf k}^{[1]}) 
\exp \{ia[{\bf k}^{[1]}.{\bf x}]\}
e^2_je_{j'}\frac{1}{2\pi^2}
\nonumber \\&&\times\frac1{(2\pi)^3}
\int d^3 l\frac{1}{l^2}
\left(\frac{1}{i\epsilon+({\bf l}+a{\bf k}^{[1]}).{\bf v}_j-{\bf l}.{\bf v}_{j'} -ak^{[1]}}
\right)
\frac{-1}{(+a{\bf k}^{[1]}.{\bf v}_j-ak^{[1]})^2}
\nonumber \\&&\times\left(
\frac{1}
{k^{[1]}}
\right)^{\frac12} 
l_v(ak_u^{[1]})\frac{\partial v_{ju}}{\partial p_{jv}}
2\pi({\bf v}_j.{\bf e}^{(\alpha)[1]})
e^{-i([{\bf l}+a{\bf k}^{[1]}].{\bf q}_j-{\bf l}.{\bf q}_{j'})}
\nonumber \\&&+
\int d^3{\bf k}^{[1]}
\sum_{\alpha=1,2} \sum_{a=\pm 1}
{k^{[1]}}^{\frac12} {\bf e}_r^{\alpha}({\bf k}^{[1]}) 
\exp \{ia[{\bf k}^{[1]}.{\bf x}]\}
(-)e^2_je_{j'}\frac{-1}{2\pi^2}
\nonumber \\&&\times\frac1{(2\pi)^3}
\int d^3 l\frac{1}{l^2}
\left(\frac{1}{i\epsilon+({\bf l}+a{\bf k}^{[1]}).{\bf v}_j-{\bf l}.{\bf v}_{j'} -ak^{[1]}}
\right)
\left(\frac{1}{a{\bf k}^{[1]}.{\bf v}_j-ak^{[1]}}
\right)
\nonumber \\&&\times\left(
\frac{1}
{k^{[1]}}
\right)^{\frac12} 
2\pi l_v\frac{\partial v_{ju}}{\partial p_{jv}}e_u^{(\alpha)[1]}
e^{-i([{\bf l}+a{\bf k}^{[1]}].{\bf q}_j-{\bf l}.{\bf q}_{j'})}.
\label{4.71}
\end{eqnarray}
This new expression is the equivalent of (\ref{4.31}) in presence of the Coulomb interaction.
It determines the field due to the accelerated particles in terms of the actual values of the position
${\bf q}_j$ and momentums ${\bf p}_j$ of the charged particles. 
Usually,  expressions of the acceleration fields are given in terms of the retardated positions.
We look for the comparison only for the radiative force, since we have illustrated in $\S$4 the
equivalence of the formalisms outside the locations of the particles.
The self-field of the particle due to the Coulomb interaction, is then given at first order by
\begin{eqnarray}
&&<{\bf E}_r^{{\bot }}({\bf q}_j)>^{e_j(1,1)}
=
-e^2_je_{j'}\frac{1}{\pi}\frac1{(2\pi)^3}
\int d^3{\bf k}^{[1]}
\sum_{\alpha=1,2} \sum_{a=\pm 1}
{\bf e}_r^{\alpha}({\bf k}^{[1]}) 
\nonumber \\&&\times\int d^3 l\frac{1}{l^2}
\left(\frac{1}{i\epsilon+({\bf l}+a{\bf k}^{[1]}).{\bf v}_j-{\bf l}.{\bf v}_{j'} -ak^{[1]}}
\right)
\frac{1}{(+a{\bf k}^{[1]}.{\bf v}_j-ak^{[1]})^2}
\nonumber \\&&\times
l_v(ak_u^{[1]})\frac{\partial v_{ju}}{\partial p_{jv}}
({\bf v}_j.{\bf e}^{(\alpha)[1]})
e^{-i{\bf l}.[{\bf q}_j-{\bf q}_{j'}]}
\nonumber \\&&+
e^2_je_{j'}\frac{1}{\pi}\frac1{(2\pi)^3}
\int d^3{\bf k}^{[1]}
\sum_{\alpha=1,2} \sum_{a=\pm 1}
{\bf e}_r^{\alpha}({\bf k}^{[1]}) 
\nonumber \\&&\times
\int d^3 l\frac{1}{l^2}
\left(\frac{1}{i\epsilon+({\bf l}+a{\bf k}^{[1]}).{\bf v}_j-{\bf l}.{\bf v}_{j'} -ak^{[1]}}
\right)
\left(\frac{1}{a{\bf k}^{[1]}.{\bf v}_j-ak^{[1]}}
\right)
\nonumber \\&&\times
l_v\frac{\partial v_{ju}}{\partial p_{jv}}e_u^{(\alpha)[1]}
e^{-i{\bf l}.[{\bf q}_j-{\bf q}_{j'}]}.
\label{4.73}
\end{eqnarray}
Using (\ref{3.11d}), that expression can be identified with the result obtained from the $\tilde \Theta$
operator (in the previous section) that leads to the usual expression for the self-force in the low
velocity limit.

\section{Finite classical electrodynamics.}   
\setcounter{equation}{0}
\subsection{General outline of the approach}   

We intend to prove that resummations, usual in the context of statistical physics, enables to get rid
of the divergences in the kinetic operator, computed from the expression of the self-field at the
location of the particle.  
These resummations involve a renormalisation of the propagators associated with the particles.
Our analysis proceeds through several steps.

In the first one, 
the previous divergent contribution for the self-field is written in an adequate way. 
A linear dependence in a cut-off wave number $K_c$ is found, as expected.
A usual diagrammatic representation, such that it can be found in
\cite{RB75} for instance, is convenient.  
The dependence of a simple cycle on a cut-off wave number $K_c$ is then analysed and established.
The two propagators of the previously divergent contribution are then
renormalised using simple cycles.
When their dependence in  $K_c$ is reported in the expression of the self-field, it 
induces a supplementary
$\frac1{K_c^2}$ and the resulting contribution vanishes!

In the second step, we use cycles that are renormalised by themselves: the propagator
present in the expression of the cycles is  a renormalised one.
That first resummation contains only disjoint or inserted cycles in arbitrary numbers, with no
overlap:  two vertices of a cycle are either disjoint or completely inserted with respect to two
vertices of another one. 
The contribution of the cycle is thus defined in a self-consistent way. 
The resulting dependence of the renormalised cycle in $K_c$ is non analytic, in $K_c^{\frac12}$. 
This result of the self consistency can be understood in a simple way.
The integral defining the bare cycle behaves as $K_c$.
The renormalised cycle involves one propagator that, by hypothesis, provides a factor
$K_c^{-\frac12}$. 
When combined, their product reproduces correctly the assumed $K_c^{\frac12}$
dependence. 
When that value is introduced into the expression of the self-field, each of the two
propagators provide a factor $K_c^{-\frac12}$ that compensates the $K_c$ dependence due to the
integration. Therefore, the limit $K_c\to \infty$ is well defined.

That property of the renormalised cycles can be extended to all contributions:
For imbricated vertices, each new wave number will be associated with two propagators, 
that can be renormalised by (renormalised) cycles.
The dependence of these new vertices with $K_c$ is then the same as the the simple cycle.
Therefore, they can also be used for a global renormalisation while preserving the finiteness of the
self-field.

\subsection{The previous divergent contribution}

The correlated elements describing the self-field are easily obtained from the expression of the
creation  operator.  
We have to focus on the divergent contribution to the the self-field.
Its computation starts from the expresson of $<11[1(s_j)]|\tilde \Sigma(t)|11[1(f)]>^{(1,1)}$ 
((\ref{4.37})) and we have to reconsider it in order to be able to add the contributions of higher order
that will ensure its finiteness.

The final dependence in the cut-off wave number $K_c$ has to be appreciated after the a last vertex,
enabling to get $\tilde{\Theta}$ or the self-field from the creation operator,  has been
introduced.   
A displacement operator involving the wave number
exchanged through the Coulomb interaction is natural and will remain in our expressions.
We keep the dependence in particle operator that has provided, in the previous ``II" contribution,
the ultraviolet divergence,  in the computation of the self-interaction.
We still restrict ourselves to the  effect of the absence of fields: all strenghts of the fields
oscillators are close to zero: the distribution function of the action variables of the field oscillator
are peaked around zero (the limit is taken afterwards).

We reconsider the contribution, affected by a $d$-index, responsible for the divergence, from the
first steps of its computation.  
In the expression (\ref{4.37}), the operator 
${\bf l}.\left(\frac{\partial}{\partial {\bf p}^{(j)}}-\frac{\partial} {\partial {\bf p}^{(j')}}\right)$ has to
act on 
$-\pi({{\bf v}}^{(j)}.{\bf e}^{(\beta)[1]})$.
Moreover, in the factor $\left( 
2 \frac{\partial}{\partial \eta_{\beta}^{[1]}}-\frac{b}{\eta_{\beta}^{[1]}}(m_{\beta}^{[1]}-b)
\right)$, only the derivative will provide a non vanishing contribution for the self-interaction.
\begin{eqnarray}
&&<11[1(s_j)]|\tilde \Sigma(t)|11[1(f)]>^{(1,1)}_{d}
\nonumber\\&&=\frac{-1}{2\pi i} \int'_c dz\,e^{-izt}
\left(\frac{1}{z-{{\bf k}}^{(j)}.{{\bf v}}^{(j)}-{{\bf k}}^{(j')}.{{\bf v}}^{(j')}
+k^{[1]}(m_{\alpha}^{[1]}+m_{\alpha'}^{[1]})}
\right)
\nonumber\\&& \times
(-i)e_je_{j'}\frac{-1}{2\pi^2}\int d^3 l\frac{1}{l^2}
e^{{\bf l}.\left(\frac{\partial}{\partial {{\bf k}}^{(j)}}-\frac{\partial}
{\partial {\bf k}^{(j')}}
\right)}
\nonumber\\&& \times
\left(\frac{1}{z-{{\bf k}}^{(j)}.{{\bf v}}^{(j)}-{\bf k}^{(j')}.{\bf v}^{(j')}
+k^{[1]}(m_{\alpha}^{[1]}+m_{\alpha'}^{[1]})}
\right)
\nonumber\\&& \times
(-i)e_j \frac1{(2\pi)^{\frac32}} 
\sum_{\beta=1,2} 
\sum_{b=\pm 1}
\left(
\frac{\eta_{\beta}^{[1]}}
{k^{[1]}}
\right)^{\frac12} 
\nonumber\\&& \times
\left[ 
-\left({\bf l}.\frac{\partial}{\partial {\bf p}^{(j)}}\pi({{\bf v}}^{(j)}.{\bf e}^{(\beta)[1]})\right)
2 \frac{\partial}
{\partial \eta_{\beta}^{[1]}}
\right]
\exp b\left\{-{\bf k}^{[1]}.\frac{\partial}{\partial {{\bf k}}^{(j)}}-\frac{\partial}{\partial
m_{\beta}^{[1]}}\right\}
\nonumber\\&& \times
\left(\frac{1}{z-{{\bf k}}^{(j)}.{{\bf v}}^{(j)}-{\bf k}^{(j')}.{\bf v}^{(j')}
+k^{[1]}(m_{\beta}^{[1]}+m_{\beta'}^{[1]})}
\right).
\label{a2.15}
\end{eqnarray}
That expression has to act on the distribution function $\tilde f^V_{[1(f)]}$ describing the absence of
field.
We change the place of the displacement operators (in variables) to obtain:
\begin{eqnarray}
&&<11[1(s_j)]|\tilde \Sigma(t)|11[1(f)]>^{(1,1)}_{d}\tilde f^V_{[1(f)]}
\nonumber\\&&=
(-i)e_j \frac1{(2\pi)^{\frac32}} 
\sum_{\beta=1,2} 
\sum_{b=\pm 1}
\left(
\frac{\eta_{\beta}^{[1]}}
{k^{[1]}}
\right)^{\frac12} 
\exp b\left\{-{\bf k}^{[1]}.\frac{\partial}{\partial {{\bf k}}^{(j)}}-\frac{\partial}{\partial
m_{\beta}^{[1]}}\right\}
\nonumber\\&& \times
\frac{-1}{2\pi i} \int'_c dz\,e^{-izt}
\left(\frac{1}{z-({{\bf k}}^{(j)}+b{\bf k}^{[1]}).{{\bf v}}^{(j)}-{\bf k}^{(j')}.{\bf v}^{(j')}
+bk^{[1]}}
\right)
\nonumber\\&& \times
(-i)e_je_{j'}\frac{-1}{2\pi^2}\int d^3 l\frac{1}{l^2}
\left[ 
-\left({\bf l}.\frac{\partial}{\partial {\bf p}^{(j)}}\pi({{\bf v}}^{(j)}.{\bf e}^{(\beta)[1]})\right)
2 \frac{\partial}
{\partial \eta_{\beta}^{[1]}}
\right]
\nonumber\\&& \times
\left(\frac{1}{z-({{\bf k}}^{(j)}+{\bf l}+b{\bf k}^{[1]}).{{\bf v}}^{(j)}
-({\bf k}^{(j')}-{\bf l}).{\bf v}^{(j')}+bk^{[1]}}
\right)
\nonumber\\&& \times
\left(\frac{1}{z-({{\bf k}}^{(j)}+{\bf l}).{{\bf v}}^{(j)}-({\bf k}^{(j')}-{\bf l}).{\bf v}^{(j')}}
\right)
\nonumber\\&& \times
e^{{\bf l}.\left(\frac{\partial}{\partial {{\bf k}}^{(j)}}-\frac{\partial}
{\partial {\bf k}^{(j')}}
\right)}
\delta(\eta_{\beta}^{[1]}-\eta_\beta)\delta(\eta_{\beta'}^{[1]}-\eta_{\beta'})
\delta^{Kr}_{m_{\beta}^{[1]},0}\delta^{Kr}_{m_{\beta'}^{[1]},0}.
\label{a2.18}
\end{eqnarray}
The previous creation operator, enabling the computation of the self-field is obtained from that
expression by picking the residue at the pole corresponding to the last propagator.
\begin{eqnarray}
&&<{\bf E}_r^{{\bot }}({\bf q}_j)>_{d}^{e_j(1,1)}
=
e^2_je_{j'}\frac{1}{\pi}\frac1{(2\pi)^3}
\int d^3{\bf k}^{[1]}
\sum_{\alpha=1,2} \sum_{a=\pm 1}
{\bf e}_r^{\alpha}({\bf k}^{[1]}) 
\nonumber\\&& \times
\int d^3 l\frac{1}{l^2}
\left(\frac{1}{i\epsilon+({\bf l}+a{\bf k}^{[1]}).{\bf v}_j-{\bf l}.{\bf v}_{j'} -ak^{[1]}}
\right)
\nonumber\\&& \times
\left(\frac{1}
{a{\bf k}^{[1]}.{\bf v}_j-ak^{[1]}}
\right)
l_v\frac{\partial v_{ju}}{\partial p_{jv}}e_u^{(\alpha)[1]}
e^{-i{\bf l}.[{\bf q}_j-{\bf q}_{j'}]}.
\label{a4.73aa}
\end{eqnarray}
Each propagator provides a decreasing $\frac1{k^{[1]}}$ factor while the jacobian to get to spherical
coordinates provide a $(k^{[1]})^2$ factor, hence the linear divergence.

\subsection{The simple cycle}

We intend to renormalise to propagators through simple cycles in a first step.
In order to be acquainted with their contribution, we explicit the  operator 
$<11[0]|\tilde{\cal R}(z)|11[1(f)]>^{(0,2)}\tilde f^V_{[1(f)]}$ that contains only the cycle. 
The two interactions involve the same particle $j$. 

We consider only contributions that will not contain
displacement operators  involving the wave number of the interacting field mode acting on the
distribution function.  Indeed, in their presence, a convergence factor for ultraviolet modes would
arise from the dependence of the involved distribution function in that wave number.
The selected terms act as $c$-number in the particles
variables, hence the $c$-index. 
Those contributions are the one responsible for the divergences since the other contributions
involve characteristics of the distribution function. 
In the contribution (\ref{3.12}), we have thus $b=-a$ and $\frac{\partial}{\partial p_r^{(j)}}$ has to
act on  $({{\bf v}}^{(j)}.{\bf e}^{(\beta)[1]})$. 
We act on the  field vacuum distribution function  $\tilde f^V_{[1(f)]}$ to get:
\begin{eqnarray}
&&<11[0]|\tilde{\cal R}(z)|11[1(f)]>_c^{(0,2)}\tilde f^V_{[1(f)]}
\nonumber\\&&=
\left(\frac{1}{z-{{\bf k}}^{(j)}.{{\bf v}}^{(j)}-{\bf k}^{(j')}.{\bf v}^{(j')}}
\right)
(-i) e_j \frac1{(2\pi)^{\frac32}}
\int d^3k^{[1]}
\int_0^{\infty}d\eta_1^{[1]}\int_0^{\infty}d\eta_2^{[1]}
\nonumber\\&& \times
\sum_{m_{1}^{[1]},m_{2}^{[1]}}
\delta_{m_{1}^{[1]},0}\delta_{m_{2}^{[1]},0}
\sum_{\alpha=1,2} 
\sum_{a=\pm 1}
\left(
\frac{\eta_{\alpha}^{[1]}}
{k^{[1]}}
\right)^{\frac12} 
\left[ 
[k^{[1]} e_r^{(\alpha)[1]}-g^{st} v_s^{(j)}
(e_t^{(\alpha)[1]} k_r^{[1]}
\right.\nonumber\\&& \left.
-e_r^{(\alpha)[1]} k_t^{[1]})]
\frac{\partial}
{\partial p_r^{(j)}}
\right]
\exp a\left\{-{\bf k}^{[1]}.\frac{\partial}{\partial {{\bf k}}^{(j)}}-\frac{\partial}{\partial
m_{\alpha}^{[1]}}\right\}
\nonumber\\&& \times
\left(\frac{1}{z-{{\bf k}}^{(j)}.{{\bf v}}^{(j)}-{\bf k}^{(j')}.{\bf v}^{(j')}
+k^{[1]}(m_{\alpha}^{[1]}+m_{\alpha'}^{[1]})}
\right)
(-i)e_j \frac1{(2\pi)^{\frac32}} 
\nonumber\\&& \times
\sum_{\beta=1,2} 
\left(
\frac{\eta_{\beta}^{[1]}}
{k^{[1]}}
\right)^{\frac12} 
\left[ 
-\pi({{\bf v}}^{(j)}.{\bf e}^{(\beta)[1]})
\left( 
2 \frac{\partial}
{\partial \eta_{\beta}^{[1]}}
-\frac{-a}{\eta_{\beta}^{[1]}}(m_{\beta}^{[1]}+a)
\right)
\right]
\nonumber\\&& \times
\exp (-a)\left\{-{\bf k}^{[1]}.\frac{\partial}{\partial {{\bf k}}^{(j)}}-\frac{\partial}{\partial
m_{\beta}^{[1]}}\right\}
\nonumber\\&& \times
\left(\frac{1}{z-{{\bf k}}^{(j)}.{{\bf v}}^{(j)}-{\bf k}^{(j')}.{\bf v}^{(j')}
+k^{[1]}(m_{\beta}^{[1]}+m_{\beta'}^{[1]})}
\right)
\nonumber\\&& \times
\delta(\eta_{\beta}^{[1]}-\eta_\beta)\delta(\eta_{\beta'}^{[1]}-\eta_{\beta'})
\delta^{Kr}_{m_{\beta}^{[1]},0}\delta^{Kr}_{m_{\beta'}^{[1]},0}.
\label{a2.12}
\end{eqnarray}
When acting on the field in its ground state, to obtain a non vanishing contribution, we need an
interaction with the same mode ($\alpha=\beta$).
The summation-integration over field variables can then be performed:
\begin{eqnarray}
&&<11[0]|\tilde{\cal R}(z)|11[1(f)]>_c^{(0,2)}\tilde f^V_{[1(f)]}
\nonumber\\&&=
\left(\frac{1}{z-{{\bf k}}^{(j)}.{{\bf v}}^{(j)}-{\bf k}^{(j')}.{\bf v}^{(j')}}
\right)
 e_j^2 \frac1{(2\pi)^3}
\int d^3k^{[1]}
\sum_{\alpha=1,2} 
\sum_{a=\pm 1}
\nonumber\\&& \times
\frac{1}
{k^{[1]}}
\left[ 
[k^{[1]} e_r^{(\alpha)[1]}-g^{st} v_s^{(j)}
(e_t^{(\alpha)[1]} k_r^{[1]}-e_r^{(\alpha)[1]} k_t^{[1]})]
\right]
\nonumber\\&& \times
\left(\frac{1}{z-{{\bf k}}^{(j)}.{{\bf v}}^{(j)}+a{\bf k}^{[1]}.{{\bf v}}^{(j)}-{\bf k}^{(j')}.{\bf v}^{(j')}
+k^{[1]}(-a)}
\right)
\nonumber\\&& \times
\left[ 
-2\pi\left(\frac{\partial }
{\partial p_r^{(j)}}({{\bf v}}^{(j)}.{\bf e}^{(\alpha)[1]})\right)
\right]
\left(\frac{1}{z-{{\bf k}}^{(j)}.{{\bf v}}^{(j)}-{\bf k}^{(j')}.{\bf v}^{(j')})}
\right).
\label{a2.14}
\end{eqnarray}
We define the cycle contribution $C(z,{{\bf v}}^{(j)})$ by
\begin{eqnarray}
&&C(z,{{\bf v}}^{(j)})
=
 e_j^2 \frac1{(2\pi)^3}
\int d^3k^{[1]}
\sum_{\alpha{[1]}=1,2} 
\sum_{a=\pm 1}
\frac{1}
{k^{[1]}}
\nonumber\\&& \times
\left[ 
[k^{[1]} e_r^{(\alpha)[1]}-g^{st} v_s^{(j)}
(e_t^{(\alpha)[1]} k_r^{[1]}-e_r^{(\alpha)[1]} k_t^{[1]})]
\right]
\nonumber\\&& \times
\left(\frac{1}{z+a{\bf k}^{[1]}.{{\bf v}}^{(j)} +k^{[1]}(-a)}
\right)
\left[ 
-2\pi\left(\frac{\partial }
{\partial p_r^{(j)}}({{\bf v}}^{(j)}.{\bf e}^{(\alpha)[1]})\right)
\right],
\label{a2.14a}
\end{eqnarray}
and we recognise $C(z-{{\bf k}}^{(j)}.{{\bf v}}^{(j)}-{\bf k}^{(j')}.{\bf v}^{(j')},{{\bf
v}}^{(j)})$ in (\ref{a2.14}).

Let us examine the dependence of $C$ in the cut off wave number $K_c$.
The first matrix element contains a contribution finite for ${k^{[1]}}\to \infty$.
If we take into account the summation over $a$, the propagator will involve a $(k^{[1]})^{-2}$
dependence. 
The whole contribution behaves therefore linearly as $K_c$.
If $z$ approaches the real axis fom above, as it is required from the construction of the
subdynamics operator, we can use $z=y+i\epsilon$ and upon replacement in (\ref{a2.14a}), 
the expression of $C(y+i\epsilon,{{\bf v}}^{(j)})$ contains a Dirac delta function, of argument 
$y+a{\bf k}^{[1]}.{{\bf v}}^{(j)} +k^{[1]}(-a)$ and a principal part.
The contribution due to the Dirac delta function is independent of the cut off while the
contribution due to the principal part provides the $K_c$ dependence.

Moreover, it is easily recognised that $C(z,{{\bf v}}^{(j)})$ vanishes for $z\to 0$.
Indeed, in that limit, the contribution that could arise from the Dirac delta function
$\delta(a{\bf k}^{[1]}.{{\bf v}}^{(j)} +k^{[1]}(-a))$ vanishes since the vanishing of its argument cannot
be satisfied (except for $k^{[1]}=0$ but the jacobian provides a factor $k^{[1]2}$) and the
contribution due to the principal part vanishes by parity in the summation over $a$.

\subsection{Renormalisation by simple cycles}

We first consider the consequences of inserting a simple cycle in the contribution (\ref{a2.15}).
For the sake of illustration, we consider the insertion of such a cycle before the vertex
corresponding to the Coulomb interaction in (\ref{a2.15}).
Details of computation can be found in appendix E.
The insertion of the cycle involves of course a supplementary propagator (see the
square on the first propagator in (\ref{a2.23})) and the previouly considered  cycle contribution (\ref{a2.14a}) with the same argument as the new propagator.
That property holds for the insertion of an arbitrary number of cycles, enabling a renormalisation
for the propagators. 
The resummation of the class of diagrams where all the cycles are at the left of the particle vertex
(Coulomb interaction) is:
\begin{eqnarray}
&&<11[1(s_j)]|\tilde \Sigma(t)|11[1(f)]>^{(1,\infty)}_{d1}\tilde f^V_{[1(f)]}
\nonumber\\&&=
(-i)e_j \frac1{(2\pi)^{\frac32}} 
\sum_{\beta=1,2} 
\sum_{b=\pm 1}
\left(
\frac{\eta_{\beta}}
{k}
\right)^{\frac12}
\exp b\left\{-{\bf k}.\frac{\partial}{\partial {{\bf k}}^{(j)}}-\frac{\partial}{\partial
m_{\beta}}\right\}
\nonumber\\&& \times
\frac{-1}{2\pi i} \int'_c dz\,e^{-izt}
\left(z-({{\bf k}}^{(j)}+b{\bf k}).{{\bf v}}^{(j)}-{\bf k}^{(j')}.{\bf v}^{(j')}+bk
\nonumber\right.\\&&\left.
-C(z-({{\bf k}}^{(j)}+b{\bf k}).{{\bf v}}^{(j)}-{\bf k}^{(j')}.{\bf v}^{(j')}+bk,{{\bf v}}^{(j)})
\right)^{-1}
\nonumber\\&& \times
(-i)e_je_{j'}\frac{-1}{2\pi^2}\int d^3 l\frac{1}{l^2}
\left(\frac{1}{z-({{\bf k}}^{(j)}+{\bf l}+b{\bf k}).{{\bf v}}^{(j)}-({\bf k}^{(j')}-{\bf l}).{\bf v}^{(j')}+bk}
\right)
\nonumber\\&& \times
\left[ 
-\left({\bf l}.\frac{\partial}{\partial {\bf p}^{(j)}}\pi({{\bf v}}^{(j)}.{\bf e}^{(\beta)})\right)
2 \frac{\partial}
{\partial \eta_{\beta}}
\right]
\left(\frac{1}{z-({{\bf k}}^{(j)}+{\bf l}).{{\bf v}}^{(j)}-({\bf k}^{(j')}-{\bf l}).{\bf v}^{(j')}}
\right)
\nonumber\\&& \times
e^{{\bf l}.\left(\frac{\partial}{\partial {{\bf k}}^{(j)}}-\frac{\partial}
{\partial {\bf k}^{(j')}}
\right)}
\delta(\eta_{\beta}^{[1]}-\eta_\beta)\delta(\eta_{\beta'}^{[1]}-\eta_{\beta'})
\delta^{Kr}_{m_{\beta}^{[1]},0}\delta^{Kr}_{m_{\beta'}^{[1]},0}.
\label{a2.24}
\end{eqnarray}

When an arbitrary number of cycles are also added in the
two other propagators, the contribution is:
\begin{eqnarray}
&&<11[1(s_j)]|\tilde \Sigma(t)|11[1(f)]>^{(1,\infty)}_{d}\tilde f^V_{[1(f)]}
\nonumber\\&&=
(-i)e_j \frac1{(2\pi)^{\frac32}} 
\sum_{\beta=1,2} 
\sum_{b=\pm 1}
\left(
\frac{\eta_{\beta}}
{k}
\right)^{\frac12}
\exp b\left\{-{\bf k}.\frac{\partial}{\partial {{\bf k}}^{(j)}}-\frac{\partial}{\partial
m_{\beta}}\right\}
\nonumber\\&& \times
\frac{-1}{2\pi i} \int'_c dz\,e^{-izt}
\left(z-({{\bf k}}^{(j)}+b{\bf k}).{{\bf v}}^{(j)}-{\bf k}^{(j')}.{\bf v}^{(j')}+bk
\nonumber\right.\\&&\left.
-C(z-({{\bf k}}^{(j)}+b{\bf k}).{{\bf v}}^{(j)}-{\bf k}^{(j')}.{\bf v}^{(j')}+bk,{{\bf v}}^{(j)})
\right)^{-1}
\nonumber\\&& \times
(-i)e_je_{j'}\frac{-1}{2\pi^2}\int d^3 l\frac{1}{l^2}
\nonumber\\&& \times
\left(z-({{\bf k}}^{(j)}+{\bf l}+b{\bf k}).{{\bf v}}^{(j)}-({\bf k}^{(j')}-{\bf l}).{\bf v}^{(j')}+bk
\nonumber\right.\\&&\left.
-C(z-({{\bf k}}^{(j)}+{\bf l}+b{\bf k}).{{\bf v}}^{(j)}-({\bf k}^{(j')}-{\bf l}).{\bf v}^{(j')}+bk,{{\bf v}}^{(j)})
\right)^{-1}
\nonumber\\&& \times
\left[ 
-\left({\bf l}.\frac{\partial}{\partial {\bf p}^{(j)}}\pi({{\bf v}}^{(j)}.{\bf e}^{(\beta)})\right)
2 \frac{\partial}
{\partial \eta_{\beta}}
\right]
\nonumber\\&& \times
\left(z-({{\bf k}}^{(j)}+{\bf l}).{{\bf v}}^{(j)}-({\bf k}^{(j')}-{\bf l}).{\bf v}^{(j')}
\nonumber\right.\\&&\left.
-C(z-({{\bf k}}^{(j)}+{\bf l}).{{\bf v}}^{(j)}-({\bf k}^{(j')}-{\bf l}).{\bf v}^{(j')},{{\bf v}}^{(j)})
\right)^{-1}
\nonumber\\&& \times
e^{{\bf l}.\left(\frac{\partial}{\partial {{\bf k}}^{(j)}}-\frac{\partial}
{\partial {\bf k}^{(j')}}
\right)}
\delta(\eta_{\beta}^{[1]}-\eta_\beta)\delta(\eta_{\beta'}^{[1]}-\eta_{\beta'})
\delta^{Kr}_{m_{\beta}^{[1]},0}\delta^{Kr}_{m_{\beta'}^{[1]},0}.
\label{a2.25}
\end{eqnarray}
For the evaluation of $\tilde \Sigma(t)$ superoperator, 
we have to compute the residues to the
(multiple) bare poles due to the last propagator.
By resummation, the procedure corresponds
(cf. Lee model \cite{dH04a}) to the computation of the dressed poles.
 In view of our previous analysis, $C(i\epsilon,{{\bf v}}^{(j)})=0$.
A solution $\theta$ of $\theta=C(\theta,{{\bf v}}^{(j)})$ is thus
$\theta=0$.

The solution of $z-({{\bf k}}^{(j)}+{\bf l}).{{\bf v}}^{(j)}-({\bf k}^{(j')}-{\bf l}).{\bf v}^{(j')}
-C(z-({{\bf k}}^{(j)}+{\bf l}).{{\bf v}}^{(j)}-({\bf k}^{(j')}-{\bf l}).{\bf v}^{(j')},{{\bf v}}^{(j)})=0$.
can thus be written as $z=({{\bf k}}^{(j)}+{\bf l}).{{\bf v}}^{(j)}+({\bf k}^{(j')}-{\bf l}).{\bf v}^{(j')}$: the
bare and dressed poles coincide. The residue at the pole disappears when the creation operator $C$
is computed from the product $CA$.
Therefore, for our purpose, we do not have to bother about a renormalisation of the vacuum
propagator and we can continue to use the simple propagator.

In place of the second term in (6.5) , we get the same expression with the cycle contribution in
the propagators
\begin{eqnarray}
&&<{\bf E}_r^{{\bot }}({\bf q}_j)>_{IIC}^{e_j(1,1)}
=
e^2_je_{j'}\frac{1}{\pi}\frac1{(2\pi)^3}
\int d^3{\bf k}^{[1]}
\sum_{\alpha=1,2} \sum_{a=\pm 1}
{\bf e}_r^{\alpha}({\bf k}^{[1]}) 
\nonumber\\&& \times
\int d^3 l\frac{1}{l^2}
\left(i\epsilon+({\bf l}+a{\bf k}^{[1]}).{\bf v}_j-{\bf l}.{\bf v}_{j'} -ak^{[1]}
\nonumber\right.\\&&\left.
-C(i\epsilon+({\bf l}+a{\bf k}^{[1]}).{\bf v}_j-{\bf l}.{\bf v}_{j'} -ak^{[1]},{\bf v}_j)
\right)^{-1}
\nonumber\\&& \times
\left(\frac{1}
{a{\bf k}^{[1]}.{\bf v}_j-ak^{[1]}-C(i\epsilon+a{\bf k}^{[1]}.{\bf v}_j-ak^{[1]},{\bf v}_j)}
\right)
\nonumber\\&& \times
l_v\frac{\partial v_{ju}}{\partial p_{jv}}e_u^{(\alpha)[1]}
e^{-i{\bf l}.[{\bf q}_j-{\bf q}_{j'}]}.
\label{a4.73a}
\end{eqnarray}
In the referentiels where the velocity ${\bf v}_{j'}$ vanishes, we get
\begin{eqnarray}
&&<{\bf E}_r^{{\bot }}({\bf q}_j)>_{IIC}^{e_j(1,1)}
=
e^2_je_{j'}\frac{1}{\pi}\frac1{(2\pi)^3}
\int d^3{\bf k}^{[1]}
\sum_{\alpha=1,2} \sum_{a=\pm 1}
{\bf e}_r^{\alpha}({\bf k}^{[1]}) 
\int d^3 l\frac{1}{l^2}
\nonumber\\&& \times
\left(\frac{1}{i\epsilon+({\bf l}+a{\bf k}^{[1]}).{\bf v}_j -ak^{[1]}
-C(i\epsilon+({\bf l}+a{\bf k}^{[1]}).{\bf v}_j -ak^{[1]},{\bf v}_j)}
\right)
\nonumber\\&& \times
\left(\frac{1}
{a{\bf k}^{[1]}.{\bf v}_j-ak^{[1]}-C(i\epsilon+a{\bf k}^{[1]}.{\bf v}_j-ak^{[1]},{\bf v}_j)}
\right)
\nonumber\\&& \times
\frac1m_j l_ve_v^{(\alpha)[1]}
e^{-i{\bf l}.[{\bf q}_j-{\bf q}_{j'}]}.
\label{a2.26}
\end{eqnarray}
In the low velocity limit $v_j\to 0$, the previous expression vanishes by symmetry due to the angular
integration. (the integral over ${\bf l}$ and ${\bf k}$ become independent but the final result could
be divergent.)

We introduce a cut-off function in (\ref{a2.26}).
The same cut-off will be used in all future expressions. 
\begin{eqnarray}
&&<{\bf E}_r^{{\bot }}({\bf q}_j)>_{IIC}^{e_j(1,1)}
=
e^2_je_{j'}\frac{1}{\pi}\frac1{(2\pi)^3}
\int d^3{\bf k}^{[1]}
\sum_{\alpha=1,2} \sum_{a=\pm 1}
{\bf e}_r^{\alpha}({\bf k}^{[1]}) 
\int d^3 l\frac{1}{l^2}
\nonumber\\&& \times
\left(\frac{1}{i\epsilon+({\bf l}+a{\bf k}^{[1]}).{\bf v}_j -ak^{[1]}
-C(i\epsilon+({\bf l}+a{\bf k}^{[1]}).{\bf v}_j -ak^{[1]},{\bf v}_j)}
\right)
\nonumber\\&& \times
\left(\frac{1}
{a{\bf k}^{[1]}.{\bf v}_j-ak^{[1]}-C(i\epsilon+a{\bf k}^{[1]}.{\bf v}_j-ak^{[1]},{\bf v}_j)}
\right)
\nonumber\\&& \times
\frac1m_j l_ve_v^{(\alpha)[1]}
e^{-i{\bf l}.[{\bf q}_j-{\bf q}_{j'}]}
\frac{K_c^2}{k^{[1]2}+K_c^2},
\label{a2.27}
\end{eqnarray}
where in the low velocity, neglecting the magnetic component of the force and simplifying the
computation of the derivative with respect to ${\bf p}_j$,
\begin{eqnarray}
&& C(z,{\bf v}_j)
=
-\frac{4\pi}{m_j} e_j^2 \frac1{(2\pi)^3}
\int d^3k
\left[
\left(\frac{1}{z+{\bf k}.{\bf v}_j-k}
\right)
+\left(\frac{1}{z-{\bf k}.{\bf v}_j+k}
\right)
\right]
\nonumber\\&& \qquad\qquad\times
\frac{K_c^2}{k^2+K_c^2}.
\label{a2.28}
\end{eqnarray}
We examine the behaviour for small velocity ${\bf v}_j$.
\begin{eqnarray}
&& C(z,0)
=
-\frac{16\pi^2}{m_j} e_j^2 \frac1{(2\pi)^3}
\int_0^{\infty} dk\,k^2
\left[
\left(\frac{1}{z-k}
\right)
+\left(\frac{1}{z+k}
\right)
\right]
\frac{K_c^2}{k^2+K_c^2}
\nonumber\\&& 
\label{a2.30}
\end{eqnarray}
For $\Im z>0$, and $\Re z=y$, we prove in appendix E that
\begin{equation}
C(y+i\epsilon,0)=
\frac{16\pi^3}{m_j} e_j^2 \frac1{(2\pi)^3}
\pi K_c^3 \frac y{(y^2+K_c^2)}
+i\frac{2e_j^2}{m_j} 
\frac{K_c^2y^2}{y^2+K_c^2}.
\end{equation}

We insert the value of $C(y+i\epsilon,0)$ in the expression (\ref{a2.27}), neglecting the dependence
of $C$ on the velocity $v_j$.
That dependence is not seen as capital to get the qualitative behaviour
while the remaining dependence in the denominator is
required for the coupling between the integrations over  the wave numbers $l$ and $k$.
We have: 
\begin{eqnarray}
&&<{\bf E}_r^{{\bot }}({\bf q}_j)>_{IIC}^{e_j(1,1)}
=
e^2_je_{j'}\frac{1}{\pi}\frac1{(2\pi)^3}
\int d^3{\bf k}^{[1]}
\sum_{\alpha=1,2} \sum_{a=\pm 1}
{\bf e}_r^{\alpha}({\bf k}^{[1]}) 
\nonumber\\&& \times
\int d^3 l\frac{1}{l^2}
\left(\frac{1}{i\epsilon+({\bf l}+a{\bf k}^{[1]}).{\bf v}_j -ak^{[1]}}
\right)
\left(\frac{1}
{a{\bf k}^{[1]}.{\bf v}_j-ak^{[1]}}
\right)
\nonumber\\&& \times
\frac{[({\bf l}+a{\bf k}^{[1]}).{\bf v}_j -ak^{[1]}]^2+K_c^2}
{[({\bf l}+a{\bf k}^{[1]}).{\bf v}_j -ak^{[1]}]^2+K_c^2
-\frac{\pi}{m_j} e_j^2K_c^3
-i\frac{2e_j^2}{m_j}K_c^2[({\bf l}+a{\bf k}^{[1]}).{\bf v}_j -ak^{[1]}] }
\nonumber\\&& \times
\frac
{[a{\bf k}^{[1]}.{\bf v}_j -ak^{[1]}]^2+K_c^2}
{[a{\bf k}^{[1]}.{\bf v}_j -ak^{[1]}]^2+K_c^2
-\frac{\pi}{m_j} e_j^2K_c^3
-i\frac{2e_j^2}{m_j}K_c^2[a{\bf k}^{[1]}.{\bf v}_j -ak^{[1]}] }
\nonumber\\&& \times
\frac1m_j l_ve_v^{(\alpha)[1]}
e^{-i{\bf l}.[{\bf q}_j-{\bf q}_{j'}]}
\frac{K_c^2}{k^{[1]2}+K_c^2}.
\label{a2.35}
\end{eqnarray}
We compare that integral (\ref{a2.35}) with the contribution arising from (6.5):
\begin{eqnarray}
&&<{\bf E}_r^{{\bot }}({\bf q}_j)>_{II}^{e_j(1,1)}
=
e^2_je_{j'}\frac{1}{\pi}\frac1{(2\pi)^3}
\int d^3{\bf k}^{[1]}
\sum_{\alpha=1,2} \sum_{a=\pm 1}
{\bf e}_r^{\alpha}({\bf k}^{[1]}) 
\nonumber\\&& \times
\int d^3 l\frac{1}{l^2}
\left(\frac{1}{i\epsilon+({\bf l}+a{\bf k}^{[1]}).{\bf v}_j -ak^{[1]}}
\right)
\left(\frac{1}
{a{\bf k}^{[1]}.{\bf v}_j-ak^{[1]}}
\right)
\nonumber\\&& \times
\frac1m_j l_ve_v^{(\alpha)[1]}
e^{-i{\bf l}.[{\bf q}_j-{\bf q}_{j'}]}
\frac{K_c^2}{k^{[1]2}+K_c^2},
\label{a2.36}
\end{eqnarray}
we note the presence of two similar supplementary factors.
Except for the domain of values where the wave numbers are of the order of $K_c^{\frac32}$, the
denominator in these factors are dominated by $-\frac{\pi}{m_j} e_j^2K_c^3$.
In the relevant range of values for the wave numbers, thanks to the cut-off function, the numerator in
the factors are of the order of $K_c^2$.
Therefore, these factors can be replaced by $\frac{m_j}{e_j^2 K_c}$ to evaluate the dependence of (\ref{a2.35}) with respect to $K_c$.
Since the resulting integral is known to diverge linearly with $K_c$, the resulting dependence
vanishes as $K_c^{-1}$.
The introduction of the (simple) cycle produces a convergence with respect to $K_c$ that is much
more too strong: we go from a divergence in $K_c$ to a convergence in $K_c^{-1}$.
The consideration of renormalised cycles will produce the correct behaviour.

\subsection{The renormalised  cycle $\tilde C$}

The renormalised  cycle $\tilde C$ is  introduced by replacing, in the definition of the cycle,
the free propagator by a fully renormalised one.
$\tilde C$ is thus defined by a self-consistent equation:
\begin{eqnarray}
&&\tilde C(z,{{\bf v}}^{(j)})
=
 e_j^2 \frac1{(2\pi)^3}
\int d^3k^{[1]}
\sum_{\alpha{[1]}=1,2} 
\sum_{a=\pm 1}
\frac{1}
{k^{[1]}}
\nonumber\\&& \times
\left[ 
[k^{[1]} e_r^{(\alpha)[1]}-g^{st} v_s^{(j)}
(e_t^{(\alpha)[1]} k_r^{[1]}-e_r^{(\alpha)[1]} k_t^{[1]})]
\right]
\nonumber\\&& \times
\left(\frac{1}{z+a{\bf k}^{[1]}.{{\bf v}}^{(j)} +k^{[1]}(-a)-\tilde C(z+a{\bf k}^{[1]}.{{\bf v}}^{(j)}
+k^{[1]}(-a),{{\bf v}}^{(j)})}
\right)
\nonumber\\&& \times
\left[ 
-2\pi\left(\frac{\partial }
{\partial p_r^{(j)}}({{\bf v}}^{(j)}.{\bf e}^{(\alpha)[1]})\right)
\right]
\frac{K_c^2}{k^{[1]2}+K_c^2}.
\label{a3.1}
\end{eqnarray}
We  analyse the behaviour of $\tilde C(z,{{\bf v}}^{(j)})$ with respect to $K_c$ and ${{\bf v}}^{(j)}$.
The self-consistent equation (\ref{a2.26}) is diagonal with respect to the velocity ${{\bf v}}^{(j)}$.
In a first investigation, we consider the situation in which ${{\bf v}}^{(j)}$ is close to zero.
In these circonstances, 
$\left(\frac{\partial }{\partial p_r^{(j)}}({{\bf v}}^{(j)}.{\bf e}^{(\alpha)[1]})\right)$ is 
$\frac1m e_r^{(\alpha)[1]}$ and the summation over $r$ can be performed and we use the unity of the 
${\bf e}^{(\alpha)[1]}$ vector.
We get:
\begin{eqnarray}
&&\tilde C(z,0)
=
(-2\pi) \frac2m e_j^2 \frac1{(2\pi)^3}
\int d^3k^{[1]}
\frac{K_c^2}{k^{[1]2}+K_c^2}
\nonumber\\&& \times
\sum_{a=\pm 1}
\left(\frac{1}{z+k^{[1]}(-a)
-\tilde  C(z+k^{[1]}(-a), 0)
}
\right).
\label{a3.4}
\end{eqnarray}
We have ($\tilde C(z) \equiv\tilde C(z,0)$)
\begin{eqnarray}
&&\tilde C(z)
=
\frac{(-4\pi) }m e_j^2 \frac1{(2\pi)^3}
\int d^3k^{[1]}
\frac{K_c^2}{k^{[1]2}+K_c^2}
\nonumber\\&& \times
\sum_{a=\pm 1}
\left(\frac{1}{z+k^{[1]}(-a)
-\tilde  C(z+k^{[1]}(-a))
}
\right)
\nonumber\\&&=
-\frac{16\pi^2}{m} e_j^2 \frac1{(2\pi)^3}
\int_0^\infty dk\, k^2
\frac{K_c^2}{k^{[1]2}+K_c^2}
\nonumber\\&& \times
\left[
\left(\frac{1}{z-k-\tilde  C(z-k)}
\right)
+\left(\frac{1}{z+k-\tilde  C(z+k)}
\right)
\right].
\label{a3.5}
\end{eqnarray}
In our units in which the velocity of light is unity, the dimension of $\tilde C(z)$ is the inverse of a
time (as $z$) or the inverse of a wave number as $k$.
We can define $K_m$ by $\frac{m\pi}{2e_j^2}$.
We can therefore write $\tilde C(z)=zc(\frac{z}{K_m},\frac{K_c}{K_m})$ and we have.
\begin{eqnarray}
&&zc(\frac{z}{K_m},\frac{K_c}{K_m})
=
-\frac{1}{K_m} 
\int_0^\infty dk\, k^2
\frac{K_c^2}{k^2+K_c^2}
\nonumber\\&&\times
\left[
\left(\frac{1}{z-k-(z-k)c(\frac{z-k}{K_m},\frac{K_c}{K_m})}
\right)
+\left(
\frac{1}{z+k-(z+k)c(\frac{z+k}{K_m},\frac{K_c}{K_m})}
\right)
\right].
\nonumber\\&&
\label{a3.6}
\end{eqnarray}
We call $\gamma$ the ratio between the cut-off wave number $K_c$ and $K_m$.
Introducing $y=\frac{z}{K_m}$ and $u=\frac{k}{K_m}$, we have the equation for $y$ real (from above):
\begin{eqnarray}
&&c(y,\gamma)
=
-\frac{1}{y} 
\int_0^\infty du\, u^2
\frac{\gamma^2}{\gamma^2+u^2}
\nonumber\\&&\times
\left[
\left(\frac{1}{(i\epsilon +y-u)[1-c(y-u,\gamma)]}
\right)
+\left(\frac{1}{(i\epsilon +y+u)[1-c(y+u,\gamma)]}
\right)
\right].
\nonumber\\&&
\label{a3.7}
\end{eqnarray}
What kind of consistency can we deduce from (\ref{a3.7}) about the dependence of $c(y,\gamma)$
with respect to $\gamma$?

Let us be more specific about the behaviour of $c(y,\gamma)$.
We assume that 
\begin{equation}
c(y,\gamma)= \alpha \gamma^r+\beta \frac{\gamma^s}{y}
\label{a3.8}
\end{equation}
in the range $\vert y \vert>>\gamma$ and the self consistency should determine the complex
parameters  $\alpha$, $\beta$ and the real parameters $r$ and $s$. The parameter $r$ is assumed
positive and $s$ has no definite sign {\it a priori}. That expression for $c(y,\gamma)$ asssumes that 
$\lim_{y\to\infty}\frac{c(y,\gamma)}{c(-y,\gamma)}=1$.
We first check that property in appendix F by computing the difference
$\Delta(y,\gamma)=c(y,\gamma)-c(-y,\gamma)$ for large positive $y$.
The asymptotic behaviour of $c(y,\gamma)$ in the domain
$y>>\gamma>>1$ is then checked in a self consistent way by the proposal (\ref{a3.8}).
In order to make predictions on the behaviour of the self-field when the contributions of the
renormalised cycles are included, we need mainly the behaviour of $c(y,\gamma)$ in
the other domains, in particularly $\gamma>>y>>1$.

From the expression
\begin{eqnarray}
&&c(y,\gamma)
=
-\frac{1}{y} 
\int_0^\infty du\, u^2
\frac{\gamma^2}{\gamma^2+u^2}
\left[
\left(\frac{1}{(i\epsilon +y-u)[1-c(y-u,\gamma)]}
\right)
\right.\nonumber\\&&\left.
+\left(\frac{1}{(i\epsilon +y+u)[1-c(y+u,\gamma)]}
\right)
\right]
\label{a4.1}
\end{eqnarray}
and our previous results,
we have to deduce first the behaviour of $c(y,\gamma)$ in the other domains: $y<<\gamma$ and $y$ in
the range of $\gamma$.

We assume that 
\begin{equation}
c(y,\gamma)= \gamma^{\frac12} g(\frac{y}{\gamma})
\label{a4.2}
\end{equation}
in the range $\vert y \vert<<\gamma$  and the intermediary range. Self consistency should hold
and determine the complex function $g(\frac{y}{\gamma})$.  $g(0)$ is assumed $\not=0$ while
$g(\frac{y}{\gamma})$ is assumed to vanish for $\frac{y}{\gamma}\to \infty$.
That expression is analysed in details in appendix F and self consistency established.

That dependence can be further used to evaluate the dependence of the electric field with respect to
the cut-off value $K_c$  ($\gamma=\frac{K_c}{K_m}) $.

\subsection{Behaviour of the self-field}

We can now look at the consequences for the self-field.
The computation of the residue to get the required creation operator can be pursued in a similar way.
The residue at the vacuum pole ${{\bf k}}^{(j)}.{{\bf v}}^{(j)}+{\bf k}^{(j')}.{\bf v}^{(j')}$ does not play a
role for the creation operator: only the value of the pole matter.
We recover therefore an expression similar to the expression (\ref{a2.27}) in which the simple cycles
are replaced by the renormalised one's:
\begin{eqnarray}
&&<{\bf E}_r^{{\bot }}({\bf q}_j)>_{II\tilde C}^{e_j(1,1)}
=
e^2_je_{j'}\frac{1}{\pi}\frac1{(2\pi)^3}
\int d^3{\bf k}^{[1]}
\sum_{\alpha=1,2} \sum_{a=\pm 1}
{\bf e}_r^{\alpha}({\bf k}^{[1]}) 
\nonumber\\&& \times
\int d^3 l\frac{1}{l^2}
\left(\frac{1}{i\epsilon+({\bf l}+a{\bf k}^{[1]}).{\bf v}_j -ak^{[1]}
-\tilde C(i\epsilon+({\bf l}+a{\bf k}^{[1]}).{\bf v}_j -ak^{[1]},{\bf v}_j)}
\right)
\nonumber\\&& \times
\left(\frac{1}
{a{\bf k}^{[1]}.{\bf v}_j-ak^{[1]}-\tilde C(i\epsilon+a{\bf k}^{[1]}.{\bf v}_j-ak^{[1]},{\bf v}_j)}
\right)
\nonumber\\&& \times
\frac1m_j l_ve_v^{(\alpha)[1]}
e^{-i{\bf l}.[{\bf q}_j-{\bf q}_{j'}]}
\frac{K_c^2}{k^{[1]2}+K_c^2}.
\label{a5.1}
\end{eqnarray}
The same manipulations as for the simple cycle lead to a global dependence as the product of a
factor $K_c$ arising from the integral and two factors $K_c^{-\frac12}$ due to the presence in the
denominators of the renormalised cycle function $\tilde C$.
Therefore, the previous expression (\ref{a5.1}) for the self electric field provides a finite result in
the limit
$K_c\to \infty$.
We note that the dependence of the self-field on the charge $e_j$ disappears also in that limit.
This property is not at all astonishing and prove that an effective cut-off for wave numbers
larger than $\frac{m_j c^2}{e_j^2}$ is present.
The final dependence is proportional to that value and the dependence in $e_j$ cancels out.
Therefore, we meet here a typical case where a non-analycity in the expansion in the charge induces
divergence on individual diagrammatic contributions but an adequate resummation enables to
obtain finite results.

\subsection{Beyond the inserted cycles}
If we consider two cycles with an overlap, with respect to the previous case, we get two
supplementary propagators.
If the propagators are renormalised by the previous considered inserted cycles, the
contribution  of the imbricated cycles implies a possible $\gamma$ factor arising from the new
integration over a wave number and two $K_c^{\frac12}$ factors arising from the denominators.
Therefore, the global behaviour of the imbricated cycles is the same as a single (renormalised) cycle.
The imbricated cycles can thus be introduced for renormalising the propagators and the previous
analysis is still valid.
Therefore, if the contributions are computed in terms of renormalised propagators with all possible
vertices (defined also in terms of renormalised propagators), the result of the resummation is a
contribution to the self-field that is finite in the limit of an infinite cut-off $K_c$.
The replacement of a bare vertex in the expression of the self-field by dressed vertices does not
change either the finiteness of the global contribution.

\section{Conclusions} 

Our present work have illustrated the feasability  of a reformulation of classical electrodynamics,
that takes explicitly into account the corrections due to the self-fields.
Moreover, the procedure avoids  the existence of runaway solutions: causality is an ingredient of the
construction of the subdynamics operator.
Therfore, our expression for the self-force is not in terms of the time derivative of the acceleration 
but involves the actual position and velocity of the charged particle.
We justify in that way the procedure proposed by several author to avoid the runaway solutions: the
replacement of the time derivative of the acceleration by the time derivative of the external force.
In the traditionnal approach, the self-force is naturally computed from the characteristics of the
trajectory and the replacement has to be added by hand.
Here, we have made the opposite step: our expression in terms of the mean field has been shown to be
equivalent with the usual expression in terms of  the time derivative of the acceleration.

The present approach constitutes a statistical description of interacting charged particles and
electromagnetic fields: we are far from the classical view of well defined values for the variables
associated to the  fields and the particles: all these variables are statistical with a joint distribution
function that evolves with time.
The use of a reduced formalism enables to treat the distribution functions that are the most relevant
for the computation of mean values of all the dynamical functions.

Two distinct ingredients are required.
The first one is a relativistic statistical description of interacting fields and charged particles in
which no unobservable potential appears as dynamical variables.
Balescu-Poulain have developed further the ideas of Bialynicki-Birula \cite{BB71}, \cite{BB75} and his
coworkers to provide such a formalism free from dynamical constraints.
The elimination of the Lorentz condition is a key element of the present work that avoids
the usual derivation of the self-forces via the Li\'enard-Wiechert potentials.
The second ingredient is the possibility, that we have developed in collaboration with C. George,
of getting rid of the self-field by defining an appropriate subdynamics.
When both elements are combined, we obtain a finite kinetics for the description of the interacting
charges and fields in which no explicit self-energy process is allowed: the kinetic operator takes into
account all the effects and its computation, althought lenghty, is straightforward.

The formalism developed in this paper offers the basis to tackle in a new way the divergences in
classical electrodynamics, through (infinite) resummations of diagrams.
The renormalised propagators are then defined in terms of them-selves (for a similar procedure,
see \cite{dH04b}) and the solutions of the resulting non-linear equations do not admit a simple
expansion in the charge, enabling the presence of the natural cut-off, linked to the classical radius of
the electron.  

The present paper illustrates only one of the multiple potentialities of the approach.
Many problems can be aborded within the present formalism, such as the charge renormalisation, for
instance, of higher order effects.
Moreover, we have considered the charged particles outside an external influence: the distribution
function corresponding to the field vacuum has been used  thoroughly in this paper.
The effect of the magnetic field has not been specifically considered: when computing the power
dissipated  in the motion, its effect disappears.
We have not taken advantage of the statistical nature of the formalism: a sharp distribution function
has been assumed for the positions and velocities of the particles.
A statistical nature for the field has also been ignored.

An irreversible extension of CED, analogous to the treatment of the Lee model in  quantal case,
requires the construction of the generators of the Lie associated with the extended dynamics.
The relevance of such an extension is still to be  established.

We dare to state that we have presented new tools for dealing with problems in classical
electrodynamics.
The approach may look tedious but is nevetheless straightforward.
It opens new perspectives, not only in classical electrodynamics but also in quantum electrodynamics
\cite{BP75}.
 Obviously, a lot of work remains to be done.

\section{Appendix A. The Lorentz force}

\def\theequation{A.\arabic{equation}}

\setcounter{equation}{0}

In this part, we consider only one particle interacting with a free transverse wave.
The  particle will be pointlike, with a specific value for the velocity.

From the expression (\ref{3.11}) for $<11[0]|\tilde \Theta|11[1(f)]>^{(0,1)}$, we deduce 
the change to the one particle distribution function due to that contribution, assuming
 the independence of the field and particle variables.
\begin{eqnarray}
&&\left. \partial_t \tilde f({\bf k},{\bf p},t)\right\vert_1=
- e\frac1{(2\pi)^{\frac32}}
\int d^3k^{[1]}
\int_0^{\infty}d\eta_1^{[1]}\int_0^{\infty}d\eta_2^{[1]}
\nonumber \\&&\times
\sum_{m_{1}^{[1]},m_{2}^{[1]}}
\delta_{m_{1}^{[1]},0}\delta_{m_{2}^{[1]},0}
\sum_{\alpha=1,2} 
\sum_{a=\pm 1}
\left(
\frac{\eta_{\alpha}^{[1]}}
{k^{[1]}}
\right)^{\frac12} 
\nonumber \\&&\times[k^{[1]} e_r^{(\alpha)[1]}-g^{st} v_s^{(j)}
(e_t^{(\alpha)[i]} k_r^{[i]}-e_r^{(\alpha)[i]} k_t^{[i]})]
\frac{\partial}
{\partial p_r}
\nonumber \\&&\times
\exp a\left\{-{\bf k}^{[1]}.\frac{\partial}{\partial {\bf k}}-\frac{\partial}{\partial
m_{\alpha}^{[1]}}\right\}
\tilde f({\bf k},{\bf p},t) \tilde f_{[1]}(\eta_{1}^{[1]},m_{1}^{[1]},\eta_{2}^{[1]}, m_{2}^{[1]};{\bf k}^{[1]}).
\nonumber \\&&
\label{3.11b}
\end{eqnarray}

If we suppose that $\tilde f$ describes a particle localized at some place ${\bf r}(t)$, $\tilde
f({\bf k},{\bf v},t)$ is proportional to $\exp{-i{\bf k}.{\bf r}(t)}$  (\ref{3.5}). The action of the displacement
operator
$\exp a\left\{-{\bf k}^{[1]}.\frac{\partial}{\partial {\bf k}}-\frac{\partial}{\partial
m_{\alpha}^{[1]}}\right\}$  can thus be performed and
we get easily:
\begin{eqnarray}
&&\left. \partial_t \tilde f({\bf k},{\bf v},t)\right\vert_1=
- \frac{e}{m}\frac1{(2\pi)^{\frac32}}
\int d^3k^{[1]}
\int_0^{\infty}d\eta_1^{[1]}\int_0^{\infty}d\eta_2^{[1]}
\sum_{\alpha=1,2} 
\sum_{a=\pm 1}
\left(
\frac{\eta_{\alpha}^{[1]}}
{k^{[1]}}
\right)^{\frac12} 
\nonumber \\&&\times
[k^{[1]} e_r^{(\alpha)[1]}-g^{st} v_s^{(j)}
(e_t^{(\alpha)[i]} k_r^{[i]}-e_r^{(\alpha)[i]} k_t^{[i]})]
\frac{\partial}
{\partial v_r}
\exp a\left\{i{\bf k}^{[1]}.{\bf r}(t)\right\}
\nonumber \\&&\times\tilde f({\bf k},{\bf v},t) 
\tilde f_{[1]}(\eta_{1}^{[1]},-a\delta_{\alpha,1},\eta_{2}^{[1]},-a\delta_{\alpha,2};{\bf k}^{[1]}).
\label{3.11c}
\end{eqnarray}
The mean values $<{\bf E}_r^{{\bot }}({\bf x})>$ and $<{\bf B}_r^{{\bot }}({\bf x})>$ of the fields can be
deduced easily from (\ref{2.2}) and (\ref{2.3}):
\begin{eqnarray}
&&<{\bf E}_r^{{\bot }}({\bf x})>=
\int d^3{\bf k}^{[1]}\int_0^{\infty}d\eta_1^{[1]}\int_0^{\infty}d\eta_2^{[1]}
\int_0^1d\xi_1^{[1]}\int_0^1d\xi_2^{[1]}
\nonumber \\ &&\times\frac1{(2\pi)^{\frac32}}\sum_{\alpha=1,2} \sum_{a=\pm 1}
{k^{[1]}}^{\frac12} {\bf e}_r^{\alpha}({\bf k}^{[1]}) \eta_{\alpha}^{\frac12}({\bf k}^{[1]})
\nonumber \\ &&\times\exp \{ia[{\bf k}^{[1]}.{\bf x}-2\pi\xi^{[1]}_{\alpha}({\bf k}^{[1]})]\}
\tilde f_{[1]}(\chi^{[1]};{\bf k}^{[1]})
\nonumber \\
&&=\int d^3{\bf k}^{[1]}\int_0^{\infty}d\eta_1^{[1]}\int_0^{\infty}d\eta_2^{[1]}
\int_0^1d\xi_1^{[1]}\int_0^1d\xi_2^{[1]}
\nonumber \\ &&\times
\frac1{(2\pi)^{\frac32}}\sum_{\alpha=1,2} \sum_{a=\pm 1}
{k^{[1]}}^{\frac12} {\bf e}_r^{\alpha}({\bf k}^{[1]}) \eta_{\alpha}^{\frac12}({\bf k}^{[1]})
\exp \{ia[{\bf k}^{[1]}.{\bf x}-2\pi\xi_{\alpha}({\bf k}^{[1]})]\}
\nonumber \\ &&\times
\sum_{m_{1}^{[1]},m_{2}^{[1]}}
e^{-2\pi i(m_{1}^{[1]}\xi_{1}^{[1]}+m_{2}^{[1]}\xi_{2}^{[1]})}
\tilde {f}_{[1]}(\eta_{1}^{[1]},
m_{1}^{[1]},\eta_{2}^{[1]}, m_{2}^{[1]};{\bf k}^{[1]})
\nonumber \\
&&=
\frac1{(2\pi)^{\frac32}}\int d^3{\bf k}^{[1]}\int_0^{\infty}d\eta_1^{[1]}\int_0^{\infty}d\eta_2^{[1]}
\sum_{\alpha=1,2} \sum_{a=\pm 1}
{k^{[1]}}^{\frac12} {\bf e}_r^{\alpha}({\bf k}^{[1]}) \eta_{\alpha}^{\frac12}({\bf k}^{[1]})
\nonumber \\ &&\times\exp \{ia{\bf k}^{[1]}.{\bf x}\}
\tilde {f}_{[1]}(\eta_{1}^{[1]},-a\delta_{\alpha,1},\eta_{2}^{[1]}, -a\delta_{\alpha,2};{\bf k}^{[1]}),
\label{3.11e}
\end{eqnarray} 
\begin{eqnarray}
&&<{\bf B}_r^{{\bot }}({\bf x})>=
\int d^3{\bf k}^{[1]}\int_0^{\infty}d\eta_1^{[1]}\int_0^{\infty}d\eta_2^{[1]}
\int_0^1d\xi_1^{[1]}\int_0^1d\xi_2^{[1]}
\nonumber \\ &&\times
\frac1{(2\pi)^{\frac32}}\sum_{\alpha=1,2} \sum_{a=\pm 1}
{k^{[1]}}^{\frac12} (-1)^{\alpha}{\bf e}_r^{\alpha}({\bf k}^{[1]}) \eta_{\alpha}^{\frac12}({\bf k}^{[1]})
\nonumber \\ &&\times
\exp \{ia[{\bf k}^{[1]}.{\bf x}-2\pi\xi^{[1]}_{\alpha}({\bf k}^{[1]})]\}
\tilde f_{[1]}(\chi^{[1]};{\bf k}^{[1]})
\nonumber \\
&&=\int d^3{\bf k}^{[1]}\int_0^{\infty}d\eta_1^{[1]}\int_0^{\infty}d\eta_2^{[1]}
\int_0^1d\xi_1^{[1]}\int_0^1d\xi_2^{[1]}
\nonumber \\ &&\times
\frac1{(2\pi)^{\frac32}}\sum_{\alpha=1,2} \sum_{a=\pm 1}
{k^{[1]}}^{\frac12} (-1)^{\alpha}{\bf e}_r^{\alpha}({\bf k}^{[1]}) \eta_{\alpha}^{\frac12}({\bf k}^{[1]})
\exp \{ia[{\bf k}^{[1]}.{\bf x}-2\pi\xi_{\alpha}({\bf k}^{[1]})]\}
\nonumber \\ &&\times
\sum_{m_{1}^{[1]},m_{2}^{[1]}}
e^{-2\pi i(m_{1}^{[1]}\xi_{1}^{[1]}+m_{2}^{[1]}\xi_{2}^{[1]})}
\tilde {f}_{[1]}(\eta_{1}^{[1]},
m_{1}^{[1]},\eta_{2}^{[1]}, m_{2}^{[1]};{\bf k}^{[1]})
\nonumber \\
&&=\frac1{(2\pi)^{\frac32}}\int d^3{\bf k}^{[1]}\int_0^{\infty}d\eta_1^{[1]}\int_0^{\infty}d\eta_2^{[1]}
\sum_{\alpha=1,2} \sum_{a=\pm 1}
{k^{[1]}}^{\frac12} (-1)^{\alpha}{\bf e}_r^{\alpha}({\bf k}^{[1]}) \eta_{\alpha}^{\frac12}({\bf k}^{[1]})
\nonumber \\ &&\times
\exp \{ia{\bf k}^{[1]}.{\bf x}\}
\tilde {f}_{[1]}(\eta_{1}^{[1]},-a\delta_{\alpha,1},\eta_{2}^{[1]}, -a\delta_{\alpha,2};{\bf k}^{[1]}),
\label{3.11ea}
\end{eqnarray}
so that we can proceed to the identification (\ref{3.11d}).

\section{Appendix B}

\def\theequation{B.\arabic{equation}}

\setcounter{equation}{0}

This appendix completes the list of the matrix elements of $<11[0]|\tilde \Theta|11[1(f)]>^{(0,2)}$.
\begin{eqnarray}
&&<11[0]|\tilde \Theta|11[1(f)]>_{==}^{(0,2)}
\nonumber \\&&=
\sum_{j=1,2}
i e_j^2 
\frac1{(2\pi)^{3}}
\frac{\partial}{\partial p_r^{(j)}}
\int d^3k^{[1]}
\int_0^{\infty}d\eta_1^{[1]}\int_0^{\infty}d\eta_2^{[1]}
\sum_{\alpha=1,2} 
\sum_{a=\pm 1}
\left(
\frac{\eta_{\alpha}^{[1]}}
{k^{[1]}}
\right)
\nonumber \\&&\times
\left[ 
k^{[1]} e_r^{(\alpha)[1]}-g^{st} v_s^{(j)}
(e_t^{(\alpha)[1]} k_r^{[1]}-e_r^{(\alpha)[1]} k_t^{[1]})
\right]
\left(\frac{1}{-a{\bf k}^{[1]}.{\bf v}^{(j)}+a k^{[1]}}\right)
\nonumber \\&&\times
\left[ 
[k^{[1]} e_{r'}^{(\alpha)[1]}-g^{s't'} v_{s'}^{(j)}
(e_{t'}^{(\alpha)[1]} k_{r'}^{[1]}-e_{r'}^{(\alpha)[1]} k_{t'}^{[1]})]
\frac{\partial}
{\partial p_{r'}^{(j)}}
\right.\nonumber \\&&\left.
-2\pi({\bf v}^{(j)}.{\bf e}^{(\alpha)[1]})
\left( 
 \frac{\partial}
{\partial \eta_{\alpha}^{[1]}}
+\frac{1}{\eta_{\alpha}^{[1]}}
\right)
\right]
\nonumber \\&&\times
\sum_{m_{1}^{[1]},m_{2}^{[1]}}
\delta_{m_{1}^{[1]},0}\delta_{m_{2}^{[1]},0}
\exp 2a\left\{-{\bf k}^{[1]}.\frac{\partial}{\partial {\bf k}^{(j)}}-\frac{\partial}{\partial
m_{\alpha}^{[1]}}\right\}
\nonumber \\&&+
\sum_{j=1,2}
(-i) e_j^2 
\frac1{(2\pi)^{3}}
\frac{\partial}{\partial p_r^{(j)}}
\int d^3k^{[1]}
\int_0^{\infty}d\eta_1^{[1]}\int_0^{\infty}d\eta_2^{[1]}
\sum_{\alpha=1,2} 
\sum_{a=\pm 1}
\left(
\frac{\eta_{\alpha}^{[1]}}
{k^{[1]}}
\right)
\nonumber \\&&\times
\left[ 
k^{[1]} e_r^{(\alpha)[1]}-g^{st} v_s^{(j)}
(e_t^{(\alpha)[1]} k_r^{[1]}-e_r^{(\alpha)[1]} k_t^{[1]})
\right]
\left(\frac{1}{-a{\bf k}^{[1]}.{\bf v}^{(j)}+a k^{[1]}}\right)^2
\nonumber \\&&\times
\left[ 
k^{[1]} e_{r'}^{(\alpha)[1]}-g^{s't'} v_{s'}^{(j)}
(e_{t'}^{(\alpha)[1]} k_{r'}^{[1]}-e_{r'}^{(\alpha)[1]} k_{t'}^{[1]})
\right]
\left( k_{s'}^{(j)}-2a k_{s'}^{[1]}
\right)
\left(\frac{\partial v_{s'}^{(j)}}{\partial p_{r'}^{(j)}}
\right)
\nonumber \\&&\times
\sum_{m_{1}^{[1]},m_{2}^{[1]}}
\delta_{m_{1}^{[1]},0}\delta_{m_{2}^{[1]},0}
\exp 2a\left\{-{\bf k}^{[1]}.\frac{\partial}{\partial {\bf k}^{(j)}}-\frac{\partial}{\partial
m_{\alpha}^{[1]}}\right\},
\label{3.52e}
\end{eqnarray} 
\begin{eqnarray}
&&<11[0]|\tilde \Theta|11[1(f)]>_{\not==}^{(0,2)}
\nonumber \\&&=
\sum_{j=1,2}
i e_j^2 \frac1{(2\pi)^{3}}
\frac{\partial}
{\partial p_r^{(j)}}
\int d^3k^{[1]}
\int_0^{\infty}d\eta_1^{[1]}\int_0^{\infty}d\eta_2^{[1]}
\sum_{\alpha=1,2} 
\sum_{a=\pm 1}
\left(
\frac{\eta_{\alpha}^{[1]}}
{k^{[1]}}
\right)^{\frac12} 
\left(
\frac{\eta_{\alpha'}^{[1]}}
{k^{[1]}}
\right)^{\frac12}  
\nonumber \\&&\times
\left(\frac{1}{-a{\bf k}^{[1]}.{\bf v}^{(j)}+ak^{[1]}}
\right)
\left[ 
k^{[1]} e_r^{(\alpha)[1]}-g^{st} v_s^{(j)}
(e_t^{(\alpha)[1]} k_r^{[1]}-e_r^{(\alpha)[1]} k_t^{[1]})
\right]
\nonumber \\&&\times
[k^{[1]} e_{r'}^{(\alpha')[1]}-g^{s't'} v_{s'}^{(j)}
(e_{t'}^{(\alpha')[1]} k_{r'}^{[1]}-e_{r'}^{(\alpha')[1]} k_{t'}^{[1]})]
\frac{\partial}
{\partial p_{r'}^{(j)}}
\nonumber \\&&\times
\sum_{m_{1}^{[1]},m_{2}^{[1]}}
\delta_{m_{1}^{[1]},0}\delta_{m_{2}^{[1]},0}
\exp \left\{-2a{\bf k}^{[1]}.\frac{\partial}{\partial {\bf k}^{(j)}}-a\frac{\partial}{\partial
m_{\alpha}^{[1]}}-a\frac{\partial}{\partial m_{\alpha'}^{[1]}}\right\}
\nonumber \\&&+
\sum_{j=1,2}
(-i) e_j^2 \frac1{(2\pi)^{3}}
\frac{\partial}
{\partial p_r^{(j)}}
\int d^3k^{[1]}
\int_0^{\infty}d\eta_1^{[1]}\int_0^{\infty}d\eta_2^{[1]}
\sum_{\alpha=1,2} 
\sum_{a=\pm 1}
\left(
\frac{\eta_{\alpha}^{[1]}}
{k^{[1]}}
\right)^{\frac12} 
\left(
\frac{\eta_{\alpha'}^{[1]}}
{k^{[1]}}
\right)^{\frac12} 
\nonumber \\&&\times
\left(\frac{1}{-a{\bf k}^{[1]}.{\bf v}^{(j)}+ak^{[1]}}
\right)^2
\left[ 
k^{[1]} e_r^{(\alpha)[1]}-g^{st} v_s^{(j)}
(e_t^{(\alpha)[1]} k_r^{[1]}-e_r^{(\alpha)[1]} k_t^{[1]})
\right]
\nonumber \\&&\times
[k^{[1]} e_{r'}^{(\alpha')[1]}-g^{s't'} v_{s'}^{(j)}
(e_{t'}^{(\alpha')[1]} k_{r'}^{[1]}-e_{r'}^{(\alpha')[1]} k_{t'}^{[1]})]
[ k_{s'}^{(j)}-2a k_{s'}^{[1]}]
\left(\frac{\partial v_{s'}^{(j)}}
{\partial p_{r'}^{(j)}}\right)
\nonumber \\&&\times
\sum_{m_{1}^{[1]},m_{2}^{[1]}}
\delta_{m_{1}^{[1]},0}\delta_{m_{2}^{[1]},0}
\exp \left\{-2a{\bf k}^{[1]}.\frac{\partial}{\partial {\bf k}^{(j)}}-a\frac{\partial}{\partial
m_{\alpha}^{[1]}}-a\frac{\partial}{\partial m_{\alpha'}^{[1]}}\right\},
\label{3.99}
\end{eqnarray}
\begin{eqnarray}
&&<11[0]|\tilde \Theta|11[1(f)]>_{\not=\not=}^{(0,2)}
\nonumber \\&&=
\sum_{j=1,2}
ie_j^2 \frac1{(2\pi)^{3}}
\frac{\partial}
{\partial p_r^{(j)}}
\int d^3k^{[1]}
\int_0^{\infty}d\eta_1^{[1]}\int_0^{\infty}d\eta_2^{[1]}
\sum_{\alpha=1,2} 
\sum_{a=\pm 1}
\left(
\frac{\eta_{\alpha}^{[1]}}
{k^{[1]}}
\right)^{\frac12} 
\left(
\frac{\eta_{\alpha'}^{[1]}}
{k^{[1]}}
\right)^{\frac12} 
\nonumber \\&&\times
\left(\frac{1}{a{\bf k}^{[1]}.{\bf v}^{(j)}-ak^{[1]}}
\right)
\left[ 
k^{[1]} e_r^{(\alpha)[1]}-g^{st} v_s^{(j)}
(e_t^{(\alpha)[1]} k_r^{[1]}-e_r^{(\alpha)[1]} k_t^{[1]})
\right]
\nonumber \\&&\times
[k^{[1]} e_{r'}^{(\alpha')[1]}-g^{s't'} v_{s'}^{(j)}
(e_{t'}^{(\alpha')[1]} k_{r'}^{[1]}-e_{r'}^{(\alpha')[1]} k_{t'}^{[1]})]
\frac{\partial}
{\partial p_{r'}^{(j)}}
\nonumber \\&&\times
\sum_{m_{1}^{[1]},m_{2}^{[1]}}
\delta_{m_{1}^{[1]},0}\delta_{m_{2}^{[1]},0}
\exp \left\{-a\frac{\partial}{\partial
m_{\alpha}^{[1]}}+a\frac{\partial}{\partial m_{\alpha'}^{[1]}}\right\}
\nonumber \\&&+ 
\sum_{j=1,2}
(-i) e_j^2 \frac1{(2\pi)^{3}}
\frac{\partial}
{\partial p_r^{(j)}}
\int d^3k^{[1]}
\int_0^{\infty}d\eta_1^{[1]}\int_0^{\infty}d\eta_2^{[1]}
\sum_{\alpha=1,2} 
\sum_{a=\pm 1}
\left(
\frac{\eta_{\alpha}^{[1]}}
{k^{[1]}}
\right)^{\frac12} 
\left(
\frac{\eta_{\alpha'}^{[1]}}
{k^{[1]}}
\right)^{\frac12} 
\nonumber \\&&\times
\left(\frac{1}{a{\bf k}^{[1]}.{\bf v}^{(j)}-ak^{[1]}}
\right)^2
\left[ 
k^{[1]} e_r^{(\alpha)[1]}-g^{st} v_s^{(j)}
(e_t^{(\alpha)[1]} k_r^{[1]}-e_r^{(\alpha)[1]} k_t^{[1]})
\right]
\nonumber \\&&\times
[k^{[1]} e_{r'}^{(\alpha')[1]}-g^{s't'} v_{s'}^{(j)}
(e_{t'}^{(\alpha')[1]} k_{r'}^{[1]}-e_{r'}^{(\alpha')[1]} k_{t'}^{[1]})]
[ k_{s'}^{(j)}]
\left(\frac{\partial v_{s'}^{(j)}}
{\partial p_{r'}^{(j)}}\right)
\nonumber \\&&\times
\sum_{m_{1}^{[1]},m_{2}^{[1]}}
\delta_{m_{1}^{[1]},0}\delta_{m_{2}^{[1]},0}
\exp \left\{-a\frac{\partial}{\partial
m_{\alpha}^{[1]}}+a\frac{\partial}{\partial m_{\alpha'}^{[1]}}\right\}.
\label{3.102}
\end{eqnarray}

\section{Appendix C}

\def\theequation{C.\arabic{equation}}

\setcounter{equation}{0}

The expression of the complete electric field $<{\bf E}({\bf x})>^{e_j(0,1)}$ (\ref{4.36}) is evaluated
explicitly, using its identification with $<{\bf E}^{{\bot }}({\bf x})>_a^{e_j(0,1)}$.
Multiplying numerator and denominator by $(k^{[1]}+{\bf k}^{[1]}.{\bf v}_j)$, we have 
\begin{eqnarray}
&&<{\bf E}^{{\bot }}({\bf x})>_a^{e_j(0,1)}
=(2\pi)e_j\frac1{(2\pi)^3}\int d^3{\bf k}^{[1]}
\sin[{\bf k}^{[1]}.({\bf x}-{\bf q}_j)]
\nonumber \\ &&\times
\left(\frac{1}{(k^{[1]})^2-({\bf k}^{[1]}.{\bf v}_j)^2}
\right)
(k^{[1]}+{\bf k}^{[1]}.{\bf v}_j)
\left({\bf k}^{[1]}-k^{[1]}{\bf v}_{j}\right)\frac1{k^{[1]}}.
\label{4.36a}
\end{eqnarray}
Let us place the $x$ axis along $({\bf x}-{\bf q}_j)$ and the $y$ axis along ${\bf v}_{\bot j}$,
defined by
${\bf v}_{\bot j}={\bf v}_{j}-\frac{[{\bf v}_{ j}.({\bf x}-{\bf q}_j)]({\bf x}-{\bf q}_j)}
{\vert{\bf x}-{\bf q}_j\vert^2}$.

\begin{eqnarray}
&&<{\bf E}^{{\bot }}({\bf x})>_{a}^{e_j(0,1)}
=(4\pi)e_j\frac1{(2\pi)^3}\int_{-\infty} ^{+\infty}dk_x\,\int_{-\infty}
^{+\infty}dk_y\,\int_{-\infty} ^{+\infty}dk_z\,
\sin k_x\vert{\bf x}-{\bf q}_j\vert
\nonumber \\&&\times
\left(\frac{1}{k^2-(v_{jx}k_x+v_{jy}k_y)^2}
\right)
(k+(v_{jx}k_x+v_{jy}k_y))
\nonumber \\&&\times
\left((k_x{\bf e}_x+k_y{\bf e}_y+k_z{\bf e}_z)-k{\bf v}_{j}\right)\frac1{k}.
\label{4.36b}
\end{eqnarray}
The contribution involving $k_z{\bf e}_z$ vanishes obviously for parity reasons.
The integrand has to be even for a simultaneously change of the sign of $k_x$ and $k_y$.
Therefore,
\begin{eqnarray}
&&<{\bf E}^{{\bot }}({\bf x})>_{a}^{e_j(0,1)}
=e_j\frac1{2\pi^2}\int_{-\infty} ^{+\infty}dk_x\,\int_{-\infty}
^{+\infty}dk_y\,\int_{-\infty} ^{+\infty}dk_z\,
\sin k_x\vert{\bf x}-{\bf q}_j\vert
\nonumber \\&&\times
\left(\frac{1}{k^2-(v_{jx}k_x+v_{jy}k_y)^2}
\right)
\nonumber \\&&\times
\left(k(k_x{\bf e}_x+k_y{\bf e}_y)-k(v_{jx}k_x+v_{jy}k_y){\bf v}_{j}\right)\frac1{k}
\nonumber \\&&=e_j\frac1{2\pi^2}\int_{-\infty} ^{+\infty}dk_x\,\int_{-\infty}
^{+\infty}dk_y\,\int_{-\infty} ^{+\infty}dk_z\,
\sin k_x\vert{\bf x}-{\bf q}_j\vert
\nonumber \\&&\times
\left(\frac{1}{k^2-(v_{jx}k_x+v_{jy}k_y)^2}
\right)
\left((k_x{\bf e}_x+k_y{\bf e}_y)-(v_{jx}k_x+v_{jy}k_y){\bf v}_{j}\right).
\label{4.36c}
\end{eqnarray}
We use dimensionless variables of integration.
We then replace $\sin k_x$ by $\frac1{2i}(e^{ik_x}-e^{-ik_x})$
and perform the integration over $k_x$ by residue at the pole of
$\frac{1}{k^2-(v_{jx}k_x+v_{jy}k_y)^2}$ in the correct half plane.
We have
\begin{eqnarray}
&&I_1=\int_{-\infty} ^{+\infty}dk_x\,\int_{-\infty}
^{+\infty}dk_y\,\int_{-\infty} ^{+\infty}dk_z\,
e^{ik_x}\left(\frac{1}{k^2-(v_{jx}k_x+v_{jy}k_y)^2}
\right)
\nonumber \\&&\times
\left[(k_x{\bf e}_x+k_y{\bf e}_y)-(v_{jx}k_x+v_{jy}k_y){\bf v}_{j}\right].
\label{4.36e}
\end{eqnarray} 
The pole is obtained by the equation
\begin{eqnarray}
&&
k^2-(v_{jx}k_x+v_{jy}k_y)^2=0,
\nonumber \\&&
k_x^2(1-v_{jx}^2)-2k_xk_yv_{jx}v_{jy}+k_y^2(1-v_{jy}^2)+k_z^2=0.
\label{4.36f}
\end{eqnarray}
Therefore, 
\begin{eqnarray}
&&k_x=\frac{k_yv_{jx}v_{jy}\pm\sqrt{(k_yv_{jx}v_{jy})^2-(1-v_{jx}^2)[k_y^2(1-v_{jy}^2)+k_z^2]}}
{(1-v_{jx}^2)}
\nonumber \\&&=
\frac{k_yv_{jx}v_{jy}\pm i\sqrt{(1-v_{jx}^2)[k_y^2(1-v_{jy}^2)+k_z^2]-(k_yv_{jx}v_{jy})^2}}
{(1-v_{jx}^2)}.
\label{4.36g}
\end{eqnarray}
Due to the factor $e^{ik_x}$, the relevant pole for $I_1$ corresponds to the plus sign and we have
\begin{eqnarray}
&&I_1=2\pi i\int_{-\infty}^{+\infty}dk_y\,\int_{-\infty} ^{+\infty}dk_z\,
e^{-\frac{\sqrt{
(1-v_{jx}^2)[k_y^2(1-v_{jy}^2)+k_z^2]-(k_yv_{jx}v_{jy})^2}}
{(1-v_{jx}^2)}}
e^{i\frac{k_yv_{jx}v_{jy}}{(1-v_{jx}^2)}}
\nonumber \\&&\times
\frac{1}{2i\sqrt{(1-v_{jx}^2)[k_y^2(1-v_{jy}^2)+k_z^2]-(k_yv_{jx}v_{jy})^2}}
\nonumber \\&&\times
\left[\frac{k_yv_{jx}v_{jy}+ i\sqrt{(1-v_{jx}^2)[k_y^2(1-v_{jy}^2)+k_z^2]-(k_yv_{jx}v_{jy})^2}}
{(1-v_{jx}^2)}
[{\bf e}_x-v_{jx}{\bf v}_{j}]
\right.\nonumber \\&&\left.
+k_y[{\bf e}_y-v_{jy}{\bf v}_{j}]\right]
\nonumber \\&&=
\pi \int_{-\infty}^{+\infty}dk_y\,\int_{-\infty} ^{+\infty}dk_z\,
e^{-\frac{\sqrt{
(1-v_{jx}^2)[k_y^2(1-v_{jy}^2)+k_z^2]-(k_yv_{jx}v_{jy})^2}}
{(1-v_{jx}^2)}}
e^{i\frac{k_yv_{jx}v_{jy}}{(1-v_{jx}^2)}}
\nonumber \\&&\times
\frac{1}{\sqrt{(1-v_{jx}^2)[k_y^2(1-v_{jy}^2)+k_z^2]-(k_yv_{jx}v_{jy})^2}}
\nonumber \\&&\times
\left[\frac{k_yv_{jx}v_{jy}+ i\sqrt{(1-v_{jx}^2)[k_y^2(1-v_{jy}^2)+k_z^2]-(k_yv_{jx}v_{jy})^2}}
{(1-v_{jx}^2)}
[{\bf e}_x-v_{jx}{\bf v}_{j}]
\right.\nonumber \\&&\left.
+k_y[{\bf e}_y-v_{jy}{\bf v}_{j}]\right].
\label{4.36h}
\end{eqnarray}
We replace the oscillating factor according to its parity in $k_y$.
\begin{eqnarray}
&&I_{1a}=
\pi i\frac1{(1-v_{jx}^2)}[{\bf e}_x-v_{jx}{\bf v}_{j}]
\int_{-\infty}^{+\infty}dk_y\,\int_{-\infty} ^{+\infty}dk_z\,
\nonumber \\&&\times
e^{-\frac{\sqrt{
(1-v_{jx}^2)[k_y^2(1-v_{jy}^2)+k_z^2]-(k_yv_{jx}v_{jy})^2}}
{(1-v_{jx}^2)}}
\cos{\left(\frac{k_yv_{jx}v_{jy}}{(1-v_{jx}^2)}\right)}.
\label{4.36i}
\end{eqnarray}
Introducing polar coordinates  $r$ and $\theta$ in the $k_y$, $k_z$ plane, we get 
\begin{eqnarray}
&&I_{1a}=
\pi i[{\bf e}_x-v_{jx}{\bf v}_{j}](1-v_{jx}^2)
\int_{0}^{\infty}dr\,r\int_{0} ^{2\pi}d\theta\,
\nonumber \\&&\times
e^{-r\sqrt{1-v_{jx}^2-v_{jy}^2\cos^2\theta}}
\cos{(r{ v_{jx}v_{jy}\cos\theta})}
\nonumber \\&&=
\frac12\pi i[{\bf e}_x-v_{jx}{\bf v}_{j}](1-v_{jx}^2)
\int_{0}^{\infty}dr\,r\int_{0} ^{2\pi}d\theta\,
\nonumber \\&&\times
\left[e^{-r[\sqrt{1-v_{jx}^2-v_{jy}^2\cos^2\theta}+iv_{jx}v_{jy}\cos\theta]}
+e^{-r[\sqrt{1-v_{jx}^2-v_{jy}^2\cos^2\theta}-iv_{jx}v_{jy}\cos\theta]}
\right].
\nonumber \\&&
\label{4.36k}
\end{eqnarray}
The integration over $r$ is readily performed.
\begin{eqnarray}
&&I_{1a}=
\frac12\pi i[{\bf e}_x-v_{jx}{\bf v}_{j}](1-v_{jx}^2)
\int_{0} ^{2\pi}d\theta\,
\nonumber \\&&\times
\left[
\frac1{[\sqrt{1-v_{jx}^2-v_{jy}^2\cos^2\theta}+iv_{jx}v_{jy}\cos\theta]^2}
\right.\nonumber \\&&\left.+
\frac1{[\sqrt{1-v_{jx}^2-v_{jy}^2\cos^2\theta}-iv_{jx}v_{jy}\cos\theta]^2}
\right]
\nonumber \\&&=
\pi i[{\bf e}_x-v_{jx}{\bf v}_{j}](1-v_{jx}^2)
\int_{0} ^{2\pi}d\theta\,
\frac{1-v_{jx}^2-v_{jy}^2\cos^2\theta-v_{jx}^2v_{jy}^2\cos^2\theta}
{[1-v_{jx}^2-v_{jy}^2\cos^2\theta+v_{jx}^2v_{jy}^2\cos^2\theta]^2}
\nonumber \\&&=
\pi i[{\bf e}_x-v_{jx}{\bf v}_{j}]\frac1{(1-v_{jx}^2)}
\int_{0} ^{2\pi}d\theta\,
\frac{1-v_{jx}^2-v_{jy}^2\cos^2\theta-v_{jx}^2v_{jy}^2\cos^2\theta}
{[1-v_{jy}^2\cos^2\theta]^2}.
\nonumber \\&&
\label{4.36ka}
\end{eqnarray}
Taking $\phi=2\theta$ as new integration variable, we get
\begin{eqnarray}
&&I_{1a}=
2\pi i[{\bf e}_x-v_{jx}{\bf v}_{j}]\frac1{(1-v_{jx}^2)}
\nonumber \\&&\times
\int_{0} ^{\pi}d\phi\,
\frac{1-v_{jx}^2-\frac12v_{jy}^2(1+v_{jx}^2)-\frac12v_{jy}^2(1+v_{jx}^2)\cos\phi}
{[1-\frac12v_{jy}^2-\frac12v_{jy}^2\cos\phi]^2}.
\label{4.36l}
\end{eqnarray}
From formulae 2.554.2 and 2.554.2, 148 of \cite{GR65}, we read
\begin{eqnarray}
&&\int\frac{A+B\cos x}{(a+b\cos x)^n}dx=\frac{1}{(n-1)(a^2-b^2)}
\left[\frac{(aB-Ab)\sin x}{(a+b\cos x)^{n-1}}
\right.\nonumber \\&&\left.
+\int\frac{(Aa-bB)(n-1)+(n-2)(aB-Ab)\cos x}{(a+b\cos x)^{n-1}}dx\right],
\label{4.36m}
\end{eqnarray}
\begin{equation}
\int\frac{A+B\cos x}{a+b\cos x}dx=\frac Bb x
+\frac{Ab-aB}{b}\int\frac{1}{a+b\cos x}dx,
\label{4.36n}
\end{equation}
with, formula 2.553.3, for $a^2>b^2$ 
\begin{equation}
\int\frac{1}{a+b\cos x}dx=\frac2{\sqrt{a^2-b^2}}\arctan\frac{\sqrt{a^2-b^2}\tan \frac x2}{a+b}
\label{4.36o}
\end{equation}
Therefore, the last integration can be performed and we get
\begin{equation}
I_{1a}=
2\pi^2 i[(1-v_{jx}^2){\bf e}_x-v_{jx}v_{jy}{\bf e}_y]\frac{[1-v_{j}^2]}{(1-v_{jx}^2)}
\frac1{(1-v_{jy}^2)^{\frac32}}.
\label{4.36q}
\end{equation}
We now turn to the second term of (\ref{4.36h}) that is evaluated in a similar way:
\begin{equation}
I_{1b}
=
2\pi^2 i\frac1{(1-v_{jy}^2)^{\frac32}}
v_{jx}v_{jy}\frac{[1-v_{j}^2)]}{(1-v_{jx}^2)}{\bf e}_y.
\label{4.36id}
\end{equation}
The sum of the contributions $I_1=I_{1a}+I_{1b}$ is the contribution along ${\bf e}_x$ of $I_{1a}$
(\ref{4.36q}) and is given by
\begin{equation}
I_1=2\pi^2 i[1-v_{j}^2]
\frac1{(1-v_{jy}^2)^{\frac32}}{\bf e}_x.
\label{4.36ie}
\end{equation}
The contribution from $I_2$ is obviously its complex conjugate and, from (\ref{4.36c}) and (\ref{4.36e}), we have for
$<{\bf E}^{{\bot }}({\bf x})>_{a}^{e_j(0,1)}$ the expression:
\begin{eqnarray}
&&<{\bf E}^{{\bot }}({\bf x})>_{a}^{e_j(0,1)}
=e_j\frac1{2\pi^2}\frac1{2i}2\frac1{\vert{\bf x}-{\bf q}_j\vert^2}I_1
\nonumber \\&&=[1-v_{j}^2]
\frac1{(1-v_{jy}^2)^{\frac32}}\frac1{\vert{\bf x}-{\bf q}_j\vert^2}{\bf e}_x.
\label{4.36if}
\end{eqnarray}

\section{Appendix D}

\def\theequation{D.\arabic{equation}}

\setcounter{equation}{0}

We evaluate first in this section the power dissipated by the radiative force $<{{\bf F}}^{(j)}.{\bf v}_j>_I$
(\ref{8.12}). The second contribution is treated afterwards.

We decompose the vector ${\bf k}^{[1]}$ into its component ${\bf k}_{\parallel}^{[1]}$ and
perpendicular ${\bf k}_{\bot}^{[1]}$ to the velocity vector ${\bf v}_{j}$.
The scalar product $({\bf l}.{\bf k}^{[1]})$ becomes the sum  $({\bf l}.{\bf k}_{\parallel}^{[1]}
+{\bf l}.{\bf k}_{\bot}^{[1]})$.
By symmetry, the last term will generate a vanishing contribution when integrated over
${\bf k}_{\bot}^{[1]}$. 
The remaining scalar product
$({\bf l}.{\bf k}_{\parallel}^{[1]})$ can be written as $p_{j}^{-2}({\bf l}.{\bf p}_{j}.)({\bf k}^{[1]}.{\bf p}_{j})$ and combined with the other contribution.
Since $\frac1{p_{j}^{2}}-\frac{1}{(m_j^2+p_j^2)}=\frac{m_j^2}{p_j^2(m_j^2+p_j^2)}$, we get
\begin{eqnarray}
&&<{{\bf F}}^{(j)}.{\bf v}_j>_I
=
-i\frac1{(2\pi)^3}\frac{e_j^3 e_{j'}}{4\pi}
\int d^3k^{[1]}\int d^3 l\,\frac{1}{l^2}
\sum_{a=\pm 1}a
e^{-i\frac{{\bf l}}{2}.({\bf q}_j-{\bf q}_{j'})}
\nonumber \\&&\times\frac{m_j^2}{p_j^2(m_j^2+p_j^2)^{\frac32}}
({\bf l}.{\bf p}_{j})({\bf p}_{j}.{\bf k}^{[1]})
\nonumber \\&&\times
\left(\frac{1}{{\bf k}^{[1]}.{\bf v}_{j}
-k^{[1]}}
\right)^2
\left(\frac{1}{i\epsilon+(\frac12{\bf l}+a{\bf k}^{[1]}).{\bf v}_{j}
-\frac12{\bf l}.{\bf v}_{j'} -ak^{[1]}}
\right),
\label{8.13}
\end{eqnarray}
\begin{eqnarray}
&&<{{\bf F}}^{(j)}.{\bf v}_j>_I
=
-i\frac1{(2\pi)^3}\frac{e_j^3 e_{j'}}{4\pi}\frac{m_j^2}{p_j^2(m_j^2+p_j^2)^{\frac32}}
\int d^3k^{[1]}\int d^3 l\,\frac{1}{l^2}
\nonumber \\&&\times
e^{-i\frac{{\bf l}}{2}.({\bf q}_j-{\bf q}_{j'})}
({\bf l}.{\bf p}_{j})({\bf p}_{j}.{\bf k}^{[1]})
\nonumber \\&&\times
\left(\frac{1}{{\bf k}^{[1]}.{\bf v}_{j}
-k^{[1]}}
\right)^2
\left[
\left(\frac{1}{i\epsilon+(\frac12{\bf l}+{\bf k}^{[1]}).{\bf v}_{j}
-\frac12{\bf l}.{\bf v}_{j'} -k^{[1]}}
\right)
\right.\nonumber \\&&-
\left.
\left(\frac{1}{i\epsilon+(\frac12{\bf l}-{\bf k}^{[1]}).{\bf v}_{j}
-\frac12{\bf l}.{\bf v}_{j'} +k^{[1]}}
\right)
\right].
\label{8.14}
\end{eqnarray}
We can consider a situation where the particle $j'$ is much more heavy that the $j$ particle.
In the referentiel in which the heavy particle is at rest at the origin of coordinates, we have:
\begin{eqnarray}
&&<{{\bf F}}^{(j)}.{\bf v}_j>_I
=
-i\frac1{(2\pi)^3}\frac{e_j^3 e_{j'}}{4\pi}\frac{m_j^2}{p_j^2(m_j^2+p_j^2)^{\frac32}}
\int d^3k^{[1]}\int d^3 l\,\frac{1}{l^2}
\nonumber \\&&\times
e^{-i\frac{{\bf l}}{2}.{\bf q}_j}
({\bf l}.{\bf p}_{j})({\bf p}_{j}.{\bf k}^{[1]})
\left(\frac{1}{{\bf k}^{[1]}.{\bf v}_{j}
-k^{[1]}}
\right)^2
\left[
\left(\frac{1}{i\epsilon+(\frac12{\bf l}+{\bf k}^{[1]}).{\bf v}_{j}-k^{[1]}}
\right)
\right.\nonumber \\&&-\left.
\left(\frac{1}{i\epsilon+(\frac12{\bf l}-{\bf k}^{[1]}).{\bf v}_{j}+k^{[1]}}
\right)
\right].
\label{8.15}
\end{eqnarray}
We consider first the case where the vectors ${\bf q}_j$ and ${\bf v}_{j}$ are orthogonal (the orbital
situation). 
We place the $x$ axis along ${\bf q}_j$ and the $y$ axis along ${\bf v}_{j}$.
We have:
\begin{eqnarray}
&&<{{\bf F}}^{(j)}.{\bf v}_j>_{Iorb}
=
-i\frac1{(2\pi)^3}\frac{e_j^3 e_{j'}}{4\pi}\frac{m_j^2}{p_j^2(m_j^2+p_j^2)^{\frac32}}
\nonumber \\&&\times
\int d^3k^{[1]}
\int_{-\infty}^{\infty}
\int_{-\infty}^{\infty}\int_{-\infty}^{\infty}dl_x\,dl_y\,dl_z\,\frac{1}{l_x^2+l_y^2+l_z^2}
\nonumber \\&&\times
e^{-i\frac{l_x}{2} q_j}
(l_y p_{j})(p_{j}k_y^{[1]})
\left(\frac{1}{k_y^{[1]}v_{j}
-k^{[1]}}
\right)^2
\left[
\left(\frac{1}{i\epsilon+(\frac12l_y+k_y^{[1]})v_{j}-k^{[1]}}
\right)
\right.\nonumber \\&&-
\left.
\left(\frac{1}{i\epsilon+(\frac12l_y-k_y^{[1]})v_{j}+k^{[1]}}
\right)
\right].
\label{8.16}
\end{eqnarray}
The integration  over $l_y$ can be performed by residue, closing the path in the upper plane $\Im
l_y>0$. Indeed, the integrand decreases at least as $l_y^{-3}$.
The only pole to be considered is $l_y=i\sqrt{l_x^2+l_z^2}$.
\begin{eqnarray}
&&<{{\bf F}}^{(j)}.{\bf v}_j>_{Iorb}
=
-i\frac1{(2\pi)^3}\frac{e_j^3 e_{j'}}{4\pi}\frac{m_j^2}{p_j^2(m_j^2+p_j^2)^{\frac32}}
\nonumber \\&&\times
\int d^3k^{[1]}
\int_{-\infty}^{\infty}\int_{-\infty}^{\infty}dl_x\,dl_z\,\frac{2\pi i}{2i\sqrt{l_x^2+l_z^2}}
e^{-i\frac{l_x}{2} q_j}
\nonumber \\&&\times
(i\sqrt{l_x^2+l_z^2} p_{j})(p_{j}k_y^{[1]})
\left(\frac{1}{k_y^{[1]}v_{j}
-k^{[1]}}
\right)^2
e^{-i\frac{l_x}{2} q_j}
\nonumber \\&&\times
\left[
\left(\frac{1}{i\epsilon+(\frac12i\sqrt{l_x^2+l_z^2}+k_y^{[1]})v_{j}-k^{[1]}}
\right)
\right.\nonumber \\&&-
\left.
\left(\frac{1}{i\epsilon+(\frac12i\sqrt{l_x^2+l_z^2}-k_y^{[1]})v_{j}+k^{[1]}}
\right)
\right].
\label{8.17}
\end{eqnarray}
The $i\epsilon$ can now be dropped since they have played their role in determining the relative
position of the poles in the complex plane. We introduce polar coordinates in the $l_x$, $l_y$ plane
to get
\begin{eqnarray}
&&<{{\bf F}}^{(j)}.{\bf v}_j>_{Iorb}
=
-i\frac1{(2\pi)^3}\frac{e_j^3 e_{j'}}{4\pi}\frac{m_j^2}{p_j^2(m_j^2+p_j^2)^{\frac32}}
\int d^3k^{[1]}
\int_{0}^{\infty}dl\,l\int_{0}^{2\pi}d\theta\,
\nonumber \\&&\times
\frac{2\pi i}{2il}
e^{-i\frac{l}{2} q_j\cos\theta}
(il p_{j})(p_{j}k_y^{[1]})
\left(\frac{1}{k_y^{[1]}v_{j}
-k^{[1]}}
\right)^2
\left[
\left(\frac{1}{(\frac12il+k_y^{[1]})v_{j}-k^{[1]}}
\right)
\right.\nonumber \\&&-
\left.
\left(\frac{1}{(\frac12il-k_y^{[1]})v_{j}+k^{[1]}}
\right)
\right].
\label{8.18}
\end{eqnarray}
Since $\left[
\left(\frac{1}{(\frac12il+k_y^{[1]})v_{j}-k^{[1]}}\right)-\left(\frac{1}{(\frac12il-k_y^{[1]})v_{j}+k^{[1]}}
\right)\right]=\frac{-2(k_y^{[1]}v_{j}-k^{[1]})}{(\frac12lv_{j})^2+(k_y^{[1]}v_{j}-k^{[1]})^2}$, we have
\begin{eqnarray}
&&<{{\bf F}}^{(j)}.{\bf v}_j>_{Iorb}
=
-\frac1{(2\pi)^3}\frac{e_j^3 e_{j'}}4\frac{m_j^2 (p_{j})^2}{p_j^2(m_j^2+p_j^2)^{\frac32}}
\int d^3k^{[1]}
\int_{0}^{\infty}dl\,l\int_{0}^{2\pi}d\theta\,
\nonumber \\&&\times
e^{-i\frac{l}{2} q_j\cos\theta}
k_y^{[1]}
\left(\frac{1}{k_y^{[1]}v_{j}
-k^{[1]}}
\right)^2
\left(\frac{2(k_y^{[1]}v_{j}-k^{[1]})}{(\frac12lv_{j})^2+(k_y^{[1]}v_{j}-k^{[1]})^2}
\right).
\label{8.19}
\end{eqnarray}
By definition, $\int_{0}^{2\pi}d\theta \cos{(y\cos\theta)}=2\pi J_0(y)$, $J_0$ being the Bessel
function. Therefore, 
\begin{eqnarray}
&&<{{\bf F}}^{(j)}.{\bf v}_j>_{Iorb}
=
-\frac{4\pi}{(2\pi)^3}\frac{e_j^3 e_{j'}}4\frac{m_j^2 (p_{j})^2}{p_j^2(m_j^2+p_j^2)^{\frac32}}
\int d^3k^{[1]}
\int_{0}^{\infty}dl\,l
\nonumber \\&&\times
J_0(\frac{l}{2}q_j)
k_y^{[1]}
\left(\frac{1}{k_y^{[1]}v_{j}
-k^{[1]}}
\right)
\left(\frac{1}{(\frac12lv_{j})^2+(k_y^{[1]}v_{j}-k^{[1]})^2}
\right).
\label{8.20}
\end{eqnarray}
From p.686 of \cite{GR65} we have  (formula 6.565.4):
\begin{equation}
\int_0^{\infty}\frac{J_\nu(bx)x^{\nu+1}}{(x^2+a^2)^{\mu+1}}dx=
\frac{a^{\nu-\mu}b^\mu}{2^\mu\Gamma(\mu+1)}K_{\nu-\mu}(ab),
\label{8.21}
\end{equation}
where $K_\nu(z)$ is a bessel function of imaginary argument ($-1<\Re \nu<\Re(2\mu+\frac32)$, $a>0$,
$b>0$).
We can apply that formula for $x=l$, with $\nu=0$, $\mu=0$, $b=\frac{1}{2}q_j$, 
$a^2=\frac{4(k_y^{[1]}v_{j}-k^{[1]})^2}{v_{j}^2}$.
The function $K_0(z)$ is represented in 8.432.1 by the integral ($\nu=0$):
$K_0(z)=\int_0^{\infty}e^{-z\cosh t}dt$.
The integral over $l$ can thus be performed:
\begin{eqnarray}
<{{\bf F}}^{(j)}.{\bf v}_j>_{Iorb}
&=&
-\frac{4\pi}{(2\pi)^3}\frac{e_j^3 e_{j'}}4\frac{m_j^2 (p_{j})^2}{p_j^2(m_j^2+p_j^2)^{\frac32}}
\frac4{v_j^2}
\int d^3k^{[1]}
\nonumber \\&&\times
k_y^{[1]}
\frac{1}{k_y^{[1]}v_{j}
-k^{[1]}}
K_0(\frac{q_j(k^{[1]}-k_y^{[1]}v_{j})}{v_{j}}).
\label{8.23}
\end{eqnarray}
We take $k_y^{[1]}=k^{[1]}\cos\theta$, $x=\cos\theta$, 
$\int d^3k^{[1]}\dots=\int_0^{\infty} dk^{[1]}
\,(k^{[1]})^2\int_{-1}^{+1}dx\,\int_{0}^{2\pi}d\phi\dots$
\begin{eqnarray}
&&<{{\bf F}}^{(j)}.{\bf v}_j>_{Iorb}
=
-\frac{4\pi}{(2\pi)^3}\frac{e_j^3 e_{j'}}4\frac{m_j^2 (p_{j})^2}{p_j^2(m_j^2+p_j^2)^{\frac32}}
\frac{8\pi}{v_j^2}
\int_0^{\infty} dk^{[1]} \,(k^{[1]})^2\int_{-1}^{+1}dx\,
\nonumber \\&&\times
\frac{x}{xv_{j}-1}
K_0(\frac{q_jk^{[1]}(1-xv_{j})}{v_{j}}).
\label{8.24}
\end{eqnarray}
The formula 6.561.16 p. 684 of \cite{GR65}  is:
\begin{equation}
\int_0^{\infty} x^\mu K_\nu(ax)dx=2^{\mu-1}a^{-\mu-1}\Gamma(\frac{1+\mu+\nu}{2})
\Gamma(\frac{1+\mu-\nu}{2}),
\label{8.25}
\end{equation}
with $\Re(\mu+1\pm \nu)>0$, $\Re a>0$.
That formula (\ref{8.25}) can be applied for $x=k^{[1]}$, with $\mu=2$, $\nu=0$,
$a=\frac{q_j(1-xv_{j})}{v_{j}}$.
\begin{eqnarray}
&&<{{\bf F}}^{(j)}.{\bf v}_j>_{Iorb}
=
-\frac{8}{\pi}
\left(\Gamma(\frac{3}{2})\right)^2
\frac{e_j^3 e_{j'}}4\frac{m_j^2}{(m_j^2+p_j^2)^{\frac32}}
\frac{1}{v_j^2}
\nonumber \\&&\times
\int_{-1}^{+1}dx\,
\frac{x}{xv_{j}-1}
\left(\frac{v_{j}}{q_j(1-xv_{j})}\right)^{3},
\label{8.26}
\end{eqnarray}
\begin{equation}
<{{\bf F}}^{(j)}.{\bf v}_j>_{Iorb}
=
\frac{8}{\pi}
\left(\Gamma(\frac{3}{2})\right)^2
\frac{e_j^3 e_{j'}}4\frac{m_j^2}{(m_j^2+p_j^2)^{\frac32}}
\frac{v_j}{q_j^3}
\int_{-1}^{+1}dx\,
\frac{x}{(1-xv_{j})^{4}}.
\label{8.27}
\end{equation}
The last integral is direct and leads to:
\begin{equation}
<{{\bf F}}^{(j)}.{\bf v}_j>_{Iorb}
=
\frac43
e_j^3 e_{j'}\frac{m_j^2}{(m_j^2+p_j^2)^{\frac32}}
\frac{v_j^2}{q_j^3}
\frac{1}{(1-v_j^2)^{3}}.
\label{8.31aaaa}
\end{equation}
In the other geometries, some integrals are not known explicitly but can be shown to be more
convergent that the orbital case that provides a finite result.  

We now turn to the second contribution.
In place of (\ref{8.16}), we have now (by a change of variables, ${\bf l}$ in this expression is  of
$\frac12{\bf l}$ in the I contribution):
\begin{eqnarray}
&&
<{\bf F}^{{\bot }}({\bf q}_j).{\bf v}_j>_{IIorb}
=
e^3_je_{j'}\frac{1}{\pi}\frac1{(2\pi)^3}
\int d^3k^{[1]}
\left(\frac{1}{k_y^{[1]}v_j-k^{[1]}}
\right)
\nonumber\\&&\times
\left\{\frac{1}{(m_j^2+(p^{(j)})^2)^{\frac12}}
\left[
v_{j}- \frac{(k_y^{([1]})(v_{j}
k_y^{([1]})}{(k^{([1]})^2}
\right]
\right.\nonumber \\&&\left.
-\frac{p_j}{(m_j^2+(p^{(j)})^2)^{\frac32}}
\left[
p_jv_j
- \frac{(p_j k_y^{([1]})(v_{j}k_y^{([1]})}{(k^{([1]})^2}
\right]
\right\} 
\nonumber\\& &\times
\int_{-\infty}^{\infty}
\int_{-\infty}^{\infty}\int_{-\infty}^{\infty}dl_x\,dl_y\,dl_z\,\frac{1}{l_x^2+l_y^2+l_z^2}
e^{-il_x q_j}l_y
\nonumber\\&& \times
\left(\frac{1}{i\epsilon+(l_y+k_y^{[1]})v_j -k^{[1]}}
-\frac{1}{i\epsilon+(l_y-k_y^{[1]})v_j +k^{[1]}}
\right).
\label{4.83}
\end{eqnarray}
The integration  over $l_y$ can be performed by residue, closing the path in the upper plane $\Im
l_y>0$. Indeed, the integrand decreases at least as $l_y^{-3}$.
The only pole to be considered is $l_y=i\sqrt{l_x^2+l_z^2}$.
Introduce polar coordinates in the $l_x$, $l_y$ plane, we get
\begin{eqnarray}
&&
<{\bf F}^{{\bot }}({\bf q}_j).{\bf v}_j>_{IIorb}=
e^3_je_{j'}\frac1{(2\pi)^3}
\frac{v_{j}m_j^2}{(m_j^2+(p^{(j)})^2)^{\frac32}}
\int d^3k^{[1]}
\left(\frac{1}{k_y^{[1]}v_j-k^{[1]}}
\right)
\nonumber\\&&\times
\left(1-\frac{(k_y^{([1]})^2}{(k^{([1]})^2}\right)
\int_{0}^{\infty}dl\,l\int_{0}^{2\pi}d\theta\,
e^{-il_x q_j}
\nonumber\\&& \times
\left[
\left(\frac{1}{(il+k_y^{[1]})v_{j}-k^{[1]}}
\right)
-\left(\frac{1}{(il-k_y^{[1]})v_{j}+k^{[1]}}
\right)
\right].
\label{4.85}
\end{eqnarray}
Since $\left[
\left(\frac{1}{(il+k_y^{[1]})v_{j}-k^{[1]}}\right)-\left(\frac{1}{(il-k_y^{[1]})v_{j}+k^{[1]}}
\right)\right]=\frac{-2(k_y^{[1]}v_{j}-k^{[1]})}{(lv_{j})^2+(k_y^{[1]}v_{j}-k^{[1]})^2}$, 
and identifying  the $J_0$ Bessel function in $\int_{0}^{2\pi}d\theta \cos{(y\cos\theta)}=2\pi J_0(y)$,
we get:
\begin{eqnarray}
&&
<{\bf F}^{{\bot }}({\bf q}_j).{\bf v}_j>_{IIorb}=
-e^3_je_{j'}\frac1{(2\pi)^2}
\frac{v_{j}m_j^2}{(m_j^2+(p^{(j)})^2)^{\frac32}}
\int d^3k^{[1]}
\nonumber\\&&\times
\left(1-\frac{(k_y^{([1]})^2}{(k^{([1]})^2}\right)
\int_{0}^{\infty}dl\,l
J_0(lq_j)
\frac{2}{(lv_{j})^2+(k_y^{[1]}v_{j}-k^{[1]})^2}.
\label{4.88}
\end{eqnarray}
Using (\ref{8.21}) and  formula 6.561.16 p. 684 of \cite{GR65} leads then to:
\begin{eqnarray}
<{\bf F}^{{\bot }}({\bf q}_j).{\bf v}_j>_{IIorb}&=&
-e^3_je_{j'}\frac2{\pi}\left(\Gamma(\frac{3}{2})\right)^2
\frac{v_{j}^2m_j^2}{(m_j^2+(p^{(j)})^2)^{\frac32}}
\frac1{q_j^3}
\nonumber\\&&\times
\int_{-1}^{+1}dx\,
(1-x^2)
\left(\frac{1}{(1-xv_{j})}\right)^{3}.
\label{4.95}
\end{eqnarray}
The last integral can be performed to provide the result (\ref{8.31a}) of the main text.

\section{{Appendix E. The simple cycle}}

\def\theequation{E.\arabic{equation}}

\setcounter{equation}{0}

\subsection{Insertion of a cycle}
The vector ${\bf k}^{[1]}$ in (\ref{a2.15}). is replaced by ${\bf k}$ with a corresponding change in the
other variables associated with the field mode.
To avoid ambiguity, $\eta_\beta$ are replaced by $\epsilon_\beta$ that has the same use of
deferring a vanishing limit. We have to act on the field vacuum to use the previous result:
\begin{eqnarray}
&&<11[1(s_j)]|\tilde \Sigma(t)|11[1(f)]>^{(1,3)}_{d1}\tilde f^V_{[1(f)]}
\nonumber\\&&=
\frac{-1}{2\pi i} \int'_c dz\,e^{-izt}
\left(\frac{1}{z-{{\bf k}}^{(j)}.{{\bf v}}^{(j)}-{\bf k}^{(j')}.{\bf v}^{(j')}
+k(m_{\alpha}+m_{\alpha'})}
\right)
\nonumber\\&& \times
 e_j^2 \frac1{(2\pi)^3}
\int d^3k^{[1]}
\sum_{\alpha{[1]}=1,2} 
\sum_{a=\pm 1}
\nonumber\\&& \times
\frac{1}
{k^{[1]}}
\left[ 
[k^{[1]} e_r^{(\alpha)[1]}-g^{st} v_s^{(j)}
(e_t^{(\alpha)[1]} k_r^{[1]}-e_r^{(\alpha)[1]} k_t^{[1]})]
\right]
\nonumber\\&& \times
\left(\frac{1}{z-{{\bf k}}^{(j)}.{{\bf v}}^{(j)}+a{\bf k}^{[1]}.{{\bf v}}^{(j)}-{\bf k}^{(j')}.{\bf v}^{(j')}
+k^{[1]}(-a)+k(m_{\alpha}+m_{\alpha'})}
\right)
\nonumber\\&& \times
\left[ 
-2\pi\left(\frac{\partial }
{\partial p_r^{(j)}}({{\bf v}}^{(j)}.{\bf e}^{(\alpha)[1]})\right)
\right]
\nonumber\\&& \times
\left(\frac{1}{z-{{\bf k}}^{(j)}.{{\bf v}}^{(j)}-{\bf k}^{(j')}.{\bf v}^{(j')}
+k(m_{\alpha}+m_{\alpha'})}
\right)
\nonumber\\&& \times
(-i)e_je_{j'}\frac{-1}{2\pi^2}\int d^3 l\frac{1}{l^2}
e^{{\bf l}.\left(\frac{\partial}{\partial {{\bf k}}^{(j)}}-\frac{\partial}
{\partial {\bf k}^{(j')}}
\right)}
\nonumber\\&& \times
\left(\frac{1}{z-{{\bf k}}^{(j)}.{{\bf v}}^{(j)}-{\bf k}^{(j')}.{\bf v}^{(j')}+k(m_{\alpha}+m_{\alpha'})}
\right)
(-i)e_j \frac1{(2\pi)^{\frac32}} 
\sum_{\beta=1,2} 
\sum_{b=\pm 1}
\nonumber\\&& \times
\left(
\frac{\eta_{\beta}}
{k}
\right)^{\frac12} 
\left[ 
-\left({\bf l}.\frac{\partial}{\partial {\bf p}^{(j)}}\pi({{\bf v}}^{(j)}.{\bf e}^{(\beta)})\right)
2 \frac{\partial}
{\partial \eta_{\beta}}
\right]
\exp b\left\{-{\bf k}.\frac{\partial}{\partial {{\bf k}}^{(j)}}-\frac{\partial}{\partial
m_{\beta}}\right\}
\nonumber\\&& \times
\left(\frac{1}{z-{{\bf k}}^{(j)}.{{\bf v}}^{(j)}-{\bf k}^{(j')}.{\bf v}^{(j')}
+k(m_{\beta}+m_{\beta'})}
\right)
\nonumber\\&& \times
\delta(\eta_{\beta}-\epsilon_\beta)\delta(\eta_{\beta'}-\epsilon_{\beta'})
\delta^{Kr}_{m_{\beta},0}\delta^{Kr}_{m_{\beta'},0}.
\label{a2.20}
\end{eqnarray}
Usual manipulations lead to:
\begin{eqnarray}
&&<11[1(s_j)]|\tilde \Sigma(t)|11[1(f)]>^{(1,3)}_{d1}\tilde f^V_{[1(f)]}
\nonumber\\&&=
(-i)e_j \frac1{(2\pi)^{\frac32}} 
\sum_{\beta=1,2} 
\sum_{b=\pm 1}
\left(
\frac{\eta_{\beta}}
{k}
\right)^{\frac12}
\exp b\left\{-{\bf k}.\frac{\partial}{\partial {{\bf k}}^{(j)}}-\frac{\partial}{\partial
m_{\beta}}\right\}
\nonumber\\&& \times
\frac{-1}{2\pi i} \int'_c dz\,e^{-izt}
\left(\frac{1}{z-({{\bf k}}^{(j)}+b{\bf k}).{{\bf v}}^{(j)}-{\bf k}^{(j')}.{\bf v}^{(j')}+bk}
\right)^2
\nonumber\\&& \times
 e_j^2 \frac1{(2\pi)^3}
\int d^3k^{[1]}
\sum_{\alpha{[1]}=1,2} 
\sum_{a=\pm 1}
\frac{1}
{k^{[1]}}
\nonumber\\&& \times
\left[ 
[k^{[1]} e_r^{(\alpha)[1]}-g^{st} v_s^{(j)}
(e_t^{(\alpha)[1]} k_r^{[1]}-e_r^{(\alpha)[1]} k_t^{[1]})]
\right]
\nonumber\\&& \times
\left(\frac{1}{z-({{\bf k}}^{(j)}+b{\bf k}).{{\bf v}}^{(j)}+bk+a{\bf k}^{[1]}.{{\bf v}}^{(j)}-{\bf k}^{(j')}.{\bf v}^{(j')} +k^{[1]}(-a)}
\right)
\nonumber\\&& \times
\left[ 
-2\pi\left(\frac{\partial }
{\partial p_r^{(j)}}({{\bf v}}^{(j)}.{\bf e}^{(\alpha)[1]})\right)
\right]
(-i)e_je_{j'}\frac{-1}{2\pi^2}\int d^3 l\frac{1}{l^2}
\nonumber\\&& \times
\left(\frac{1}{z-({{\bf k}}^{(j)}+{\bf l}+b{\bf k}).{{\bf v}}^{(j)}-({\bf k}^{(j')}-{\bf l}).{\bf v}^{(j')}+bk}
\right)
\nonumber\\&& \times
\left[ 
-\left({\bf l}.\frac{\partial}{\partial {\bf p}^{(j)}}\pi({{\bf v}}^{(j)}.{\bf e}^{(\beta)})\right)
2 \frac{\partial}
{\partial \eta_{\beta}}
\right]
\left(\frac{1}{z-({{\bf k}}^{(j)}+{\bf l}).{{\bf v}}^{(j)}-({\bf k}^{(j')}-{\bf l}).{\bf v}^{(j')}}
\right)
\nonumber\\&& \times
e^{{\bf l}.\left(\frac{\partial}{\partial {{\bf k}}^{(j)}}-\frac{\partial}
{\partial {\bf k}^{(j')}}
\right)}
\delta(\eta_{\beta}-\epsilon_\beta)\delta(\eta_{\beta'}-\epsilon_{\beta'})
\delta^{Kr}_{m_{\beta},0}\delta^{Kr}_{m_{\beta'},0}.
\label{a2.21}
\end{eqnarray}
From the definition of the cycle, we have
\begin{eqnarray}
&&<11[1(s_j)]|\tilde \Sigma(t)|11[1(f)]>^{(1,3)}_{d1}\tilde f^V_{[1(f)]}
\nonumber\\&&=
(-i)e_j \frac1{(2\pi)^{\frac32}} 
\sum_{\beta=1,2} 
\sum_{b=\pm 1}
\left(
\frac{\eta_{\beta}}
{k}
\right)^{\frac12}
\exp b\left\{-{\bf k}.\frac{\partial}{\partial {{\bf k}}^{(j)}}-\frac{\partial}{\partial
m_{\beta}}\right\}
\nonumber\\&& \times
\frac{-1}{2\pi i} \int'_c dz\,e^{-izt}
\left(\frac{1}{z-({{\bf k}}^{(j)}+b{\bf k}).{{\bf v}}^{(j)}-{\bf k}^{(j')}.{\bf v}^{(j')}+bk}
\right)^2
\nonumber\\&& \times
C(z-({{\bf k}}^{(j)}+b{\bf k}).{{\bf v}}^{(j)}-{\bf k}^{(j')}.{\bf v}^{(j')}+bk,{{\bf v}}^{(j)})
\nonumber\\&& \times
(-i)e_je_{j'}\frac{-1}{2\pi^2}\int d^3 l\frac{1}{l^2}
\left(\frac{1}{z-({{\bf k}}^{(j)}+{\bf l}+b{\bf k}).{{\bf v}}^{(j)}-({\bf k}^{(j')}-{\bf l}).{\bf v}^{(j')}+bk}
\right)
\nonumber\\&& \times
\left[ 
-\left({\bf l}.\frac{\partial}{\partial {\bf p}^{(j)}}\pi({{\bf v}}^{(j)}.{\bf e}^{(\beta)})\right)
2 \frac{\partial}
{\partial \eta_{\beta}}
\right]
\nonumber\\&& \times
\left(\frac{1}{z-({{\bf k}}^{(j)}+{\bf l}).{{\bf v}}^{(j)}-({\bf k}^{(j')}-{\bf l}).{\bf v}^{(j')}}
\right)
\nonumber\\&& \times
e^{{\bf l}.\left(\frac{\partial}{\partial {{\bf k}}^{(j)}}-\frac{\partial}
{\partial {\bf k}^{(j')}}
\right)}
\delta(\eta_{\beta}^{[1]}-\eta_\beta)\delta(\eta_{\beta'}^{[1]}-\eta_{\beta'})
\delta^{Kr}_{m_{\beta}^{[1]},0}\delta^{Kr}_{m_{\beta'}^{[1]},0}.
\label{a2.23}
\end{eqnarray}

\subsection{Evaluation of the simple cycle}

For $\Im z>0$, and $\Re z=y$, we have 
\begin{eqnarray}
&& \Im C(y+i\epsilon,0)
=
\frac{16\pi^3}{m_j} e_j^2 \frac1{(2\pi)^3}
\int_0^{\infty} dk\,k^2
\left[\delta(y-k)+\delta(y+k)
\right]
\frac{K_c^2}{k^2+K_c^2}
\nonumber\\&&=
\frac{2e_j^2}{m_j} 
\frac{K_c^2y^2}{y^2+K_c^2}.
\label{a2.31}
\end{eqnarray}
Let us note that
the sign of $\Im C(y+i\epsilon,0)$ is  unusually positive  and  $\Im C(y+i\epsilon,0)$ vanishes at
$y=0$, in accordance with the solution $z=0$ of $z-C(z)=0$.
For very large $y$ ($y>>K_c$), it behaves as a constant $\frac{2e_j^2K_c^2}{m_j} $.
Since the imaginary part of $C(z)$ vanishes for $z=0$, in the denominators in (\ref{a4.73a}) and (\ref{a2.26}), that property does not involve a new  $i\epsilon$ rule in the computation of the
subdynamics operator.
For the real part of $C(y+i\epsilon,0)$, involving the principal part, we have
\begin{eqnarray}
 &&\Re C(y+i\epsilon,0)
\nonumber\\&& =
\frac{16\pi^3}{m_j} e_j^2 \frac1{(2\pi)^3}
\int_0^{\infty} dk\,k^2\frac{K_c^2}{k^2+K_c^2}
\left[{\cal P}(\frac1{(y-k)})+{\cal P}(\frac1{(y+k)})
\right]
\nonumber\\&&=
\frac{16\pi^3}{m_j} e_j^2 \frac1{(2\pi)^3}(I_1+I_2),
\label{a2.32}
\end{eqnarray}
\begin{eqnarray}
&&I_1
=
\lim_{R\to \infty}\int_0^{R} dk
\left[{\cal P}(\frac1{(y-k)})+{\cal P}(\frac1{(y+k)})
\right]
\nonumber\\&&=
\lim_{R\to \infty}[-\ln(R-y)+\ln(\vert y\vert)+\ln(R+y)-\ln(\vert y\vert)]=0,
\label{a2.33}
\end{eqnarray}
\begin{eqnarray}
&&I_2
=
\int_0^{\infty} dk\,K_c^2
\left[{\cal P}(\frac1{(y-k)})+{\cal P}(\frac1{(y+k)})
\right]
\frac{K_c^2}{k^2+K_c^2}
\nonumber\\&&=
\Re\int_0^{\infty} dk\,K_c^2
\left[(\frac1{(i\epsilon+y-k)}+(\frac1{(i\epsilon+y+k)})
\right]
\frac{K_c^2}{k^2+K_c^2}
\nonumber\\&&=
\Re\int_{-\infty}^{\infty} dk\,
\frac1{(i\epsilon+y-k)}
\frac{K_c^4}{k^2+K_c^2}
= \pi K_c^3 \frac y{(y^2+K_c^2)}.
\label{a2.34}
\end{eqnarray}
$I_2$ vanishes linearly for $y\to 0$ and behaves as $\pi K_c^3 \frac 1y$ for large $y$.

\section{{Appendix F. Self-consistency analysis}}

\def\theequation{F.\arabic{equation}}

\setcounter{equation}{0}

\subsection{Consistency check on $\Delta(y,\gamma)$}

From the expression (\ref{a3.7}), we have:
\begin{eqnarray}
&&\Delta(y,\gamma)=c(y,\gamma)-c(-y,\gamma)
\nonumber\\&&=
-\frac{1}{y} 
\int_0^\infty du\, u^2
\frac{\gamma^2}{\gamma^2+u^2}
\nonumber\\&&\times
\left[
\left(\frac{1}{(i\epsilon +y-u)[1-c(y-u,\gamma)]}
\right)
+\left(\frac{1}{(i\epsilon +y+u)[1-c(y+u,\gamma)]}
\right)
\right.\nonumber\\&&\left.
+\left(\frac{1}{(i\epsilon -y-u)[1-c(-y-u,\gamma)]}
\right)
+\left(\frac{1}{(i\epsilon -y+u)[1-c(-y+u,\gamma)]}
\right)
\right] .
\nonumber\\&&
\label{a3.9}
\end{eqnarray}
As for (2.30), we separates the contributions arising from the $\delta$ functions from the contribution
arsing from the principal parts.
That separation does no longer correspond into a separation between a real and an imaginary part.
The contribution $\Delta_{\delta}(y,\gamma)$ from the $\delta$ functions  is:
\begin{eqnarray}
&&\Delta_{\delta}(y,\gamma)=
-\frac{1}{y} (-\pi i)
\int_0^\infty du\, u^2
\frac{\gamma^2}{\gamma^2+u^2}
\nonumber\\&&\times
\left\{\delta(y-u)
\left[\frac1{[1-c(y-u,\gamma)]}+\frac1{[1-c(-y+u,\gamma)]}
\right]
\right.\nonumber\\&&\left.
+\delta(y+u)
\left[\frac1{[1-c(y+u,\gamma)]}+\frac1{[1-c(-y-u,\gamma)]}
\right]
\right\} 
\nonumber\\&&=
\frac{1}{y} (2\pi i)
\frac1{[1-c(0,\gamma)]}
y^2\frac{\gamma^2}{\gamma^2+y^2}.
\label{a3.9a}
\end{eqnarray}
$\Delta_{\delta}(y,\gamma)$ contributes to the second term of (\ref{a3.8}) since it decreases as
$\frac1y$.
The parameter $s$ is thus related to the behaviour in $\gamma$ of
$\frac1{[1-c(0,\gamma)]}$.

The contribution  $\Delta_{\cal P}(y,\gamma)$ from the principal parts is:
\begin{eqnarray}
&&\Delta_{\cal P}(y,\gamma)
=
-\frac{1}{y} 
\int_0^\infty du\, u^2
\frac{\gamma^2}{\gamma^2+u^2}
\nonumber\\&&\times
\left\{{\cal P}\left(\frac{1}{(y-u)}\right)
\left[\frac1{[1-c(y-u,\gamma)]}-\frac1{[1-c(-y+u,\gamma)]}
\right]
\right.\nonumber\\&&\left.
+{\cal P}\left(\frac{1}{(y+u)}\right)
\left[\frac1{[1-c(y+u,\gamma)]}-\frac1{[1-c(-y-u,\gamma)]}
\right]
\right\}. 
\label{a3.10}
\end{eqnarray}
To determine the behaviour for $y>>\gamma$, it is too simple to say that we have two $y$ factors in
the denominators in that expression and thus a behaviour as $\frac1{y^2}$. We have indeed to make
sure that the convergence of the integral is not affected by the limit.
Obviously, if we neglect $u$ in front of $y$ in that expression, the remaining integral over $u$
diverges linearly.
We analyse therefore the behaviour of the integrand $I(u)$.
For the first contribution, the form (\ref{a3.8}) cannot be used since the argument $y-u$ of
$c(y-u,\gamma)$ is not large for all values of $u$ inside the domain of integration.
For that term $\Delta_{{\cal P} 1}(y,\gamma)$, we separate the domains $0<u<2y$ and
$u>2y$  in the conditions $y>>\gamma>>1$. 
In the first domain, the domain of integration over $u$ is finite and we have:
\begin{eqnarray}
&&\Delta_{{\cal P} 1a}(y,\gamma)
=
-\frac{1}{y} 
\int_0^{2y} du\, u^2
\frac{\gamma^2}{\gamma^2+u^2}
\nonumber\\&&\times
{\cal P}\left(\frac{1}{y-u}\right)
\left[\frac1{[1-c(y-u,\gamma)]}-\frac1{[1-c(-y+u,\gamma)]}
\right],
\label{a3.10b}
\end{eqnarray}
\begin{eqnarray}
&&\Delta_{{\cal P} 1a}(y,\gamma)
=
-\frac{1}{y} 
\int_0^{2y} du\, \gamma^2
{\cal P}\left(\frac{1}{y-u}\right)
\nonumber\\&&\times
\left[\frac1{[1-c(y-u,\gamma)]}-\frac1{[1-c(-y+u,\gamma)]}
\right]
\nonumber\\&&
+\frac{1}{y} 
\int_0^{2y} du\, 
\frac{\gamma^4}{\gamma^2+u^2}
{\cal P}\left(\frac{1}{y-u}\right)
\nonumber\\&&\times
\left[\frac1{[1-c(y-u,\gamma)]}-\frac1{[1-c(-y+u,\gamma)]}
\right].
\label{a3.10c}
\end{eqnarray}
The first term $\Delta_{{\cal P} 1aa}(y,\gamma)$ can be written as:
\begin{equation}
\Delta_{{\cal P} 1aa}(y,\gamma)
=
\frac{\gamma^2}{y} 
\int_{-y}^{y} dv\,
{\cal P}\left(\frac{1}{v}\right)
\left[\frac1{[1-c(-v,\gamma)]}-\frac1{[1-c(v,\gamma)]}
\right],
\label{a3.10ca}
\end{equation}
where the principal part symbol can be dropped.
For large $y$, we can replace $\pm y$ as limit of integration by $\pm\infty$,
providing the convergence of the integral.
For large $v$, 
\begin{eqnarray}
&&\left[\frac1{[1-c(-v,\gamma)]}-\frac1{[1-c(v,\gamma)]}
\right] =
\left[\frac1{[1-\alpha \gamma^r+\beta \frac{\gamma^s}{v}]}
-\frac1{[1-\alpha \gamma^r-\beta \frac{\gamma^s}{v}]}
\right]
\nonumber\\&&=
\left[\frac{v}{[v(1-\alpha \gamma^r)+\beta \gamma^s]}
-\frac{v}{[v(1-\alpha \gamma^r)-\beta \gamma^s]}
\right]
\nonumber\\&&=
-2\beta \gamma^s\frac{v}{[v^2(1-\alpha \gamma^r)^2+\beta^2 \gamma^{2s}]}.
\label{a3.10cc}
\end{eqnarray}
That factor behaves as $\frac1v$ and the integrand in (\ref{a3.10ca}) behaves as $\frac1{v^2}$,
ensuring the convergence.
$\Delta_{{\cal P} 1aa}(y,\gamma)$ contributes also to the second term of (\ref{a3.8}).

The second term $\Delta_{{\cal P} 1ab}(y,\gamma)$ of (\ref{a3.10c}) behaves as least as $\frac1{y^2}$
since no convergence problem can arise. 

For the second contribution $\Delta_{{\cal P} 2}(y,\gamma)$ of (\ref{a3.10}), the asymptotic behaviour (\ref{a3.8}) can be used in all the integration domain and we have, keeping  the two terms:
\begin{eqnarray}
&&\Delta_{{\cal P} 2}(y,\gamma)
=
-\frac{1}{y} 
\int_0^{\infty} du\,u^2
\frac{\gamma^2}{\gamma^2+u^2}
{\cal P}\left(\frac{1}{y+u}\right)
\nonumber\\&&\times
\left[\frac1{[1-\alpha \gamma^r-\beta \frac{\gamma^s}{y+u}]}
-\frac1{[1-\alpha \gamma^r-\beta \frac{\gamma^s}{-y-u}]}
\right].
\label{a3.10d}
\end{eqnarray} 
Elementary manipulations lead to:
\begin{equation}
\Delta_{{\cal P} 2}(y,\gamma)
=
\frac{1}{y} 
\int_0^{\infty} du\,u^2
\frac{\gamma^2}{\gamma^2+u^2}
\frac{-2\beta\gamma^s}{[(y+u)^2(1-\alpha \gamma^r)^2-\beta^2 \gamma^{2s}]}.
\label{a3.10h}
\end{equation} 
$\Delta_{{\cal P} 2}(y,\gamma)$ can be split as (cf. (\ref{a3.10c})):
\begin{eqnarray}
&&\Delta_{{\cal P} 2}(y,\gamma)
=
\frac{\gamma^2}{y} 
\int_0^{\infty} du\,
\frac{-2\beta\gamma^s}{[(y+u)^2(1-\alpha \gamma^r)^2-\beta^2 \gamma^{2s}]}
\nonumber\\&&
-\frac{\gamma^4}{y} 
\int_0^{\infty} du\,
\frac{1}{\gamma^2+u^2}
\frac{-2\beta\gamma^s}{[(y+u)^2(1-\alpha \gamma^r)^2-\beta^2 \gamma^{2s}]}.
\label{a3.10i}
\end{eqnarray} 
In the first contribution, the integral converges and behaves as $\frac1y$, providing a global
behaviour at least as $\frac1{y^2}$.
In the second term, the factor $\frac{1}{\gamma^2+u^2}$ provides an effective cut to the values of $u$
that contributes in the integral.
We can in the integrand neglect $u$ with respect to $y$ and the global behaviour is in $\frac1{y^3}$.
Therefore, the form (\ref{a3.8}) is compatible with our analysis of $\Delta(y,\gamma)$.

\subsection{First consistency check on $c(y,\gamma)$}

We now turn to the analysis of $c(y,\gamma)$ itself.
We can choose a positive sign to $y$ to fix the ideas since $\Delta(y,\gamma)$ can then provide the
behaviour for negative $y$.
We start with (cf. (\ref{a3.9}))
\begin{eqnarray}
&&c(y,\gamma)
=
-\frac{1}{y} 
\int_0^\infty du\, u^2
\frac{\gamma^2}{\gamma^2+u^2}
\left[
\frac{1}{(i\epsilon +y-u)[1-c(y-u,\gamma)]}
\right.\nonumber\\&&\left.
+\frac{1}{(i\epsilon +y+u)[1-c(y+u,\gamma)]}
\right].
\label{a3.11}
\end{eqnarray}
For $y>0$, we have
\begin{eqnarray}
&&c_{\delta}(y,\gamma)=
-\frac{1}{y} (-\pi i)
\int_0^\infty du\, u^2
\delta(y-u)
\frac1{[1-c(y-u,\gamma)]}
\frac{\gamma^2}{\gamma^2+u^2}
\nonumber\\&&=
\frac{1}{y} (\pi i)
\frac1{[1-c(0,\gamma)]}
y^2\frac{\gamma^2}{\gamma^2+y^2}.
\label{a3.11a}
\end{eqnarray}
The contribution  $c_{\cal P}(y,\gamma)$ from the principal parts is:
\begin{eqnarray}
&&c_{\cal P}(y,\gamma)
=
-\frac{1}{y} 
\int_0^\infty du\, u^2
\frac{\gamma^2}{\gamma^2+u^2}
\left\{{\cal P}\left(\frac{1}{y-u}\right)
\frac1{[1-c(y-u,\gamma)]}
\right.\nonumber\\&&\left.
+{\cal P}\left(\frac{1}{y+u}\right)
\frac1{[1-c(y+u,\gamma)]}
\right\}. 
\label{a3.12}
\end{eqnarray}
For the first contribution to $c_{\cal P}(y,\gamma)$, the form (\ref{a3.8}) cannot be used since the
argument $y-u$ of $c(y-u,\gamma)$ is not large for all values of $u$ inside the domain of
integration. 
That integral cannot be written as a sum of integrals involving different principal
terms since only the global integrand in (\ref{a3.12}) decreases enough to ensure the convergence
at infinity. The following decomposition still holds (\ref{a3.10c}):
\begin{eqnarray}
&&c_{\cal P}(y,\gamma)
=
-\frac{\gamma^2}{y} 
\int_0^{\infty} du\, 
\left[{\cal P}\left(\frac{1}{y-u}\right)
\frac1{[1-c(y-u,\gamma)]}
\right.\nonumber\\&&\left.
+{\cal P}\left(\frac{1}{y+u}\right)
\frac1{[1-c(y+u,\gamma)]}
\right]
\nonumber\\&&
+\frac{\gamma^4}{y} 
\int_0^{\infty} du\, 
\frac{1}{\gamma^2+u^2}
\left[{\cal P}\left(\frac{1}{y-u}\right)
\frac1{[1-c(y-u,\gamma)]}
\right.\nonumber\\&&\left.
+{\cal P}\left(\frac{1}{y+u}\right)
\frac1{[1-c(y+u,\gamma)]}
\right].
\label{a3.12a}
\end{eqnarray}
For the second term, the factor $\frac{1}{\gamma^2+u^2}$ ensures the convergence and cut the
integral for values of $u$ limited by $\gamma$.
For that part, the dominant behaviour in $y$ is obtained by neglecting $u$ in front of $y$.
For the first term in (\ref{a3.12a}), the convergence in $u$ is ensured by the other factors.
For studying the convergence, we take $u$ much larger that $y$ and we can replace the $c$ by their
value (\ref{a3.8}) and the integrand $I_1(u,y)$ is
\begin{equation}
I_1(u,y)
=
-\frac{\gamma^2}{y} 
\left[\frac{1}{y-u}
\frac1{[1-\alpha \gamma^r-\beta \frac{\gamma^s}{y-u}]}
+\frac{1}{y+u}
\frac1{[1-\alpha \gamma^r-\beta \frac{\gamma^s}{y+u}]}
\right],
\label{a3.13}
\end{equation}
\begin{equation}
I_1(u,y)
=
-\frac{\gamma^2}{y} 
\left[
\frac1{(y-u)(1-\alpha \gamma^r)-\beta \gamma^s}
+
\frac1{(y+u)(1-\alpha \gamma^r)-\beta \gamma^s}
\right],
\label{a3.13a}
\end{equation}
\begin{equation}
I_1(u,y)
=
\frac{\gamma^2}{y} 
2[y(1-\alpha \gamma^r)-\beta \gamma^s]
\frac1{u^2(1-\alpha \gamma^r)^2-[y(1-\alpha \gamma^r)-\beta \gamma^s]^2}.
\label{a3.13b}
\end{equation}
Therefore, the factor that ensures the convergence of the integrand in $u$ contains as a factor $y$.
The integral for $c_{{\cal P} }(y,\gamma)$ contains therefore a contribution independent of $y$ for
large $y$. It corresponds  to (and determines) the first term of (\ref{a3.8}).
The $\gamma$ dependence factor  in the $y$ independent contribution for very large $y$ is thus 
$\gamma^{2-r}$, that should be identified with $\gamma^{r}$.
The self consistency possibility is thus $r=1$.

\subsection{Second consistency check on $c(y,\gamma)$}

In the expression (\ref{a4.1}) for $c(y,\gamma)$, in the right hand side, three domains for the $u$
integration variables can be distinguished.
In the first one, the argument of $c(y-u,\gamma)$  and $c(y+u,\gamma)$ are inside the conditions
asssumed for (\ref{a4.2}): $\vert y \vert<<\gamma$.
In another domain, we are in the conditions studied in the preceding section: $\vert y \vert>>\gamma$.
We have also the transition domain where $y$ is of the order of $\gamma$.

We separate anew in (\ref{a4.2}) the $c_{\delta}(y,\gamma)$ and $c_{\cal P}(y,\gamma)$ contributions.
If we consider $y>0$, we have still the form (\ref{a3.11a}):
\begin{eqnarray}
&&c_{\delta}(y,\gamma)=
\frac{1}{y} (\pi i)
\frac1{[1-c(0,\gamma)]}
y^2\frac{\gamma^2}{\gamma^2+y^2}
\propto\frac{\pi iy}{c(0,\gamma)}.
\label{a4.3}
\end{eqnarray}
Assuming the form (\ref{a4.2}) in $c_{\delta}(y,\gamma)$ provides  a behaviour as $y
\gamma^{-\frac12}$ as dominant contribution, much smaller that the assumed (\ref{a4.2})
behaviour.

For the second term $c_{\cal P}(y,\gamma)$, we can use (\ref{a3.12}):
\begin{eqnarray}
&&c_{\cal P}(y,\gamma)
=
-\frac{1}{y} 
\int_0^\infty du\, u^2
\frac{\gamma^2}{\gamma^2+u^2}
\left\{{\cal P}\left(\frac{1}{y-u}\right)
\frac1{[1-c(y-u,\gamma)]}
\right.\nonumber\\&&\left.
+{\cal P}\left(\frac{1}{y+u}\right)
\frac1{[1-c(y+u,\gamma)]}
\right\}. 
\label{a4.4}
\end{eqnarray}
Combining the behaviour (\ref{a4.2}) and the behaviour (\ref{a3.8}) where $r=1$, we assume the
following behaviour for $c_{\cal P}(y,\gamma)$ for large $\gamma$.
\begin{equation}
c_{\cal P}(y,\gamma)=
 \gamma^{\frac12} g(\frac{y}{\gamma})
+\gamma g_1(\frac{y}{\gamma}),
\label{a4.5}
\end{equation}
where the function $g_1$ has the qualitative features of $\frac{y^2}{y^2+(l\gamma)^2}$ ($l$ is a very
large number): it is negligible for $y<<l\gamma$ and becomes 1 in the other limit $y>>l\gamma$.
This expression reproduces qualitatively the behaviour of $c_{\cal P}(y,\gamma)$ for
all values of $y$ with respect to $\gamma$.
To check the consistency, we introduce that expression in (4.4)
 for $y<<\gamma$. We have (1 is negligible in front of $\gamma$)
\begin{eqnarray}
&& \gamma^{\frac12} g(\frac{y}{\gamma})
=
-\frac{1}{y} 
\int_0^\infty du\, u^2
\frac{\gamma^2}{\gamma^2+u^2}
\left\{{\cal P}\left(\frac{1}{y-u}\right)
\frac1{[ \gamma^{\frac12} g(\frac{y-u}{\gamma})
+\gamma g_1(\frac{y-u}{\gamma})]}
\right. \nonumber\\&& \left.
+{\cal P}\left(\frac{1}{y+u}\right)
\frac1{[ \gamma^{\frac12} g(\frac{y+u}{\gamma})
+\gamma g_1(\frac{y+u}{\gamma})]}
\right\}. 
\label{a4.6}
\end{eqnarray}
If we introduce variables $t=\frac{y}{\gamma}$, $v=\frac{u}{\gamma}$, we get
\begin{eqnarray}
&& \gamma^{\frac12} g(t)
=
-\frac{\gamma}{t} 
\int_0^\infty du\, u^2
\frac{1}{1+v^2}
\left\{{\cal P}\left(\frac{1}{t-v}\right)
\frac1{[ \gamma^{\frac12} g(t-v)
+\gamma g_1(t-v)]}
\right. \nonumber\\&& \left.
+{\cal P}\left(\frac{1}{t+v}\right)
\frac1{[ \gamma^{\frac12} g(t+v)
+\gamma g_1(t+v)]}
\right\} 
\label{a4.7}
\end{eqnarray}
The front factor $\gamma^{\frac12}$ can be simplified and we have
\begin{eqnarray}
 &&g(t)
=
-\frac{1}{t} 
\int_0^\infty du\, u^2
\frac{1}{1+v^2}
\left\{{\cal P}\left(\frac{1}{t-v}\right)
\frac1{[  g(t-v)
+\gamma g_1(t-v)]}
\right. \nonumber\\&& \left.
+{\cal P}\left(\frac{1}{t+v}\right)
\frac1{[  g(t+v)
+\gamma g_1(t+v)]}
\right\}. 
\label{a4.8}
\end{eqnarray}
The remaining dependence of $\gamma$ plays a role only for very large values of $t-v$ or 
$t+v$.
Its consequence is to diminish the contribution for that part of the domain of integration.
Since that expression has no problem of convergence due to the factor $\frac{1}{1+v^2}$, that
dependence does not play a sensitive role in the evaluation of $g$.
It can therefore qualitatively be dropped 
 and we can deduce that $c_{\cal P}(y,\gamma)$ behaves as $\gamma^{\frac12}$ for the the relevant
domain of values of its argument.

\bibliographystyle{unsrt}

\end{document}